\documentclass[aps,twocolumn,nofootinbib,amssymb,eqsecnum,epsfig,floatfix,nobibnotes]{revtex4-1}

\usepackage{amssymb,amsfonts,amsmath}
\usepackage{graphicx}
\usepackage{epsfig}
\usepackage{epstopdf}
\usepackage[all]{xy}
\usepackage{bm}
\usepackage{amsthm}
\usepackage[latin1]{inputenc}
\usepackage[english]{babel}
\usepackage{psfrag}

\newcommand{\be}{\begin{equation}}
\newcommand{\ee}{\end{equation}}
\newcommand{\bea}{\begin{eqnarray}}

\newcommand{\eea}{\end{eqnarray}}

\def\inbar{\,\vrule height1.5ex width.4pt depth0pt}
\def\IR{\relax{\rm I\kern-.18em R}}
\def\IC{\relax\hbox{$\inbar\kern-.3em{\rm C}$}}

\begin{document}
\title{The Newtonian and relativistic theory of orbits and the emission of
gravitational waves}

\author{Mariafelicia De Laurentis}
\affiliation{Dipartimento di Scienze Fisiche, Universit\'a di Napoli " Federico II", INFN Sez. di Napoli,\\ Compl. Univ. di
Monte S. Angelo, Edificio G, Via Cinthia, I-80126, Napoli, Italy~}
\email{felicia@na.infn.it}


\maketitle
\tableofcontents

\section{Introduction}
\label{uno}

Dynamics of standard and compact astrophysical bodies
could be aimed to unravelling the information contained in the
gravitational wave (GW) signals. Several situations are possible
such as GWs emitted by  coalescing   compact binaries--systems of
neutron stars (NS), black holes (BH) driven into  coalescence by
emission of gravitational radiation (considered in the
inspiralling, merging and ring-down phases, respectively), hard
stellar encounters, and other high-energy phenomena where GW
emission is expected.  Furthermore the signature of GWs can be
always determined by the relative motion of the sources. In this
review paper, we want to discuss the problem of how the waveform
and the emission of GWs depend on the relative motions in
Newtonian, Relativistic and post-Relativistic regimes as for
example situations where gravitomagnetic corrections have to be
considered in the orbital motion.

As a first consideration, we have to say that the problem of
motion, {\it i.e.} the problem of describing the dynamics of
gravitationally interacting bodies, is the cardinal problem of any
theory of gravity. From the publication of Newton's {\it
Principia} to the beginning of the twentieth century, this problem
has been thoroughly investigated within the framework of Newton's
dynamics. This approach  led to the formulation of many concepts
and theoretical tools which have been applied to other fields of
physics. As a consequence, the relationship between Einstein's and
Newton's theories of gravity has been, and still is, very
peculiar. On the one hand, from a theoretical point of view, the
existence of Newton's theory facilitated the early development of
Einstein's theory by suggesting an approximation method (called
post-Newtonian (PN)) which allowed  to draw very soon some
observational consequences of General Relativity (GR). Indeed, the
PN approximation method, developed by Einstein himself
\cite{einstein}, Droste and de Sitter \cite{droste,desitter}
within one year after the publication of GR,  led to the
predictions of $i)$ the relativistic advance of  perihelion of
planets, $ii)$ the gravitational redshift, $iii)$ the deflection
of light, $iv)$ the relativistic precession of the Moon orbit,
that are the so--called "classical" tests of GR.

On the other hand, as emphasized  by Eisenstaedt
\cite{eisenstaedt}, the use of
 PN approximation method has had, from a conceptual point of view,
the adverse side-effect of introducing implicitly a
'neo-Newtonian' interpretation of GR. Indeed, technically this
approximation method follows  the Newtonian way of tackling
gravitational problems as closely as possible. But this technical
reduction of Einstein's theory into the Procrustean bed of
Newton's theory surreptitiously entails a corresponding conceptual
reduction: the Einstenian problem of motion is conceived within
the Newtonian framework of an "absolute" coordinate space and an
"absolute" coordinate time. However, some recent developments
oblige us to reconsider the problem of motion within Einstein's
theory. On the other hand, the discovery of the binary pulsar PSR
1913+16 by Hulse and Taylor in 1974 \cite{hulse}, and its
continuous observation by Taylor and coworkers (see references
\cite{taylor,weisberg}), led to an impressively accurate tracking
of the  orbital motion of a NS in a binary system. This means that
it is  worth reconsidering in detail, i.e. at its foundation, the
problem of motion also in relation to the problem of generation
and detection of GWs. In other words, the motion of sources could
give further signatures to GWs and then it has to be carefully
reconsidered.
 
 The first part of this review paper is devoted to the theory of
orbits. The most natural way to undertake this task is starting
with the discussion of the Newtonian problem of motion then we
consider the relativistic problem of motion, in particular the PN
approximation and the further gravitomagnetic corrections.

The theory of orbits can be connected to GWs since studies of
binary systems prove, beyond  reasonable doubts, that such a form
of radiation has to exist. Detecting the waves directly and
exploiting  them could result  a very impressive way to study
astrophysical objects.  In other words, the detection of GWs could
give rise to the so-called {\it Gravitational Astronomy}.

In view of this achievement, it is relevant to stress that GW
science has entered a new era. Experimentally~\footnote{GW
experiments started with the pioneering work of Joseph Weber at
Maryland in the 60s}, several ground-based laser-interferometer GW
detectors ($ 10 \mbox{--} 1$ kHz) have been built in the United
States (LIGO)~\cite{LIGO}, Europe (VIRGO and GEO)
~\cite{VIRGO,GEO} and Japan (TAMA)~\cite{TAMA}, and are now taking
data at designed sensitivity.

 A laser
interferometer space antenna (LISA)~\cite{LISA}
($10^{-4} \mbox{--} 10^{-2}$ Hz) might fly within
the next decade. 

From a theoretical point of view,  last years have witnessed
numerous major advances. Concerning the most promising GW sources
for ground-based and space-based detectors, i.e. binary systems
composed of NS, BHs, our understanding of the relativistic
two-body problem, and the consequent GW-generation problem, has
improved significantly.

Knowledge has also progressed on the problem of motion of a point
particle in curved spacetime when the emission of GWs is taken
into account (non-geodesic motion)~\cite{RR,others}. Solving this
problem is of considerable importance for predicting very accurate
waveforms emitted by extreme mass-ratio binaries, which are among
the most promising sources for LISA~\cite{emri}.

The GW community working at the interface between
the theory and the experiment has provided
{\it templates}~\cite{templates,DIS98,EOB} for binaries
and developed robust algorithms~\cite{DA,algorithms} for
pulsars and other GW-sources observable
with ground-based and space-based interferometers. The joined work of
data analysts and experimentalists has established
astrophysically significant upper limits for several GW sources
~\cite{lsc,lscpulsar,lscstoch} and
is now eagerly waiting for the first detection.

In this research framework, searching for  criteria to classify
how sources collide and interact is of fundamental importance.  A
first rough criterion can be the classification of stellar
encounters in {\it collisional}, as in the globular clusters, and
in {\it collisionless}  as in the galaxies \cite{binney}.  A
fundamental parameter is the richness and the density of the
stellar system and then, obviously, we expect a large production
of GWs in rich and dense systems.

Systems with these features are the globular clusters and the
galaxy centers. In particular, one can take into account  the
stars (early-type and late-type) which are around our Galactic
Center, e.g. Sagittarius $A^{*}$ ($Sgr A^{*}$) which could be very
interesting targets for the above mentioned ground-based and
space-based detectors.

In recent years, detailed information has been achieved for
kinematics and dynamics of stars moving in the gravitational field
of such a central object. The statistical properties of spatial
and kinematical distributions are of particular interest (see e.g.
\cite{Genzel,Sellgreen,CapozIov}). Using them, it is possible to
give a quite accurate estimate of the mass and the size of the
central object: we have  $(2.61\pm0.76)\times10^6M_{\odot}$
concentrated within a radius of $0.016 pc$ (about $30$
light-days)\cite{Ghez,Thatte}. More precisely, in \cite{Ghez}, it
is described a campaign of observations where velocity
measurements in the central $arcsec^{2}$ are extremely accurate.
Then  from this bulk of data, considering  a field of resolved
stars whose proper motions are accurately known, one can classify
orbital motions and deduce, in principle,  the rate of production
of GWs according to the different types of orbits.

These issues motivate this review paper in which, by a
classification of orbits in accordance with the conditions of
motion, we want to calculate the GW luminosity for  different
types of stellar encounters and orbits (see also \cite{CDDIN,SF}).

Following the method outlined in  \cite{pm1,pm2}, we investigate
the GW emission by  binary systems  in the quadrupole
approximation considering bounded   (circular or elliptical) and
unbounded  (parabolic or hyperbolic) orbits. Obviously, the main
parameter is the approaching energy of the stars in the system
(see also \cite{schutz} and references therein).  We expect that
gravitational waves are emitted with a "peculiar" signature
related to the encounter-type: such a signature has to be a
"burst" wave-form with a maximum in correspondence of the
periastron distance. The problem of {\it bremsstrahlung}-like
gravitational wave emission has been studied in detail by Kovacs
and Thorne \cite{kt} by considering stars interacting on unbounded
and bounded orbits. In this review paper, we face this problem discussing
in detail the dynamics of such a phenomenon which could greatly
improve the statistics of possible GW sources.
For further
details see also
\cite{landau,gravitation,weinberg,BW,BS,SC,maggiore, KT87,BA,M00,CT02,AB03,FH05,KTcaltech}. 

The review is organized as follows. In Part I, as we said, we
discuss the theory of orbits. In Sec.2, we start with the
Newtonian theory of orbits and discuss the main features of
stellar encounters by classifying trajectories. Sec.3 is devoted
to orbits with relativistic corrections.  A method for solving the
equations of motion of  binary systems at the first
PN-approximation is reviewed. The aim is to express the solution
in a quasi-Newtonian form. In the Sec.4, we study  higher order
relativistic corrections to the orbital motion  considering
gravitomagnetic effects. We discuss in details how such
corrections come out by taking into account  "magnetic" components
in the weak field limit of gravitational field. Finally, the
orbital structure and the stability conditions are discussed
giving numerical examples. Beside the standard periastron
corrections of GR, a  new nutation effect have  to be considered
thanks to ${\displaystyle c^{-3}}$ corrections. The transition to
a chaotic behavior strictly depends on the initial conditions. The
orbital phase space portrait is discussed.

Part II is devoted to the production and signature of
gravitational waves. We start, in Sec.5, by deriving the wave
equation in linearized gravity and discuss the gauge properties of
GWs. Sec.6 is devoted to the problems of  generation, emission and
interaction of GWs with a detector. In Sect.7, we discuss the
problem of GW luminosity and emission  from binary systems moving
along Newtonian orbits. The quadrupole approximation is assumed.
Sect.8 is devoted to the same problem taking into account
relativistic motion. In Sect.9,  also gravitomagnetic effects on
the orbits and the emission are considered. In Sect.10, as an
outstanding application of the above considerations, we derive the
expected rate of events from the Galactic Center. Due to the
peculiar structure of this stellar system, it can be considered a
privileged  target from where GWs could be detected and
classified. Discussion, concluding remarks and perspectives are
given in Sect.11.

\part{\large Theory of orbits}

\section{Newtonian orbits}
\label{due}

We want  to describe, as accurately as possible, the dynamics of a
system of two bodies, gravitationally interacting, each one having
finite dimensions. Each body exerts a conservative, central force
on the other and no other external forces are considered assuming
the system  as isolated from the rest of the universe. Then, we
first take into account the non-relativistic theory of orbits
since stellar systems, also if at high densities and constituted
by compact objects, can be usually assumed in Newtonian regime. In
most cases, the real situation is more complicated. Nevertheless,
in all cases, it is an excellent starting approximation to treat
the two bodies of interest as being isolated from  outside
interactions. We give here a self-contained summary of the
well-known orbital types \cite{binney,landau} which will be
extremely useful for the further discussion.
\subsection{Equations of motion and conservation laws}
\label{uno1}
Newton's equations of motion for two particles  of masses $m_1$
and $m_2$, located at ${\bf r_1}$ and ${\bf r_2}$, respectively,
and interacting by gravitational attraction are, in the absence of
external forces,

\begin{eqnarray}
\frac{d{\bf p_1}}{dt} &=& -G\frac{m_1 m_2}{|{\bf r_1}-{\bf r_2}|^3}({\bf r_1}-{\bf r_2})\, ,\nonumber\\
\frac{d{\bf p_2}}{dt} &=& +G\frac{m_1 m_2}{|{\bf r_1}-{\bf
r_2}|^3}({\bf r_1}-{\bf r_2})\, , \label{1}\end{eqnarray}
where $\displaystyle{{\bf p_i}=m_{i}\frac{\bf dr_{i}}{\bf dt}}$
is the momentum of particle $i$,  $ (i = 1,2)$, and $G$ is
Newtonian gravitational constant.
\begin{displaymath}
\frac{d}{dt}({\bf p_1}+{\bf p_2})=0\, ,\nonumber\\
\end{displaymath}
or, with ${\bf P}={\bf p_1}+{\bf p_2}$ denoting the total momentum of the two body system,
\begin{displaymath}
{\bf P}=const\,.
\end{displaymath}

\begin{figure}[ht]
\includegraphics[scale=0.4]{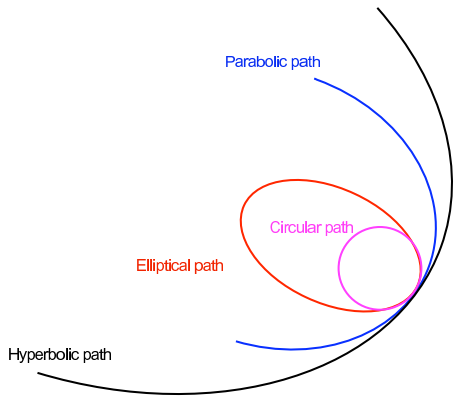}
\caption{Newtonian paths: in black line we have hyperbolic path,
in blue line we have parabolic path, in red line the elliptical
path and in ciano the circular path} \label{fig:orbits}
\end{figure}

Thus we have found a  first {\it conservation law}, namely the conservation of the total momentum
of a two-body system in the absence of external forces. We can make use of this by carrying
out a Galilei transformation to another inertial frame in which the total momentum is equal to
zero. Indeed, let us apply the transformation
\begin{displaymath}
{\bf r_i}\rightarrow{\bf r'_i}={\bf r_i}-{\bf v}t,\qquad i=1,2
\end{displaymath}
hence $\displaystyle{{\bf p_i}\rightarrow {\bf p'_i}={\bf p_i}-m_i{\bf v}}$ and hence, with $M = m_1 + m_2$,
\begin{displaymath}
{\bf P}\rightarrow{\bf P'}={\bf P}-M{\bf v}\, ,\nonumber\\
\end{displaymath}
and if we choose $\displaystyle{{\bf v} = \frac{\bf P}{M}}$, then
the total momentum is equal to zero in the primed frame. We also
note that the gravitational force is invariant under the Galilei
transformation, since it depends only on the difference
$\displaystyle{{\bf r_1}-{\bf r_2}}$. Thus let us from now on work
in the primed frame, but drop the primes for convenience of
notation. We can now replace the original equations of motion with
the equivalent ones,
\begin{displaymath}
{\bf P}=0,\qquad\frac{d\bf p}{dt}=-G\frac{m_1 m_2}{r^3}{\bf r}\, ,\nonumber\\
\end{displaymath}
where $\displaystyle{{\bf r}= {\bf r_1}-{\bf r_2}}$,
$\displaystyle{r = |{\bf r}|}$, and $\displaystyle{{\bf p} = {\bf
p_1}-{\bf p_2}}$. Next we introduce the position vector ${\bf R}$
of the center of mass of the system:
\begin{displaymath}
{\bf R}=\frac{m_1{\bf r_1}+m_2{\bf r_2}}{m_1+m_2}\, ,\nonumber\\
\end{displaymath}
hence
\begin{displaymath}
{\bf P}=M\frac{{d\bf R}}{dt}\, ,
\end{displaymath}
and hence from ${\bf P}=0$ we have
\begin{displaymath}
{\bf R}=const\, .
\end{displaymath}
and we can carry out a translation of the origin of our coordinate frame such that ${\bf R}= 0$. The
coordinate frame we have arrived at is called {\it center-of-mass frame} (CMS). We can also see now
that
\begin{displaymath}
{\bf p}={\bf p_1}=m_1\frac{d{\bf r_1}}{dt}=\mu\frac{d{\bf r}}{dt}\, ,
\end{displaymath}
where $\displaystyle{\mu=\frac{m_1 m_2}{m_1+m_2}}$ is the {\it reduced mass} of the system, and hence the equation of
motion can be cast in the form
\begin{equation}
\mu\frac{d^2{\bf r}}{dt^2}=-G\frac{\mu M}{r^2}{\bf \hat {r}}\, ,
\label{10}\end{equation} where we have defined the radial unit
vector $\displaystyle{{\bf \hat {r}}=\frac{{\bf r}}{r}}$. We can
get two more conservation laws if  we take the scalar product of
Eq. (\ref{10}) with $\displaystyle{\frac{d{\bf r}}{dt}}$ and its
vector product with ${\bf r}$. The scalar product with
$\displaystyle{\frac{d{\bf r}}{dt}}$ gives on the left-hand side
\begin{displaymath}
\frac{d{\bf r}}{dt}\cdot\frac{d^2{\bf r}}{dt^2}=\frac{1}{2}\frac{d}{dt}\left(\frac{d{\bf r}}{dt}\right)^2\, ,
\end{displaymath}
and on the right-hand side we have
\begin{displaymath}
\frac{{\bf \hat{r}}}{r^2}\cdot\frac{d{\bf r}}{dt}=-\frac{d}{dt}\left(\frac{1}{r}\right)\, ,
\end{displaymath}
hence
\begin{displaymath}
\frac{d}{dt}\left(\frac{\bf p^2}{2\mu}-\frac{\gamma}{r}\right)=0\, ,
\end{displaymath}
where $\displaystyle{\gamma=G\mu M}$. This implies that the expression in brackets is conserved, {\it i.e.}
\begin{equation}
\frac{\bf p^2}{2\mu}-\frac{\gamma}{r}=E=const\, .
\label{12}\end{equation}
Here the first term is the kinetic energy, the second term is the potential energy, and the sum of
kinetic energy and potential energy is the total energy $E$, which is a constant of motion. Now
take the cross product of Eq. (\ref{10}) with ${\bf r}$: on the right-hand side, we get the cross product of
collinear vectors, which is equal to zero, hence
\begin{displaymath}
{\bf r}\times \mu\frac{d}{dt}\left({\bf r}\times\frac{d{\bf r}}{dt}\right)=\frac{d}{dt}({\bf r }\times {\bf p})=0\, ,
\end{displaymath}
and hence, if we define the {\it angular momentum} ${\bf L}$ by
\begin{displaymath}
{\bf L}={\bf r}\times{\bf p}\, ,
\end{displaymath}
we get the result
\begin{displaymath}
\frac{d{\bf L}}{dt}=0\, ,
\end{displaymath}
or
\begin{displaymath}
{\bf L}=const\, ,
\end{displaymath}
{\it i.e.} conservation of angular momentum. An immediate consequence of this conservation law is
that the radius vector ${\bf r}$ always stays in one plane, namely the plane perpendicular to ${\bf L}$. This
implies that we can without loss of generality choose this plane as the $(xy)$ coordinate plane.
The vector  ${\bf r}$ is then a two-dimensional vector,
\begin{displaymath}
{\bf r}=(x,y)=r {\bf \hat{r}}. \qquad {\bf \hat{r}}=(\cos\phi,\sin\phi)\, ,
\end{displaymath}
where we have defined the polar angle $\phi$. With this notation we can express the magnitude of
angular momentum as
\begin{equation}
L=r^2\frac{d\phi}{dt} \label{eq:momang1}\, ,
\end{equation}
The conservation of angular momentum can be used to simplify the equation of motion
(\ref{12}). To do this we note that
\begin{displaymath}
{\bf L}^2=({\bf r}\times{\bf p})^2={\bf r^2} {\bf p^2}-({\bf r}\cdot {\bf p})^2\, ,
\end{displaymath}
hence
\begin{displaymath}
{\bf p^2}=\frac{{\bf L^2}}{r^2}+p_{r}^{2}\, ,
\end{displaymath}
where $\displaystyle{p_r={\bf \hat{r}}\cdot  {\bf p}}$ is the radial component of momentum. Substituting into Eq. (\ref{12}) then gives
\begin{displaymath}
\frac{p_{r}^{2}}{2\mu}+\frac{{\bf L^2}}{2\mu r^2}-\frac{\gamma}{r}=E\, ,
\end{displaymath}
or, with $\displaystyle{p_r=\frac \mu {dr}{dt}}$,
\begin{equation}
\frac{1}{2}{\mu\left(\frac{dr}{dt}\right)}^{2}+\frac{\mathbb{L}^{2}}{2\mu
r^{2}}-\frac{\gamma}{r}=E\, . \label{eq:energia}\end{equation}
Looking back at our starting point,  Eq. (\ref{1}), we reduce the
dimensionality of our problem: from the simultaneous differential
equations of six functions of time, namely the six components of
the position vectors ${\bf r_1}$ and ${\bf r_2}$, we reduce to a
pair of simultaneous differential equations for the polar
coordinates $r(t)$ and $\phi(t)$  these equations contain two
constants of motion, the total energy $E$ and angular momentum
$L$. Then a mass $m_1$ is moving in the gravitational potential
$\Phi$ generated by a second mass $m_2$. The  vector radius and
the polar angle depend on the time as a consequence of the star
motion, i.e. $\textbf{r}=\textbf{r}(t)$ and $\phi=\phi(t)$. With
this choice, the velocity $\textbf{v}$ of the mass $m_1$ can be
parameterized as
\begin{displaymath}
\textbf{v}=v_r\widehat{r}+v_{\phi}\widehat{\phi}~,
\end{displaymath}
where the radial and tangent components of the velocity are,
respectively,
\begin{displaymath}
v_r=\frac{dr}{dt}\, , ~~~~~~~~v_{\phi}=r \frac{d\phi}{dt}~.
\end{displaymath}
We can split the kinetic energy into two terms where, due to the
conservation of angular momentum, the second one is a function of
$r$ only. An effective potential energy $V_{eff}$,
\begin{displaymath}
V_{eff}=\frac{\mathbb{L}^{2}}{2\mu r^{2}}-\frac{\gamma}{r}\, ,\label{eq:energpot}
\end{displaymath}
is immediately defined. The first  term corresponds to a repulsive
force, called the angular momentum barrier. The second term is the
gravitational attraction. The interplay between attraction and
repulsion is such that the effective potential energy has a
minimum. Indeed, differentiating with respect to $r$ one finds
that the minimum lies at $\displaystyle{r_0=\frac{L^{2}}{\gamma\mu}}$ and that
\begin{displaymath}
V_{eff}^{min}=-\frac{\mu\gamma^{2}}{2L^{2}}\label{eq:enrgpotmin}\,.\end{displaymath}
Therefore, since the radial part of kinetic energy,
\begin{displaymath}
K_{r}=\frac{1}{2}\mu\left(\frac{dr}{dt}\right)^{2}\, ,
\end{displaymath},
is non-negative, the total energy must be not less than
$V_{eff}^{min}$, i.e.
\begin{displaymath}
E\geq
E_{min}=-\frac{\mu\gamma^{2}}{2L^{2}}\label{eq:emin}\,.\end{displaymath}
The equal sign corresponds to the radial motion. For
$E_{min}<E<0$, the trajectory lies between a smallest value
$r_{min}$ and greatest value $r_{max}$ which can be found from the
condition $E=V_{eff}$, i.e.
\begin{displaymath}
r_{\{min,max\}}=-\frac{\gamma}{2E}\pm\sqrt{\left(\frac{\gamma}{2E}\right)^{2}+\frac{L^{2}}{2\mu
E}}\, ,\label{eq:rminmax}\end{displaymath} where the upper (lower) sign
corresponds to $r_{max}$ ($r_{min}$). Only for $E>0$, the upper
sign gives an acceptable value; the second root is negative and
must be rejected.
Let us now proceed in solving the  differential equations (\ref{eq:momang1})
 and (\ref{eq:energia}). We have
 \begin{equation}
\frac{dr}{dt}=\frac{dr}{d\phi}\frac{d\phi}{dt}=\frac{L}{\mu r^{2}}\frac{dr}{d\phi}=
-\frac{L}{\mu}\frac{d}{d\phi}\left(\frac{1}{r}\right)\label{eq:diff}\, ,\end{equation}

and  defining, as standard,  the auxiliary variable $u=1/r$,  Eq.
(\ref{eq:energia}) takes the form
\begin{equation}
u'^{2}+u^{2}-\frac{2\gamma\mu}{L^{2}}u=\frac{2\mu E}{L^{2}}\, ,\label{eq:diffenerg}\end{equation}

where $\displaystyle{u'=du/d\phi}$ and we have divided by $\displaystyle{L^{2}/2\mu}$. Differentiating
with respect to $\phi$, we get

\begin{displaymath}
u'\left(u''+u-\frac{\gamma\mu}{L^{2}}\right)=0\, ,\end{displaymath}

hence either $u'=0$, corresponding to the circular motion, or

\begin{equation}
u''+u=\frac{\gamma\mu}{L^{2}}\, ,\label{eq:moto}\end{equation}

which has the solution
\begin{displaymath}
u=\frac{\gamma\mu}{L^{2}}+C\cos\left(\phi+\alpha\right)\, ,
\end{displaymath}

or,  reverting the variable,

\begin{equation}
r=\left[\frac{\gamma\mu}{L^{2}}+C\cos\left(\phi+\alpha\right)\right]^{-1}\, ,\label{eq:solution}\end{equation}
which is the canonical form of conic sections in polar coordinates
\cite{smart}. The constant $C$ and $\alpha$ are two integration
constants of the second order differential equation
(\ref{eq:moto}). The solution (\ref{eq:solution}) must satisfy the
first order differential equation (\ref{eq:diffenerg}).
Substituting (\ref{eq:solution}) into (\ref{eq:diffenerg}) we
find, after a few algebra,

\begin{equation}
C^{2}=\frac{2\mu E}{L^{2}}+\left(\frac{\gamma\mu}{L^{2}}\right)^{2}\, ,\label{eq:C}
\end{equation}

and therefore, taking account of Eq. (\ref{eq:emin}), we get
$C^{2}\geq 0$. This implies the four kinds of orbits given in
Table I and in Fig. \ref{fig:orbits}.

\begin{table}
\begin{center}
\begin{tabular}{|c|c|c|cl}
\hline $C=0$& $E=E_{min}$& circular orbits\tabularnewline \hline
$0<\left|C\right|<\frac{\gamma\mu}{L^{2}}$& $E_{min}<E<0$&
elliptic orbits\tabularnewline \hline
$\left|C\right|=\frac{\gamma\mu}{L^{2}}$& $E=0$& parabolic
orbits\tabularnewline \hline
$\left|C\right|>\frac{\gamma\mu}{L^{2}}$& $E>0$,& hyperbolic
orbits\tabularnewline \hline
\end{tabular}
\end{center}
\caption{Orbits in Newtonian regime classified by the approaching
energy.}
\end{table}
\subsection{Circular Orbits}
\label{uno2}

{\bf Circular motion} corresponds to the condition $u'=0$ from which one
find $r_{0}=L^{2}/\mu\gamma$ where $V_{eff}$ has its minimum. We
also note that the expression for $r_{0}$ together with
Eq.(\ref{eq:emin}) gives

\begin{equation}
r_{0}=-\frac{\gamma}{2E_{min}}\, .\label{eq:rzero}\end{equation}

Thus the two bodies move in concentric circles with radii, inversely
proportional to their masses and are always in opposition.

\subsection{Elliptical  Orbits}
\label{uno3}

For $0<\left|C\right|<\mu\gamma/L^{2}$, $r$ remains finite for all
values of $\phi$. Since  $r(\phi+2\pi)=r(\phi)$, the trajectory is
closed and it is an {\bf ellipse}. If one chooses $\alpha=0$,  the major
axis of the ellipse corresponds to $\phi=0$. We get

\begin{displaymath}
r_{\left|\phi=0\right.}=r_{min}=\left[\frac{\gamma\mu}{L^{2}}+C\right]^{-1}\, ,\label{eq:rphi}\end{displaymath}

and

\begin{displaymath}
r_{\left|\phi=\pi\right.}=r_{max}=\left[\frac{\gamma\mu}{L^{2}}-C\right]^{-1}\, ,\label{eq:rpi}\end{displaymath}

and since $r_{max}+r_{min}=2a$, where $a$ is the semi-major axis
of the ellipse, one obtains

\begin{displaymath}
a=r_{\left|\phi=0\right.}=r_{min}=\frac{\gamma\mu}{L^{2}}\left[\left(\frac{\gamma\mu}{L^{2}}\right)^{2}+C^{2}\right]^{-1}\, ,\end{displaymath}

$C$ can be eliminated from the latter equation and Eq. (
\ref{eq:C}) and then

\begin{equation}
a=-\frac{\gamma}{2E}\, ,\label{eq:a}\end{equation}

Furthermore, if we denote the distance
$r_{\left|\phi=\pi/2\right.}$ by $l$, the so-called {\it
semi-latus rectum} or the parameter of the ellipse,  we get

\begin{equation}
l=\frac{L^{2}}{\gamma\mu}\, ,\label{eq:latusrectum}\end{equation}

and hence the equation of the trajectory

\begin{equation}
r=\frac{l}{1+\epsilon\cos\phi}\, ,\label{eq:traiettoria}\end{equation}

where $\displaystyle{\epsilon=\sqrt{\frac{1-l}{a}}}$ is the
eccentricity of the ellipse. If we consider the major  semiaxis of
orbit $a$ Eq. (\ref{eq:a}) and the eccentric anomaly ${\cal E}$,
the orbit can be written also as (see \cite{roy})
\begin{displaymath}\label{Ellisse}
r = a (1 - \epsilon \cos{\cal E})\, ,
\end{displaymath}
 this equation, known as Kepler's equation, is transcendental
in ${\cal E}$, and the solution for this quantity cannot expressed
in a finite numbers of terms. Hence, there is the following
relation between the eccentric anomaly and the angle $\phi$:
\begin{displaymath}\label{camb_E}
\cos \phi\,=\,\frac{\cos {\cal E} - \epsilon}{1-\epsilon \cos {\cal E}}\, .
\end{displaymath}

\subsection{Parabolic and Hyperbolic Orbits}
\label{uno4}
These solutions can be dealt together. They correspond to $E\geq
0$ which is the condition to obtain unbounded orbits.
Equivalently, one has $\left|C\right|\geq\gamma\mu/L^{2}$.

The trajectory is
\begin{equation}
r=l\left(1+\epsilon\cos\phi\right)^{-1}\, ,\label{eq:traie}\end{equation}

where $\epsilon\geq1$. The equal sign corresponds to $E=0$ .
Therefore, in order to ensure positivity of $r$, the polar angle
$\phi$ has to be restricted to the range given by

\begin{displaymath}
1+\epsilon\cos\phi>0 \label{eq:cosphi}\, .\end{displaymath}

This means $\cos\phi>-1$, i.e. $\phi\in(-\pi,\pi)$ and the
trajectory is  not closed any more. For $\phi\rightarrow\pm\pi$,
we have $r\rightarrow\infty$. The curve (\ref{eq:traie}), with
$\epsilon=1$, is a {\bf parabola}. For  $\epsilon>1$, the allowed
interval of polar angles is smaller than $\phi\in(-\pi,\pi)$, and
the trajectory is a {\bf hyperbola}. Such  trajectories correspond to
non-returning objects.
Let us consider a  semi-axis $a$ and ${\cal F}$ as
variable, analogous to the elliptic eccentric anomaly ${\cal E}$. The
 hyperbolic orbit is defined also by
\begin{displaymath}\label{Iperbole}
r = a(\epsilon \cosh {\cal F} - 1)\, ,
\end{displaymath}
hence, there is the following relation between $F$ and the angle
$\phi$:
\begin{displaymath} \label{camb_F}
\cos \phi\,=\,\frac{l-a(\epsilon \cosh {\cal F} - 1)}{\epsilon
a (\epsilon \cosh {\cal F} - 1)}\, .
\end{displaymath}
Finally  the parabolic orbit can be defined by the another
relation (see \cite{roy})
\begin{displaymath}\label{Parabola}
r \,=\, \frac{P^2}{2}\, ,
\end{displaymath}
where P is a parameter. In this case
\begin{displaymath} \label{camb_P}
\cos \phi \,=\, \frac{2 l - P^2}{P^2}\, .
\end{displaymath}
As we will discuss below, this classification of orbital motions
can reveal extremely useful for the waveform signature  of
gravitational radiation. Let us now take into account the
relativistic theory of orbits.

\section{Relativistic Orbits}
\label{tre}
 As we have seen in the above Section, the  non-relativistic two bodies problem consists in two sub-problems:
 \begin{enumerate}
 \item deriving the equations of orbital motion for two gravitationally interacting extended bodies, 
 \item solving these equations of motion.
 \end{enumerate}
In the  case of widely separated objects, one can simplify the
sub-problem by neglecting the contribution of the quadrupole and
higher multipole momenta of the bodies to their external
gravitational field, thereby approximating the equations of
orbital motion of two extended bodies by the equations of motion
of two point masses located at the Newtonian center of mass of the
extended objects. Then the sub-problem can be exactly solved as
shown in the above Section. The two body problem in GR is more
complicated: because of the non-linear hyperbolic structure of
Einstein's field equations, one is not sure of the good
formulation of boundary conditions at infinity, so that the
problem is not even well posed \cite{ehlers}. Moreover, since in
Einstein's theory the local equations of motion are contained in
the gravitational field equations, it is {\it a priori} difficult
to separate the problem in two sub-problems, as in the
non-relativistic case, where one can compute the gravitational
field as a linear functional of the matter distribution
independently of its motion. Furthermore, even when one can
achieve such separation and derive some equations of orbital
motion for the two bodies, these equations will {\it a priori} not
be ordinary differential equations but, because of the finite
velocity of propagation of gravity, will consist in some kind of
retarded integro-differential system \cite{deruelle}. However, all
these difficulties can be somehow dealt with if one resorts to
approximation procedures and breaks the general covariance by
selecting special classes of coordinates systems \cite{dixon}.

Two physically different situations, amenable to perturbation
treatments, have been considered in the literature:
   \begin{enumerate}
   \item the problem of two {\it weakly self-gravitating, slowly moving, widely separated}
   fluid bodies which has been treated by the so-called PN approximation schemes
   (for references see \cite{breuer,caporali,spyrou,maggiore,gravitation}),
   \item the problem of two {\it strongly self-gravitating, widely separated}
   bodies which has been treated by matching a strong field "internal" approximation
    scheme in and near the objects to a weak field "external" approximations scheme outside the objects.
    \end{enumerate}

The approach has been pursued both for slowly moving objects,
either BHs \cite{death}  or in general strongly self-gravitating
objects \cite{kates}, and for strongly self-gravitating objects
moving with arbitrary velocities \cite{damour,bel}. In the latter
case,  equations of orbital motion were considered  in the form of
a retarded-integro-differential system which however could  be
transformed into ordinary differential equations and which, when
attention was restricted to slowly moving bodies, were expanded in
power series of $\displaystyle{\frac{v}{c}}$ \cite{DD,damour}.
When keeping only the first relativistic corrections to Newton's
law (first post-Newtonian approximation), it turns out that the
equations of orbital motion of widely separated, slowly moving,
{\it strongly} self-gravitating objects depend only on two
parameters (the Schwarzschild masses) and are identical to the
equations of motion of weakly self-gravitating objects (when
using, in both cases, a coordinate system which is harmonic at
lowest order). This is, in fact, a non-trivial consequence of the
structure of Einstein's theory \cite{deruelle}. Then, in the next
subsections, we consider the PN motion including secular and
periodic effects at first order approximation and we shall show
that the equations of motion can be written in a quasi-Newtonian
form.

\subsection{Relativistic Motion and conservation laws}
The relativistic case can be seen as a correction to the Newtonian
theory of orbits \cite{deruelle,DD}. In GR, the time is incorporate as a mathematical
dimension, so that the four-dimensional rectangular perifocal
coordinates are $(x,y,z,t)$ and the four dimensional polar
coordinates are $(r,\theta,\phi,t) $. The first post Newtonian
equations of orbital motion of a binary system constrain the
evolution in coordinate time $t$ of the positions ${\bf r_1}$ and
${\bf r_2}$ of the two objects. These positions represent the
center of mass in the case of weakly self-gravitating objects (see
e.g.\cite{spyrou,deruelle}) and the center of field in the case of
strongly self gravitating objects (see \cite{damour}). They can be
derived from Lagrangian which is function of the {\it
simultaneous} position ${\bf r_1}(t),{\bf r_2}(t)$, and velocities
$\displaystyle{{\bf v_1}(t)=\frac{d{\bf r}}{dt}}$ and
$\displaystyle{{\bf v_2}(t)=\frac{d{\bf r_2}}{dt}}$  in a given
harmonic coordinate system, and of two constant parameters, the
Schwarzschild masses of the objects $m_1$ and $m_2$:

 \begin{equation}
 {\cal L}_{PN}\left({\bf r_1}(t),{\bf r_2}(t),{\bf v_1}(t),{\bf v_2}(t)\right)={\cal L}_N+\frac{1}{c^2}{\cal L}_2\, ,
\label{2.1a} \end{equation}
 with
  \begin{equation}
 {\cal L}_N=\frac{1}{2}m_1v_1^2+\frac{1}{2}m_2v_2^2+\frac{Gm_1 m_2}{R}\, ,
 \label{2.1b} \end{equation}
 and
 \begin{eqnarray}
 {\cal L}_2 &=& \frac{1}{8}m_1v_1^4+\frac{1}{8}m_2v_2^4+\nonumber\\ && +\frac{Gm_1 m_2}{2R}\left[3(v_1^2+v_2^2)-7(v_1v_2)-N^2v_1v_2-G\frac{M}{R}\right]\,,\nonumber\\ 
 \label{2.1c} \end{eqnarray}
 
 where we have introduced the  instantaneous relative position
vector ${\bf R}={\bf r_1}-{\bf r_2}$ and $R=|{\bf R}|$ while
$\displaystyle{{\bf N}=\frac{{\bf R}}{R}}$. In (\ref{2.1a}) and
(\ref{2.1b})  we used the short notations: $\displaystyle{{\bf
v_1}\cdot {\bf v_1}= |{\bf v_1}|=v_1^2}$, $\displaystyle{{\bf
v_1}\cdot {\bf v_2}= v_1v_2}$ for the ordinary Euclidean scalar
products, and $c$ is the velocity of light. The invariance, at the
PN approximation, and modulo an exact time derivative, of $ {\cal
L}_{PN}$ under spatial translations and Lorentz boosts implies,
via Noether's theorem, the conservation of the total linear
momentum of the system:

 \begin{displaymath}
  {\bf P}_{PN}=\frac{\partial  {\cal L}_{PN}}{\partial {\bf v_1}}+\frac{\partial  {\cal L}_{PN}}{\partial {\bf v_2}}\,,
 \label{2.2} \end{displaymath}
 and of the relativistic center of mass integral
 
  \begin{eqnarray*}
 {\bf K}_{PN} & = & {\bf G}_{PN}-t{\bf P}_{PN}\, ,\\
 {\bf G}_{PN} & = &\sum \left(m_1+\frac{1}{2}\frac{m_1v^2}{c^2}-\frac{1}{2}\frac{Gm_1m_2}{R c^2}\right){\bf r}\, ,\label{2.3ab}
 \end{eqnarray*}
 the sum is over the two objects \cite{spyrou,DD}.
 By a Poincar\'e transformation it is possible to get a PN center of mass frame where $\displaystyle{{\bf P}_{PN}= {\bf K}_{PN}= 0}$. In this frame one has:
  \begin{eqnarray}
  {\bf r_1} & = & \frac{\mu}{m_1}{\bf R}+\frac{\mu(m_1-m_2)}{2M^2c^2}\left(V^2-\frac{GM}{R}\right){\bf R}\, ,\nonumber\\
  {\bf r_2} & = & -\frac{\mu}{m_2}{\bf R}+\frac{\mu(m_1-m_2)}{2M^2c^2}\left(V^2-\frac{GM}{R}\right){\bf R}\, ,\nonumber\\ \label{2.4ab}
 \end{eqnarray}
 where $\displaystyle{{\bf V}=\frac{d{\bf R}}{dt}={\bf v_1}-{\bf v_2}}$ is the istantaneous relative velocity. The problem of solving the motion of the binary system is then reduced to the simpler problem of solving the relative motion in the PN center of mass frame. For the sake of completeness, let us write down these equations of motion derived from (\ref{2.1a})-(\ref{2.1c}), and where, after variation, the positions and velocities are replaced by their center of mass expressions (\ref{2.4ab}):

 \begin{eqnarray}
&& \frac{d{\bf V}}{dt}= \frac{G M}{2 c^2 R^3}\left[4 G M {\bf N} (\nu +2)-
 R \left(2 {\bf N}  c^2+\right.\right.\nonumber\\ &&+4 (N V) {\bf V}  (\nu -2)-\left.\left.3 (NV)^2 {\bf N}
    \nu +2 {\bf N}  V^2 (3 \nu +1)\right)\right], \nonumber\\
   \label{2.5}
   \end{eqnarray}

where we have introduced a mass parameter
$\displaystyle{\nu=\frac{\mu}{M}=\frac{m_1m_2}{(m_1+m_2)^2}}$ with
 $\displaystyle{\left(0\leq\nu\leq\frac{1}{4}\right)}$. At this
point it is worth to notice that in spite of the fact that it is in
general incorrect to use, {\it before variation}, in a Lagrangian
a consequence, like Eq.(\ref{2.4ab}), of the equations of motions,
which are obtained only {\it after variation}, it turns out that
the relative motion in PN center of mass frame, Eq. (\ref{2.5}),
can be correctly derived from a Lagrangian obtained by replacing
in the total Lagrangian (divided by $\mu$)
$\displaystyle{\frac{1}{\mu} {\cal L}_{PN}\left({\bf r_1},{\bf
r_2},{\bf v_1}, {\bf v_2}\right)}$ the positions and velocities by
their PN center of mass expressions obtained from (\ref{2.4ab}) and
that moreover it is  even sufficient to use the {\it
non-relativistic} center of mass expressions:

  \begin{eqnarray}
 {\bf r_1}_N & = &\frac{\mu}{m_1}{\bf R}\, ,\nonumber\\
 {\bf r_2}_N & = &\frac{\mu}{m_2}{\bf R}\, ,\nonumber\\
 {\bf v_1}_N & = &\frac{\mu}{m_1}{\bf V}\, ,\nonumber\\
 {\bf v_2}_N & = &\frac{\mu}{m_2}{\bf V}\, .\nonumber\\
 \label{2.6abcd}
  \end{eqnarray}
  The proof goes as follows \cite{deruelle}.
 Let us introduce the following linear change spatial variables in the PN Lagrangian $\displaystyle{{\cal L}_{PN}\left( {\bf r_1}- {\bf r_2},\frac{ {\bf r_1}}{dt},\frac{ {\bf r_2}}{dt}\right)}$ :
  $\displaystyle{( {\bf r_1}, {\bf r_2})\rightarrow({\bf R},{\bf X})}$ with $\displaystyle{{\bf R}={\bf r_1}-{\bf r_2}}$ and $\displaystyle{{\bf X}=\frac{(m_1{\bf r_1}+m_2{\bf r_2})}{M}}$, that is:
 \begin{eqnarray*}
 {\bf r_1} & =&  {\bf r_1}_N+{\bf X}\, ,\\
  {\bf r_2} & =&  {\bf r_2}_N+{\bf X}\, ,\\
  \label{2.7ab} \end{eqnarray*}
  which implies (denoting $\displaystyle{\frac{d{\bf X}}{dt}={\bf W}}$):
   \begin{eqnarray*}
 {\bf v_1} & =&  {\bf v_1}_N+{\bf W}\, ,\\
  {\bf v_2} & =&  {\bf v_2}_N+{\bf W}\, .\\
  \label{2.7cd} \end{eqnarray*}
Expressing 
   \begin{eqnarray*}
{\cal L}_{PN}={\cal L}_{N}\left(  {\bf r_1}- {\bf r_2}, {\bf v_1}, {\bf v_2}\right)+\left(\frac{1}{c^2}\right) {\cal L}_{2}\left(  {\bf r_1}- {\bf r_2}, {\bf v_1}, {\bf v_2}\right),
 \end{eqnarray*}
given by Eq. (\ref{2.1a})-(\ref{2.1c}) in terms of the new variables one finds:
\begin{eqnarray}
{\cal L}_{PN}&=&\frac{1}{2}MW^2+\frac{1}{2}\mu V^2+\frac{G\mu M}{R}+\nonumber\\ &&+\frac{1}{c^2}{\cal L}_2 \left( {\bf R}\frac{\mu {\bf V}}{m_1}+{\bf W}, -\frac {\bf V}{m_2}+{\bf W}\right)\, .
\label{2.8}\end{eqnarray}
Hence one obtains as a consequence of the equations of the PN motion:
 \begin{eqnarray}
 {\cal O} & = & \frac{1}{\mu}\frac{\delta{\cal L}_{PN}}{\delta{\bf R} }
      =  \left(\frac{\partial}{\partial {\bf R}} - \frac{d}{dt}\frac{\partial}{\partial {\bf R}}\right)
      \left[\frac{1}{2}{\bf V^2}+\frac{G M}{R} +\right.\nonumber\\ &&+\left.\frac{1}{\mu c^2}{\cal L}_2 \left( {\bf R},\frac{\mu}{m_1}{\bf V}+{\bf W}, -\frac{\mu}{m_2} {\bf V}+{\bf W}\right)\right]\, ,\nonumber\\
 \label{2.9} \end{eqnarray}
 where in the last bracket we have discarded $\displaystyle{\frac{1}{2}MW^2}$ which gives no contribution. The first two terms in the  ({\it rhs})  of Eq. (\ref{2.9}) yield the Newtonian relative motion. We wish to evaluate the relativistic corrections to the relative motion: $\displaystyle{\left(\frac{\delta}{\delta{\bf R}}\right)\left(\frac{{\cal L}}{\mu c^2}\right)}$ in the PN center of mass frame. Now ${\cal L}_2$ is a polynomial in the velocities and therefore a polynomial in ${\bf W}$, and from Eq. (\ref{2.4ab}) one sees that in the PN center of mass frame $\displaystyle{{\bf W}={\cal O}\left(\frac{1}{c^2}\right)}$. Therefore as $\displaystyle{\frac{\delta}{\delta{\bf R}}}$ does not act on ${\bf W}$, we see that the contributions coming from  ${\bf W}$ to the {\it rhs} of Eq. (\ref{2.9}) are of the second PN order $\displaystyle{{\cal O}\left(\frac{1}{c^4}\right)}$ that we shall consistently neglect throughout  this work. In other words one obtains as a consequence of the equations of the PN motion in the PN center of mass frame:
 \begin{eqnarray*}
&& \frac{\delta}{\delta{\bf R}}\left[\frac{1}{2}V^2+ \frac{G M}{R}+ \frac{1}{\mu c^2}{\cal L}_2 \left( {\bf R},\frac{\mu}{m_1}{\bf V}, -\frac{\mu}{m_2} {\bf V}\right)  \right]\nonumber\\ &&
 ={\cal O} \left(\frac{1}{c^4}\right)\, .\nonumber\\
 \label{2.10}\end{eqnarray*}
 This shows that the equations of the relative motion in the PN center of mass frame derive from the following Lagrangian:
  \begin{eqnarray*}
  {\cal L}_{PN}^R({\bf R},{\bf V})&=& \frac{1}{2}V^2+\frac{G M}{R}+\frac{1}{\mu c^2}{\cal L}_2 \left( {\bf R},\frac{\mu}{m_1}{\bf V}, -\frac{\mu}{m_2} {\bf V}\right)\, ,
  \label{2.11}\end{eqnarray*}
  which happens to be obtained by replacing in the full PN Lagrangian, see Eq. (\ref{2.8}) above, $\bf X$ and $\bf W$ by zero, i. e. the original variables by Eq. (\ref{2.6abcd}) and by dividing by $\mu$ \cite{deruelle}. The explicit expression of $  {\cal L}_{PN}^R$ reads:
   \begin{eqnarray}
  {\cal L}_{PN}^R({\bf R},{\bf V}&)=&\frac{1}{2}V^2+\frac{G M}{R}+\frac{1}{8}(1-3\nu)\frac{V^4}{c^2}+\nonumber\\ &&\frac{G M}{2Rc^2}\left[(3+\nu)V^2+\nu(NV)^2-\frac{G M}{R}\right]\,.\nonumber\\  \label{2.12}\end{eqnarray}
  The Lagrangian (\ref{2.12}) was obtained in \cite{infeld}. The integration of the equations (\ref{2.5}) can be done in several different ways.
  \begin{itemize}
 \item A standard approach: Lagrange's method of variation the osculating elements. \item The Hamilton-Jacobi equation approach which, takes advantage of the existence of the PN  Lagrangian is the route which has been taken by Landau and Lifshits \cite{landau}, who worked out only the secular precession of the periastron. 
 \item Another approach, based on the Maupertuis principle \footnote {In classical mechanics, Maupertuis' principle is an integral equation that determines the path followed by a physical system without specifying the time parameterization of that path. It is a special case of the more generally stated principle of least action. More precisely, it is a formulation of the equations of motion for a physical system not as differential equations, but as an integral equation, using the calculus of variations.}, which reduces the PN problem to a simple auxiliary Newtonian problem.
   \end{itemize}
  To describe the motion, it is convenient to use the standard method to solve the non-relativistic two-bodies problem and which consists in exploiting the symmetries of the relative Lagrangian $\  {\cal L}_{PN}^R$. The invariance $\displaystyle{ {\cal L}_{PN}^R}$ under time translations and space rotations implies the existence of four first integrals:\\ $\displaystyle{ E={\bf V} \cdot \frac{\partial {\cal L}_{PN}^R}{\partial {\bf V}}-{\cal L}_{PN}^R} $ and $\displaystyle{ {\bf J} = {\bf R}\times  \frac{\partial {\cal L}_{PN}^R}{\partial {\bf V}}}$:
  
  \begin{eqnarray}
   E&=&\frac{1}{2}V^2-\frac{G M}{R}+\frac{3}{8}(1-3\nu)\frac{V^4}{c^2}+\nonumber\\ &&\frac{G M}{2Rc^2}\left[(3+\nu)V^2+\nu(NV)^2-\frac{G M}{R}\right] \, ,
  \label{2.13}\\
  {\bf J}&=& {\bf R}\times {\bf V}\left[ 1+\frac{1}{2}(1-3\nu)\frac{V^2}{c^2}+(3+\nu)\frac{G M}{2Rc^2}\right]\, .
    \label{2.14}\end{eqnarray}
  It is checked that these quantities coincide respectively with $\mu^{-1}$ times the total Noether energy and the total Noether angular momentum of the binary system when computed in the PN center of mass frame \cite{wagoner}. Eq. (\ref{2.14}) implies that the motion takes place in a coordinate plane, therefore one can introduce polar coordinates $R=r$ and $\phi$ in the plane (i.e. $r_x=r\cos\phi$,\quad$r_y=r\sin\phi$, $r_z=0$). Then starting from the first integrals (\ref{2.13})-(\ref{2.14}) and using the identities: $\displaystyle{ V^2=\left(\frac{dr}{dt}\right)^2+r^2\left(\frac{d\phi}{dt}\right)^2}$, $\displaystyle{ |{\bf R}\times{\bf V}|= r^2\frac{d\phi}{dt}}$, $\displaystyle{NV=\frac{dr}{dt}}$, we obtain by iteration \footnote{In these and the following equations we neglect terms of the second PN order $\displaystyle{ {\cal O}\left(\frac{1}{c^4}\right).}$}
  \begin{equation}
 \left(\frac{dr}{dt}\right)^2 = A+\frac{2B}{r}+\frac{C}{r^2}+\frac{D}{r^3}\, ,
   \label{2.15}\end{equation}
     \begin{equation}
 \frac{d\phi}{dt}=\frac{H}{r^2}+\frac{I}{r^3}\, ,
  \label{2.16}\end{equation}
  where the coefficients $A,B,C,D,H,I$ are the following polynomials in $E$ and $J=|{\bf J}|$:
  \begin{eqnarray}
  A & = & 2  E\left(1+\frac{3}{2}(3\nu-1)\frac{E}{c^2}\right)\, ,\nonumber\\
  B & = & GM\left(1+(7\nu-6)\frac{ E}{c^2}\right)\, ,\nonumber\\
  C & = & -J^2\left(1+2(3\nu-1)\frac{ E}{c^2}\right)+(5\nu-10)\frac{G^2M^2}{c^2}\, ,\nonumber\\
  D &=& (-3\nu+8)\frac{GMJ^2}{c^2}\, ,\nonumber\\
  H &=& J\left(1+(3\nu-1)\frac{  E}{c^2}\right)\, ,\nonumber\\
  I &=& (2\nu-4)\frac{GMJ}{c^2}\, .
   \label{2.17}\end{eqnarray}

 The relativistic "relative motion", {\it i.e.} the solution of Eq. (\ref{2.15}) can be simply reduced to the integration of auxiliary {\it non-relativistic} radial motion. Indeed let us consider the following change of the radial variable:
  \begin{equation}
  r={\bar r}+\frac{D}{2C_0}\, , \label{3.1}
  \end{equation}

  where $C_0$ is the limit of $C$ when $c^{-1}\rightarrow0$ with $(C_0=-J^2)$. Geometrically, the transformation which is expressed in polar coordinates by the equation: $\displaystyle{r'=r+cons}t$, $\displaystyle{\phi'=\phi}$, is called a {\bf conchoidal transformation} \cite{deruelle}. Taking into account the fact that $D$ is $\displaystyle{{\cal O}\left(\frac{1}{c^2}\right)}$ and that we can neglect all terms of order $\displaystyle{{\cal O}\left(\frac{1}{c^4}\right)}$, we find that replacing Eq. (\ref{3.1}) in Eqs. (\ref{2.1a})-(\ref{2.1c}), leads to:
   \begin{equation}
 \left(\frac{d{\bar r}}{dt}\right) ^2=A+\frac{2B}{{\bar r}}+\frac{{\bar C}}{{\bar r}^2}\, ,
  \label{3.2} \end{equation}
   with
    \begin{displaymath}
  {\bar C}=C-\frac{BD}{C_0}\, .
     \end{displaymath}
  Then, in the case of {\bf quasi-elliptical} motion $( E<0; A<0)$, ${\bar r}$ is a linear function of $\displaystyle{\cos {\cal E}}$, ${\cal E}$ being an eccentric anomaly and the same is true of $\displaystyle{r={\bar r}+\frac{D}{2C_0}}$. We then obtain the PN radial motion in quasi-Newtonian parametric form ($t_0$ being a constant of integration):
   \begin{eqnarray}
   n(t-t_0) & = &  {\cal E}-\epsilon_t\sin {\cal E}\, ,
     \label{3.3}\end{eqnarray}
   \begin{eqnarray}
   r & = & a_r(1-\epsilon_r\cos  {\cal E})\, ,
   \label{3.4}\end{eqnarray}
   with
   \begin{eqnarray*}
   n & = &\frac{(-A)^{3/2}}{B}\, , \nonumber\\
   \epsilon_t & = &\left[ 1\frac{A}{B^2}\left(C-\frac{BD}{C_{0}}\right)\right]^{1/2}\, , \nonumber\\
   a_r & = &-\frac{A}{B}+\frac{D}{2C_{0}}\, , \nonumber\\
   \epsilon_r & = & \left(1+\frac{AD}{2BC_0}\right)\epsilon_t \, .\nonumber\\
   \label{3.5}\end{eqnarray*}
   The main difference between the relativistic radial motion and the non-relativistic one is the appearence of two eccentricities: the {\it time eccentricity} $\epsilon_t$ appearing in the Kepler equation (\ref{3.3}) and the {\it relative radial eccentricity} $\epsilon_r$. Using (\ref{2.17}) we can express $a_r,\epsilon_r,\epsilon_t$ and $n$ in terms of $E$ and $J$:
    \begin{eqnarray}
    a_r & = & \frac{GM}{ E}\left[1-\frac{1}{2}(\nu-7)\frac{E}{c^2}\right]\, , \nonumber\\
    \epsilon_r & = & \left\{1+\frac{2E}{G^2M^2}\left[1+\left(\frac{5}{2}\nu-\frac{15}{2}\right )\frac{E}{c^2}\right]\right. \nonumber\\ &&
    \left.\left[J^2+(\nu-6)\frac{G^2M^2}{c^2}\right] \right\}^{1/2}\, , \nonumber\\
    \epsilon_t & = &\left \{1+\frac{2E}{G^2M^2}\left[1+\left(-\frac{7}{2}\nu-\frac{17}{2}\right )\right] \right. \nonumber\\ &&
    \left. \left[J^2+(-2\nu+2)\frac{G^2M^2}{c^2}\right] \right\} \, ,\nonumber\\
    n & = & \frac{(-2  E)^{3/2}}{GM}\left[1+\frac{1}{4}(\nu-15)\frac{E}{c^2}\right] \, .
     \label{3.6}\end{eqnarray}
  It is remarkable that a well known result of the Newtonian motion is still valid at PN level: both the relative semi-major axis $a_r$ the mean motion $n$ depend only on the center of mass energy ${\cal E}$. The same is true for the time of return to periastron period $\displaystyle{P=\frac{2\pi}{n}}$.
  As a consequence we can also express $n$ in term of $a_r$:
  \begin{displaymath}
  n=\left(\frac{GM}{a_r^3}\right)^{1/2}\left[1+\frac{GM}{2a_r c^2}(-9+\nu)\right]\, .
  \label{3.7}\end{displaymath}
   Let us note also that the relationships between $e_r$ and $e_t$ are:
      \begin{eqnarray*}
      \frac{\epsilon_r}{\epsilon_t} &=& 1+(3\nu-8)\frac{ E}{c^2}\, ,\nonumber\\
       \frac{\epsilon_r}{\epsilon_t}&=&1+\frac{GM}2{a_rc^2}\left(4-\frac{3}{2}\nu\right)\, .
      \label{3.8}\end{eqnarray*}
      
  The relativistic angular motion, i.e. the solution of Eq. (\ref{2.16}) can also be simply reduced to the integration of an auxiliary non relativistic angular motion. Let us first make, at
$\displaystyle{{\cal O}\left(\frac{1}{c^2}\right)}$ order, the
following  conchoidal transformation:
    \begin{equation}
    r= {\tilde r}+\frac{I}{2H}\, ,
      \label{4.1}\end{equation}
      which transforms Eq.(\ref{2.16}) into
         \begin{displaymath}
      \frac{d\phi}{dt}=\frac{H}{{\tilde r}^2}\, ,
       \label{4.2}\end{displaymath}
      where $ {\tilde r}$ can be expressed as
        \begin{equation}
       {\tilde r}={\tilde a}(1- {\tilde \epsilon}\cos {\cal E})\,.
           \label{4.3}\end{equation}
          
            Let us note also the relationship between $e_r$ and $e_t$:
      \begin{displaymath}
         {\tilde a} =a_r- \frac{I}{2H}\, ,
            \label{4.4}\end{displaymath}
              \begin{equation}
         {\tilde \epsilon} = \epsilon_r\left(1-  \frac{AI}{2BH}\right)\, .
        \label{4.5}\end{equation}
  The  differential time is given, from Eq. (\ref{3.3}) by:
    \begin{displaymath}
    dt= n^{-1}(1-\epsilon_t\cos {\cal E})d{\cal E}\, .
      \label{4.6}\end{displaymath}
      Hence we get
        \begin{displaymath}
    d\phi= \frac{H}{n   {\tilde a}^2}\frac{(1-\epsilon_t\cos {\cal E})}{(1-    {\tilde \epsilon}\cos {\cal E})^2}d{\cal E}\, .
      \label{4.7}\end{displaymath}
      As can be seen from Eq. (\ref{3.3}) and Eq. (\ref{4.5}) $\epsilon_t$ and ${\tilde \epsilon}$ differ by only small terms of order $\displaystyle{\frac{1}{c^2}}$. Now if we introduce any new eccentricity say $\displaystyle{\epsilon_\phi}$ also very near $\displaystyle{\epsilon_t}$ so that we can write:
      $\displaystyle{\epsilon_t=\frac{1}{(\epsilon_t+\epsilon_\phi)}{2}+\varepsilon}$, $\displaystyle{\epsilon_\phi=\frac{(\epsilon_t+\epsilon_\phi)}{2}-\varepsilon}$, with $\displaystyle{\varepsilon={\cal O}\left(\frac{1}{c^2}\right)}$ then

   \begin{displaymath}
    (1-\epsilon_t\cos {\cal E})  (1-\epsilon_\phi\cos {\cal E})=\left(1-\frac{(\epsilon_t+\epsilon_\phi)}{2}\cos {\cal E}\right)^2-\varepsilon^2\cos^2 {\cal E}\, .
      \label{4.8}\end{displaymath}

   Therefore if we choose $\epsilon_\phi$ such that the average of $\epsilon_t$ and $\epsilon_\phi$ is equal to $ {\tilde \epsilon}$ i.e. $\epsilon_\phi=2{\tilde \epsilon}-\epsilon_t$ we have
    \begin{displaymath}
  \frac{(1-\epsilon_t\cos {\cal E})}{(1-  {\tilde \epsilon}\cos {\cal E})}^{2}=\frac{1}{1- \epsilon_\phi \cos {\cal E}}+  {\cal O}\left(\frac{1}{c^4}\right)\, ,
    \label{4.9}\end{displaymath}
 which transforms Eq.   (\ref{4.7}) into a Newtonian like angular motion equation
   \begin{displaymath}
   d\phi=\frac{H}{n{\tilde a}^2}\frac{d{\cal E}}{1- \epsilon_\phi \cos {\cal E}}\, ,
       \label{4.10}\end{displaymath}
 which is easily integrated to give
  \begin{equation}
  \phi-\phi_0=KA_{\epsilon_{\phi}}( {\cal E})\, ,
     \label{4.11a}\end{equation}
     $\phi_0$ being a constant of integration and where for the sake of simplicity we have introduced the notations:
     \begin{equation}
  A_{\epsilon_{\phi}}(E)=2\arctan \left[\left(\frac{1+\epsilon}{1-\epsilon}\right)^{\frac{1}{2}} \tan\frac{ {\cal E}}{2} \right] \, ,
     \label{4.11b}\end{equation}
     and
       \begin{equation}
  K=\frac{H}{n{\tilde a^2}(1-\epsilon_{\phi}^{2})^\frac{1}{2}}\, .
     \label{4.11c}\end{equation}
     From Eq. (\ref{4.5}) and  (\ref{3.5}) and the definition of $\epsilon_\phi=2{\tilde \epsilon}-\epsilon_t$ we have:

   \begin{displaymath}
  \epsilon_{\phi}= \epsilon_t\left(1+\frac{AD}{BC_0}-\frac{AI}{BH}\right)=\epsilon_r\left(1+\frac{AD}{2BC_0}-\frac{AI}{BH}\right)
     \label{4.12}\end{displaymath}
     then, as shown by straightforward calculations:
     \begin{eqnarray}
   &&\epsilon_{\phi}=  \epsilon_r\left(1+\frac{G\mu}{2a_r c^2}\right)=\nonumber\\ && \left \{1+ \frac{2E}{G^2M^2}\left[1+\left(\frac{1}{2}\nu-\frac{15}{2}\right)\frac{E}{c^2}\right]  \left[J^2-6\frac{G^2 M^2}{c^2}\right]  \right \}^\frac{1}{2}\, ,\nonumber\\
\label{4.13}\end{eqnarray}
and
    \begin{equation}
 K=\frac{J}{\left(J^2-\frac{6G^2M^2}{c^2}\right)^\frac{1}{2}}\, .
     \label{4.14}\end{equation}
     As it is clear from Eqs. (\ref{4.1}) and (\ref{4.3}), the radial variable $r$ reaches its successive minima "periastron passages" for ${\cal E}=0,2\pi,4\pi,...$The periastron therefore precesses at each turn by the angle $\displaystyle{\Delta\phi=2\pi(L-1)}$, which if $\displaystyle{J>>\frac{GM}{c}}$ reduces to the well-known result \cite{robertson}:
     \begin{displaymath}
     \Delta\phi=6\pi\frac{G^2M^2}{J^2c^2}+{\cal O} \left(\frac{1}{c^4}\right)=\frac{6\pi GM}{a_R(1-\epsilon_r)c^2}+{\cal O} \left(\frac{1}{c^4}\right)\, .
  \label{4.15}   \end{displaymath}
  Contrarily to the usual approach which derives first the orbit by eliminating the time between Eq. (\ref{2.15}) and (\ref{2.16}) before working out the motion on the orbit we find the orbit by eliminating $ {\cal E}$ between Eq. (\ref{3.4}) and  (\ref{4.11a})-(\ref{4.11c}). With the aim to simplify the formulae we introduce the notation $f$ for the polar angle counted from a periastron and corrected for the periastron precession i.e.:
    \begin{displaymath}
    f=\frac{\phi-\phi_0}{K}\, ,
   \label{5.1} \end{displaymath}
   We must eliminate $\cal E$ between:
     \begin{displaymath}
    r=a_r(1-\epsilon_r\cos {\cal E})\, ,
   \label{5.2} \end{displaymath}
   and
     \begin{displaymath}
     f=  A_{\epsilon_{\phi}}( {\cal E})\, .
   \label{5.3} \end{displaymath}
 In order to get, it is convenient to play a new conchoidal
transformation on $r$  writing:
     \begin{equation}
   r=\frac{\epsilon_r}{\epsilon_\phi}a_r(1-\epsilon_\phi\cos  {\cal E})+a_r\left(1-\frac{\epsilon_r}{\epsilon_\phi}\right)\, .
      \label{5.4} \end{equation}
      From the definition of $\displaystyle{A_{\epsilon_{\phi}}({\cal E})}$ we have:

   \begin{displaymath}
  1-\epsilon_\phi\cos {\cal E}=\frac{1-\epsilon_{\phi}^2}{1+e_\phi A_{\epsilon_{\phi}}( {\cal E})} =\frac{1-\epsilon_{\phi}^2}{1+\epsilon_{\phi} \cos f}\, .
         \label{5.5} \end{displaymath}
      Moreover we find from Eq. (\ref{4.13}) that the radial displacement appearing in Eq.  (\ref{5.4}) is simply
  \begin{displaymath}
   a_r   \left(1-\frac{\epsilon_r}{\epsilon_\phi}\right)=\frac{G\mu}{2c^2}\, ,
   \label{5.6} \end{displaymath}
  so that we find the polar equation of the relative orbit as:
  \begin{displaymath}
   r=   \left(a_r-\frac{G\mu}{2c^2}\right)\frac{1-\epsilon_{\phi}^2}{1+\epsilon_{\phi} \cos f}+\frac{G\mu}{2c^2}
   \label{5.7} \end{displaymath}
This equations means that the relative orbit is the {\it conchoid} of a {\it precessing ellipse}, which means that it is obtained from an ellipse: $r'=l(1+e\cos\phi')^{-1}$ by a radial displacemnet $r=r'+const$ together with an angular homothetic transformation: $\phi=const\cdot\phi'$.
Let us finally note that the relative orbit con also be written as:
  \begin{displaymath}
  r=\frac{a_r(1-\epsilon_{r}^2)}{1+\epsilon_r \cos  f'}\, ,
   \label{5.10a} \end{displaymath}
   with
    \begin{displaymath}
   f'= f+2\left(\frac{\epsilon_{2 f}}{\epsilon_r}\right)\sin f\, .
   \label{5.10b} \end{displaymath}
   The conservation laws and the coordinate transformations which we
have obtained here will reveal particularly useful to characterize
the relativistic orbits, as we will see below.
  \subsection{Relativistic quasi-Elliptical orbits}
   The relativistic motions of each body are obtained by replacing the solutions for the relative motion, $t({\cal E}),r({\cal E}),\phi({\cal E})$, in the PN center of mass formulae Eq. (\ref{2.4ab}) (see\cite{deruelle}). We see first that the polar angle of the first object (of mass $m_1$) is the same as the relative polar angle and that the polar angle of the second object (mass $m_2$) is simply $\phi+\pi$. Therefore it is sufficient to work out the radial motions. From Eq. (\ref{2.4ab})  we have by replacing $V^2$ in the relativistic corrections with $\displaystyle{\frac{2GM}{R}+2E\simeq \frac{2GM}{R}-\frac{GM}{a_r}}$:
 \begin{displaymath}
r=\frac{m_2R}{M}+\frac{G\mu(m_1-m_2)}{2mc^2}\left(1-\frac{R}{a_R}\right)
   \label{6.1} \end{displaymath}
   (and similar results for the second object by exchanging $m_1$ and $m_2$) which shows remarkably enough,
   that $r$ can also be written in a quasi-Newtonian form:
 \begin{displaymath}
 r=a_{r'}(1-\epsilon_{r'}\cos {\cal E}) \, ,
       \label{6.2} \end{displaymath}
       with
       \begin{eqnarray*}
       a_{r'} &=& \frac{m_2}{M}a_r\, ,\\
     \epsilon_{r'} &=& e_R\left[1-\frac{Gm_1(m_1-m_2}{2Ma_r c^2}\right] \, ,
      \label{6.3}   \end{eqnarray*}
 and where as before:
\begin{eqnarray*}
n(t-t_0) &=& {\cal E}-\epsilon_t \sin {\cal E}\, ,\\
\phi-\phi_0 &=& K A_{\epsilon_{\phi}}({\cal E})\, .
    \label{6.4}\end{eqnarray*}
 The orbit in space of the first object can be written  by using
the same method as in the preceding Section for the relative
orbit, that is:
  \begin{displaymath}
 r=\frac{\epsilon_{r'}}{\epsilon_\phi}a_{r'}(1-\epsilon_\phi \cos {\cal E})+a_{r'}\left(1-\frac{\epsilon_{r'}}{\epsilon_\phi}\right)\, .
       \label{6.5} \end{displaymath}
       One finds:
 \begin{displaymath}
 a_{r'}\left(1-\frac{\epsilon_{r'}}{\epsilon-\phi}\right)  =\frac{Gm_{1}^2 m_2}{2M^2c^2}\, ,
       \label{6.6} \end{displaymath}
   hence we find that the orbits is conchoid of a precessing ellipse with
 \begin{displaymath}
 r'=\left(a_{r'}-\frac{Gm_{1}^2m_2}{2M^2c^2}\right) \frac{1-\epsilon_{\phi}^2}{1+\epsilon_\phi\cos\left(\frac{\phi-\phi_0}{L}\right)}+   \frac{Gm_{1}^2m_2}{2M^2c^2}
   \label{6.7} \end{displaymath}

Summarizing  then gathering our results for the {\bf
elliptic-like} ($E<0$) PN motion in the PN center of mass frame,
we have:
\begin{eqnarray}
&&n(t-t_0) = {\cal E}-\epsilon_t\sin {\cal E}\, ,\nonumber\\
&&  \phi-\phi_0 = K2\arctan \left[\left(\frac{1+\epsilon_\phi}{1-\epsilon_\phi}\right)^\frac{1}{2}\tan \frac{{\cal E}}{2} \right]\, , \nonumber\\
 &&r = a_r(1-\epsilon_r\cos {\cal E})\, ,\nonumber\\
     && r ' = a_{r'}(1-e_{r'}\cos {\cal E})\, ,\nonumber\\
     \label{7.1}\end{eqnarray}
 with
\begin{displaymath}
a_r=\frac{GM}{2E}\left[1-\frac{1}{2}(\nu-7)\frac{E}{c^2}\right]\, ,
\end{displaymath}
\begin{equation}
n=\frac{(-2E)^\frac{3}{2}}{GM}   \left[1-\frac{1}{4}(\nu-15)\frac{E}{c^2}\right]\, .
\label{7.2}\end{equation}
 and $\displaystyle{K,\epsilon_t,e_\phi,\epsilon_r,e_r,a_r,e_{r'},a_{r'}}$ given in terms of the total angular momentum by unit reduced mass in the center of mass frame, $E$ and $J$, by Eq. (\ref{3.6}), (\ref{4.14}), (\ref{4.13}),
and interchange of $m_1$ and $m_2$ for $\epsilon_{r'},a_ {r'}$.
The above equations are very similar to the standard Newtonian
solutions of the non-relativistic two-body problem.
\subsection{Relativistic quasi-Hyperbolic orbits}
The simplest method to obtain the Post-Newtonian motion in the
{\bf hyperbolic-like} case ($E>0$) consists simply in making, in
Eqs.(\ref{7.1})-(\ref{7.2}), the analytic continuation 
from $E<0$ to $E>0$, passing below $E=0$ in the complex $E$ plane
and replacing the parameter $\cal E$ by $i\cal F$ ($i^2=-1$). The
proof that this yields to a correct parametric solution 
consists in noticing, on one hand, that $K$ and the various
eccentricities are analytic in $E$, near $E=0$, and that if one
replaces the parametric solution (\ref{7.1})-(\ref{7.2}) and the
corresponding expressions of
$\epsilon_t,e_\phi,\epsilon_r,e_r,a_r,e_{r'},a_{r'}$ in terms of
$E$ and $J$ in $\displaystyle{\left(\frac{dr}{dt}\right)^2}$ and
in  $\displaystyle{\left(\frac{d\phi}{dt}\right)^2}$, one finds
that Eq. (\ref{2.15}) and the square of Eq. (\ref{2.16}) are
satisfied identically, modulo $\displaystyle{{\cal
O}\left(\frac{1}{c^4}\right)}$, and that the resulting identities
are {\it analytic} in $E$ and $\cal E$ as purely imaginary ones.
One finds:
\begin{eqnarray}
\bar{n}(t-t_0) &=&\epsilon_t\sinh {\cal F}- {\cal F}\, ,\nonumber\\
  \phi-\phi_0 &=& K 2\arctan \left[\left(\frac{\epsilon_\phi+1}{\epsilon_\phi-1}\right)^\frac{1}{2}\tanh \frac{{\cal F}}{2} \right]\, , \nonumber\\
  r &=& \bar {a}_r(1-\epsilon_r\cos {\cal F})\, ,\nonumber\\
       r ' &=& \bar{a}_{r'}(1-\epsilon_{r'}\cos {\cal F})\, ,\nonumber\\
\label{7.3}\end{eqnarray}
where  $K,\epsilon_t,\epsilon_\phi,\epsilon_r,\epsilon_{r'}$ are functions of $E$ and $J$, as before, but where it has been conveninet to introduce the opposites of analytic continuations of the semi-major axes:
\begin{displaymath}
\bar {a}_{r'}=\frac{GM}{2E}\left[1-\frac{1}{2}(\nu-7)\frac{E}{c^2}\right]\, ,
\label{7.4}\end{displaymath}
and $\displaystyle{\bar{a}_{r'}=\frac{m_1\bar {a}_{r}}{M}}$ and the modulus of the analytic continuation of the mean motion:
\begin{displaymath}
\bar n=\frac{(2M)^\frac{3}{2}}{GM}\left[1-\frac{1}{4}(\nu-15)\frac{E}{c^2}\right]\,.
\label{7.5}\end{displaymath}

\subsection{Relativistic quasi-Parabolic orbits}
The {\bf quasi-parabolic} post-Newtonian motion ($E=0$)  can be obtained as a limit when $E\rightarrow0$. For istance, let us start from the quasi-elliptic solution in Eq.(\ref{7.1}) and pose
\begin{displaymath}
{\cal E}=\left(\frac{-2E}{G^2M^2}\right)^\frac{1}{2}x\, .
\label{7.6}\end{displaymath} Taxing now the limit
$E\rightarrow0^-$, holding $x$ fixed, yields the following
parametric representation of the quasi parabolic motion:
\begin{equation}
t-t_0 = \frac{1}{2G^2M^2}\left[ \left(J^2+(2-2\nu)\frac{G^2M^2}{c^2}x+\frac{1}{3}x^3\right)\right]\, ,
\label{7.7a}\end{equation}
\begin{equation}
\phi-\phi_0=\frac{J}{(J^2-6)\frac{G^2M^2}{c^2})^\frac{1}{2}}2\arctan\frac{x}{(J^2-\frac{6G^2M^2}{c^2})^\frac{1}{2}}\, ,
\label{7.7b}\end{equation}
\begin{equation}
r =  \frac{1}{2GM}\left[
\left(J^2+(\nu-6)\frac{G^2M^2}{c^2}x+x\right)\right]\, .
\label{7.7c}\end{equation}
 Moreover let us point out that our
solutions (for the three cases $E<0$, $E>0$ and $E=0$) have been
written in a suitable form  when
$\displaystyle{J^2>6\left(\frac{GM}{c}\right)^2}$. However the
validity of our solutions can be extended to the range
$\displaystyle{J^2\leq 6\left(\frac{GM}{c}\right)^2}$ by first
replacing in the solutions for the angular motion , the second equation of
(\ref{7.1}) and (\ref{7.3}), and considering the Eqs. (\ref{7.7a})-(\ref{7.7c}), the function
$\arctan$ by $\cot$ (at the price of a simple modifiation of
$\phi_0$) and then by making an analytic continuation in
$J$. One ends up with an angular motion expressed by an $\arg\coth$
which can also be approximatively replaced by its asymptotic
behaviour for large arguments: $\arg\coth(
\displaystyle{X})\sim\frac{1}{X}$. The case of purely radial
motion ($J=0$) is also obtained by taking the limit
$J\rightarrow0$ ( at $\cal E,F$ or respectively $x$ fixed).
Finally a parametric representation of the general post-Newtonian
motion in an arbitrary (post-Newtonian harmonic) coordinate system
is obtained from our preceding center of mass solution by applying
a general transformation of the Poincar\'e group
$\displaystyle{x'^a=L_b^ax^b+T^a}$  \cite{deruelle}.

\section{Relativistic orbits with gravitomagnetic corrections}
Using the orbital theory developed up to now for  relativistic
orbit, we have neglected terms of  order
$\displaystyle{\frac{v}{c}^3}$. However, we succeed in explaining,
for instance, the perihelion precession of Mercury. In  cases
where $\displaystyle{10^{-2}\le\frac{v}{c}\le10^{-1}}$, higher
order terms like $\displaystyle{\frac{v}{c}^3}$ cannot be
discarded in order to discuss consistently the problem of motion
(see for example \cite{SMFL}). In this situations, we are dealing
with   gravitomagnetic corrections. Before discussing the theory
of  orbits under the gravitomagnetic effects, let us give some
insight into gravitomagnetism and derive the corrected metric.
Theoretical and experimental aspects of gravitomagnetism are
discussed in \cite{iorio9,ruffini}.

A remark is in order at this point: any theory  combining, in a
consistent way,  Newtonian gravity together with Lorentz
invariance  has to include a gravitomagnetic field generated by
the mass-energy currents. This is the case, of course, of GR: it
was shown by Lense and Thirring
\cite{Thirring,barker,ashby,iorio10,Tartaglia}, that a rotating
mass generates a gravitomagnetic field, which, in turn, causes a
precession of planetary orbits. In the framework of the linearized
weak-field and slow-motion approximation of GR, the
gravitomagnetic effects are induced by the off-diagonal components
of the space-time metric tensor which are proportional to the
components of the matter-energy current density of the source. It
is possible to take into account two types of mass-energy
currents.  The former is induced by the matter source rotation
around its center of mass: it generates the intrinsic
gravitomagnetic field which is closely related to the angular
momentum (spin) of the rotating body. The latter is due to the
translational motion of the source: it is responsible of the
extrinsic gravitomagnetic field \cite{pascual,kopeikin2}. Let us
now discuss the gravitomagnetic effects  in order to see how they
affect the orbits.

\subsection{Gravitomagnetic effects}
Starting from the Einstein field equations in the weak field
approximation, one obtain the gravitoelectromagnetic equations and
then the corrections on the metric\footnote{Notation: Latin
indices run from 1 to 3, while Greek indices run from 0 to 3; the
flat space-time metric tensor is
$\eta_{\mu\nu}=diag(1,-1,-1,-1)$.} \cite{SMFL}. The weak field approximation
can be set as

\begin{equation}
g_{\mu\nu}(x)=\eta_{\mu\nu}+h_{\mu\nu}(x),\qquad\left|h_{\mu\nu}(x)\right|<<1,\label{eq:g_muni}\end{equation}
where $\eta_{\mu\nu}$ is the Minkowski metric tensor and
$\left|h_{\mu\nu}(x)\right|<<1$ is a small deviation from it
\cite{weinberg}.

The stress-energy tensor for perfect - fluid matter is given by

\begin{displaymath}
T^{\mu\nu}=\left(p+\rho
c^{2}\right)u^{\mu}u^{\nu}-pg^{\mu\nu}\, ,\end{displaymath}
which, in the
weak field approximation $p\ll\rho c^{2}$, is

\begin{displaymath}
T^{00}\simeq\rho c^{2},\qquad T^{0j}\simeq\rho cv^{j},\qquad
T^{ij}\simeq\rho v^{i}v^{j}\,.\end{displaymath} From the Einstein
field equations $G_{\mu\nu}=(8\pi G/c^4)T_{\mu\nu}$, one finds

\begin{equation}
\bigtriangledown^{2}h_{00}=\frac{8\pi
G}{c^{2}}\rho\,,\label{eq:nabla_00}\end{equation}

\begin{equation}
\bigtriangledown^{2}h_{ij}=\frac{8\pi
G}{c^{2}}\delta_{ij}\rho\,,\label{eq:nabla_ij}\end{equation}

\begin{equation}
\bigtriangledown^{2}h_{0j}=-\frac{16\pi G}{c^{2}}\delta_{jl}\rho
v^{l}\,,\label{eq:nabla_0j}\end{equation}

where $\bigtriangledown^{2}$ is the standard Laplacian operator
defined on the flat spacetime. To achieve Eqs.
(\ref{eq:nabla_00})-(\ref{eq:nabla_0j}),  the harmonic condition

\begin{displaymath}
g^{\mu\nu}\Gamma_{\mu\nu}^{\alpha}=0\;,\end{displaymath} has been
used.

By integrating  Eqs. (\ref{eq:nabla_00})-(\ref{eq:nabla_0j}), one
obtains

\begin{equation}
h_{00}=-\frac{2\Phi}{c^{2}}\;,\label{eq:h_00}\end{equation}

\begin{equation}
h_{ij}=-\frac{2\Phi}{c^{2}}\delta_{ij}\;,\label{eq:h_ij}\end{equation}

\begin{equation}
h_{0j}=\frac{4}{c^{3}}\delta_{jl}V^l\;.\label{eq:h_0j}\end{equation}
The metric is determined by the gravitational Newtonian potential

\begin{equation}
\Phi(x)=-G\int\frac{\rho}{\left|\mathbf{x}-\mathbf{x}'\right|}d^{3}x'\;,\label{eq:fi_x}\end{equation}

and by the vector potential $V^{l}$,

\begin{equation}
V^{l}=-G\int\frac{\rho
v^{l}}{\left|\mathbf{x}-\mathbf{x}'\right|}d^{3}x'\;.\label{eq:Vl}\end{equation}
given by the  matter current density $\rho v^{l}$ of the moving
bodies. This last potential gives rise to the gravitomagnetic
corrections.

From Eqs. (\ref{eq:g_muni}) and (\ref{eq:h_00})-(\ref{eq:Vl}),  the
metric tensor in terms of Newton and gravitomagnetic potentials is

\begin{eqnarray}
ds^{2}&=&\left(1+\frac{2\Phi}{c^{2}}\right)c^{2}dt^{2}-
\frac{8\delta_{lj}V^{l}}{c^{3}}cdtdx^{j}+\nonumber\\ &&-\left(1-\frac{2\Phi}{c^{2}}\right)\delta_{lj}dx^{i}dx^{j}\;.
\label{eq:ds_DUE}\end{eqnarray}

From Eq. (\ref{eq:ds_DUE}) it is possible to construct a
variational principle from which the geodesic equation follows.
Then we can derive the orbital equations. As standard, we have

\begin{displaymath}
\ddot{x}^{\alpha}+\Gamma_{\mu\nu}^{\alpha}\dot{x}^{\mu}\dot{x}^{\nu}=0\;,\label{eq:geodedica_uno}\end{displaymath}

where the dot indicates the differentiation with respect to the
affine parameter. In order to put in evidence the gravitomagnetic
contributions, let us explicitly calculate the Christoffel symbols
at lower orders. By some straightforward calculations, one gets
\begin{eqnarray}
\Gamma^0_{00} &=&0\\
\Gamma^0_{0j} &=&\frac{1}{c^2}\frac{\partial\Phi}{\partial x^j} \\
\Gamma^0_{ij} &=&-\frac{2}{c^3}\left(\frac{\partial V^i}{\partial x^j}+\frac{\partial V^j}{\partial x^i}\right) \\
\Gamma^k_{00} &=& \frac{1}{c^2}\frac{\partial\Phi}{\partial x^k}\\
\Gamma^k_{0j} &=&\frac{2}{c^3}\left(\frac{\partial V^k}{\partial x^j}-\frac{\partial V^j}{\partial x^k}\right) \\
\Gamma^k_{ij} &= &-\frac{1}{c^2}\left(\frac{\partial \Phi}{\partial
x^j}\delta^k_i+\frac{\partial \Phi}{\partial
x^i}\delta^k_j-\frac{\partial \Phi}{\partial
x^k}\delta_{ij}\right)\end{eqnarray}

In the
approximation which we are going to consider, we are retaining
terms up to the orders $\displaystyle{\Phi/c^2}$ and $\displaystyle{V^j/c^3}$. It is important
to point out that we are discarding terms like
$\displaystyle{(\Phi/c^4)\partial\Phi/\partial x^k}$,
$\displaystyle{(V^j/c^5)\partial\Phi/\partial x^k}$, $\displaystyle{(\Phi/c^5)\partial
V^k/\partial x^j}$, $\displaystyle{(V^k/c^6)\partial V^j/\partial x^i}$ and of
higher orders. Our aim is to show that, in several  cases like in
tight binary stars, it is not correct to discard higher order
terms in $\displaystyle{v/c}$ since physically interesting effects could come
out.
The geodesic equations up to $c^{-3}$ corrections are then

\begin{eqnarray}
&& c^{2}\frac{d^{2}t}{d\sigma^{2}}+\frac{2}{c^{2}}\frac{\partial\Phi}{\partial
x^{j}}c\frac{dt}{d\sigma}\frac{dx^{j}}{d\sigma}\nonumber\\ &&-\frac{2}{c^{3}}\left(\delta_{im}\frac{\partial
V^{m}}{\partial x^{j}}+\delta_{jm}\frac{\partial V^{m}}{\partial
x^{i}}\right)\frac{dx^{i}}{d\sigma}\frac{dx^{j}}{d\sigma}=0\;,\nonumber\\ \label{time}
\end{eqnarray}
for the time component, and

\begin{eqnarray}
&& \frac{d^{2}x^{k}}{d\sigma^{2}}+\frac{1}{c^{2}}\frac{\partial\Phi}{\partial
x^{j}}\left(c\frac{dt}{d\sigma}\right)^{2}+
\frac{1}{c^{2}}\frac{\partial\Phi}{\partial x^{k}}\delta_{ij}\frac{dx^{i}}{d\sigma}\frac{dx^{j}}{d\sigma}\nonumber\\
& &-\frac{2}{c^{2}}\frac{\partial\Phi}{\partial
x^{l}}\frac{dx^{l}}{d\sigma}\frac{dx^{k}}{d\sigma}+\frac{4}{c^{3}}\left(\frac{\partial
V^{k}}{\partial x^{j}}-\delta_{jm}\frac{\partial V^{m}}{\partial
x^{k}}\right)c\frac{dt}{d\sigma}\frac{dx^{i}}{d\sigma}=0\;,\label{eq:dduexk}\nonumber\\
\end{eqnarray}
for the spatial components.

In the case of a null-geodesic, it results  $ds^{2}=d\sigma^{2}=0$. Eq.
(\ref{eq:ds_DUE}) gives, up to the order $c^{-3}$,

\begin{equation}
cdt=\frac{4V^{l}}{c^{3}}dx^{l}+\left(1-\frac{2\Phi}{c^{2}}\right)dl_{euclid}\;,\label{eq:c_dt}\end{equation}

where $dl_{euclid}^{2}=\delta_{ij}dx^{i}dx^{j}$ is the Euclidean
length interval. Squaring Eq. (\ref{eq:c_dt}) and keeping terms up
to order $c^{-3}$, one finds

\begin{eqnarray}
c^{2}dt^{2}=\left(1-\frac{4\Phi}{c^{2}}
\right)dl_{euclid}^{2}+\frac{8V^{l}}{c^{3}}dx^{l}dl_{euclid}\;.\label{eq:cdue_dtdue}\end{eqnarray}

Inserting Eq.(\ref{eq:cdue_dtdue}) into Eq. (\ref{eq:dduexk}), one
gets, for the spatial components,

\begin{eqnarray}
&&\frac{d^{2}x^{k}}{d\sigma^{2}}+\frac{2}{c^{2}}\frac{\partial\Phi}{\partial
x^{k}}\left(\frac{dl_{euclid}}{d\sigma}\right)^{2}-\frac{2}{c^{2}}\frac{\partial\Phi}{\partial
x^{l}}\frac{dx^{l}}{d\sigma}\frac{dx^{k}}{d\sigma}+\nonumber\\ && \frac{4}{c^{3}}\left(\frac{\partial
V^{k}}{\partial x^{j}}-\delta_{jm}\frac{\partial V^{m}}{\partial
x^{k}}\right)\frac{dl_{euclid}}{d\sigma}\frac{dx^{j}}{d\sigma}=0\;.\label{eq:ddue_dsigma}\end{eqnarray}

Such an equation can be seen as a differential equation for
$dx^k/d\sigma$ which is the tangent 3-vector to the trajectory. On
the other hand, Eq.  (\ref{eq:ddue_dsigma}) can be expressed in
terms of $l_{euclid}$ considered as a parameter. In fact, for null
geodesics and taking into account the lowest order in $v/c$,
$d\sigma$ is proportional to $dl_{euclid}$. From Eq. (\ref{time})
multiplied for ${\displaystyle \left(1+\frac{2}{c^2}\Phi\right)}$,
we have
\begin{displaymath}
\frac{d}{d\sigma}\left(\frac{dt}{d\sigma}+\frac{2}{c^2}\Phi\frac{dt}{d\sigma}-
\frac{4}{c^4}\delta_{im}V^m\frac{dx^i}{d\sigma}\right)=0\,,
\end{displaymath}
and then
\begin{equation}
\frac{dt}{d\sigma}\left(1+\frac{2}{c^2}\Phi\right)-
\frac{4}{c^4}\delta_{im}V^m\frac{dx^i}{d\sigma}=1\,,\label{constant}
\end{equation}
where, as standard, we have defined the affine parameter so that
the integration constant is equal to 1 \cite{weinberg}.
Substituting Eq. (\ref{eq:c_dt}) into Eq. (\ref{constant}), at
lowest order in $v/c$, we find
\begin{equation} \frac{dl_{euclid}}{c d\sigma}=1\,.\end{equation}
In the weak field regime, the spatial 3-vector, tangent to a given
trajectory, can be expressed as
\begin{equation} \frac{dx^k}{ d\sigma}=\frac{cdx^k}{dl_{euclid}}\,.\end{equation}
Through the definition
\begin{displaymath} e^k=\frac{dx^k}{dl_{euclid}}\,,\end{displaymath}
Eq. (\ref{eq:ddue_dsigma}) becomes
\begin{eqnarray*}
&& \frac{de^{k}}{dl_{euclid}}+\frac{2}{c^{2}}\frac{\partial\Phi}{\partial
x^{k}}-\frac{2}{c^{2}}\frac{\partial\Phi}{\partial
x^{l}}e^{l}e^{k}+\nonumber\\ &&+\frac{4}{c^{3}}\left(\frac{\partial
V^{k}}{\partial x^{j}}-\delta_{jm}\frac{\partial V^{m}}{\partial
x^{k}}\right)e^{j}=0\;,\label{eq:e_dsigma}\end{eqnarray*}

 which can
be expressed in a vector form as
\begin{equation}
\frac{d\mathbf{e}}{dl_{euclid}}=-\frac{2}{c^2}\left[\nabla\Phi-\mathbf{e}(\mathbf{e}\cdot\nabla\Phi)\right]+\frac{4}{c^3}
\left[\mathbf{e}\wedge(\nabla\wedge\mathbf{V})\right]\label{vector}\,.
\end{equation}\\

The gravitomagnetic term is the second one in Eq. (\ref{vector})
and it is usually discarded since considered not relevant. This is
not true if $v/c$ is quite large as in the cases of tight binary
systems or point masses approaching to black holes.

Our task is now to  achieve explicitly the trajectories, in
particular the orbits, corrected by such  effects.
\vspace{0.5cm}
\subsection{Gravitomagnetically corrected orbits}
Orbits with gravitomagnetic effects can be obtained  starting from
the classical Newtonian theory and then correcting it by
successive relativistic terms. Starting from the above
considerations (see Sec. \ref{due}, and \ref{tre}) we can see how
gravitomagnetic corrections affect the problem of orbits.
Essentially, they act as a further $v/c$ correction leading to
take into account terms up to $c^{-3}$, as shown above. Let us
start from the line element (\ref{eq:ds_DUE}) which can be written
in spherical coordinates. Here we assume the motion of point-like
bodies and then we can work in the simplified hypothesis
${\displaystyle \Phi=-\frac{GM}{r}}$ and $V^{l}=\Phi v^{l}$. It is

\begin{eqnarray*}
ds^{2} & = &
\left(1+\frac{2\Phi}{c^{2}}\right)cdt^{2}-\left(1-\frac{2\Phi}{c^{2}}\right)\\ &&\left[dr^{2}+r^{2}d\theta^{2}+
r^{2}\sin^{2}\theta d\phi^{2}\right]\\ &&
 -\frac{8\Phi}{c^{3}}cdt \left[\cos\theta+\sin\theta\left(\cos\phi+\sin\phi\right)\right]dr
  \\ &&+\frac{8\Phi}{c^{3}}cdt\left[\cos\theta\left(\cos\phi+\sin\phi\right)-\sin\theta\right]rd\theta
 \\ && +\frac{8\Phi}{c^{3}}cdt\left[\sin\theta\left(\cos\phi-\sin\phi\right)\right]rd\phi\,.
 \label{totalmetricnostrpola}\end{eqnarray*}

As in the Newtonian and relativistic cases, from the line element
(\ref{totalmetricnostrpola}), we can construct the Lagrangian

\begin{eqnarray}
\mathcal{L} & = &
\left(1+\frac{2\Phi}{c^{2}}\right)\dot{t}-\left(1-\frac{2\Phi}{c^{2}}\right)\left[{\dot r}+r^{2}{\dot\theta^{2}}+
r^{2}\sin^{2}\theta {\dot\phi^{2}}\right]\nonumber\\ &&
 -\frac{8\Phi}{c^{3}}{\dot t} \left[\cos\theta+\sin\theta\left(\cos\phi+\sin\phi\right)\right]{\dot r}
  \nonumber\\ &&+\frac{8\Phi}{c^{3}}{\dot t}\left[\cos\theta\left(\cos\phi+\sin\phi\right)-\sin\theta\right]r{\dot\theta}
 \nonumber\\ && +\frac{8\Phi}{c^{3}}{\dot t}\left[\sin\theta\left(\cos\phi-\sin\phi\right)\right]r{\dot\phi}\,. \nonumber\\ \label{Lagrangiannostra}\end{eqnarray}
 Being,
$\mathcal{L}=1$, one can multiply both  members for
${\displaystyle \left(1+\frac{2\Phi}{c^{2}}\right)}$. In the
planar motion condition $\theta=\pi/2$ , we  obtain

\begin{eqnarray*}
&&E^{2}-\left(1+\frac{2\Phi}{c^{2}}\right)\left(1-\frac{2\Phi}{c^{2}}\right)\left(\dot{r}^2+\frac{L^{2}}{r^{2}}\right)\\ &&
-\frac{8\Phi
E}{c^{3}}\left[\left(\cos\phi+\sin\phi\right)\dot{r}-\left(\cos\phi-\sin\phi\right)\dot{\phi}\right]
 \nonumber\\ &&= \left(1+\frac{2\Phi}{c^{2}}\right)\,,\end{eqnarray*} and then,
being ${\displaystyle \frac{2\Phi}{c^2}=-\frac{R_{s}}{r}}$ (where  $R_{s}$ is
the Schwarzschild radius) and
${\displaystyle u=\frac{1}{r}}$ it is

\begin{eqnarray*}
&&E^{2}-h^{2}\left(1-R_{s}^{2}u^{2}\right)\left(u'^{2}+u^{2}\right)+\\ &&
\frac{4R_{S}uE}{c}\left[\left(\cos\varphi+\sin\varphi\right)u'+\left(\cos\varphi-\sin\varphi\right)u^{2}\right]
 \nonumber\\ &&=  \left(1-R_{S}u\right)\label{eq:L4}\,.\end{eqnarray*}
By deriving such an equation, it is easy to show that, if the
relativistic and gravitomagnetic terms are discarded, the
Newtonian theory is recovered, being
\begin{displaymath}
u''+u=\frac{R_s}{2L^2}\,.\end{displaymath} This result probes the
self-consistency of the problem. However, it is nothing else but a
particular case since we have assumed the planar motion. This
planarity condition does not hold in general  if  gravitomagnetic
corrections are taken into account.
From the above Lagrangian (\ref{Lagrangiannostra}), it is
straightforward to derive the equations of motion

  \begin{eqnarray}
 \ddot{r}&=&\frac{1}{c r \left(r c^2+2 G M\right)}
 \Big[c \left(r c^2+G M\right) \left(\dot{\theta}^2+\sin ^2\theta \dot{\phi}^2\right) r^2+
 \nonumber\\
 && -4 G M \dot{t} \left((\cos\theta(\cos\phi+\sin\phi)-\sin \theta)\dot{\theta}+\right. \nonumber\\ &&\left.
 \sin\theta (\cos\phi-\sin \phi)\dot{\phi} \right) r+c G M \dot{r}^2-c G M
 \dot{t}^2\Big]\,,\nonumber\\
 \label{ddr}
  \end{eqnarray}

 \begin{eqnarray}
\ddot{\phi} &=&-\frac{2}{r^2 \left(r c^3+2 G M
   c\right)}  c \cot \theta \left(r c^2+2 G M\right) \dot{\theta}\dot{ \phi} r^2\nonumber\\ &&+\dot{r}
   \left(2 G M \csc \theta (\sin\phi-\cos\phi) \dot{t}+c r \left(r c^2+G
   M\right) \dot{\phi}\right)\,,\nonumber\\ &&
   \label{ddphi}
   \end{eqnarray}

\begin{eqnarray}
 \ddot{\theta}&=&\frac{1}{r^2 \left(r c^3+2 G M
   c\right)}c \cos\theta r^2 \left(r c^2+2 G M\right) \sin\theta \dot{\phi}^2\nonumber\\ &&+\dot{r}
   \left(4 G M (\cos\theta (\cos \phi+\sin\phi)-\sin\theta)\dot{t}+\right. \nonumber\\ &&\left.-
   2 c r \left(r c^2+G M\right) \dot{\theta}\right)\,,\nonumber\\
   \label{ddtheta}
   \end{eqnarray}
   
corresponding to the spatial components of the geodesic Eq.
(\ref{eq:ddue_dsigma}). The time component $\ddot{t}$  is not
necessary for the discussion of orbital motion. Being the
Lagrangian (\ref{Lagrangiannostra}) $\mathcal{L}=1$ it is easy to
achieve a first integral for $\dot{r}$ which is a natural
constrain equation related to the energy.

Our aim is to show how gravitomagnetic effects modify the orbital
motion and what the parameters determining the stability of the
problem are. As we will see, the energy and the mass, essentially,
determine the stability. Beside the standard periastron precession
of GR, a nutation effect is genuinely induced by gravitomagnetism
and stability greatly depends on it. A fundamental issue for this
study is to achieve the orbital phase space portrait.

\begin{figure}[ht]
\begin{tabular}{|c|c|}
\hline
\tabularnewline
\includegraphics[scale=0.35]{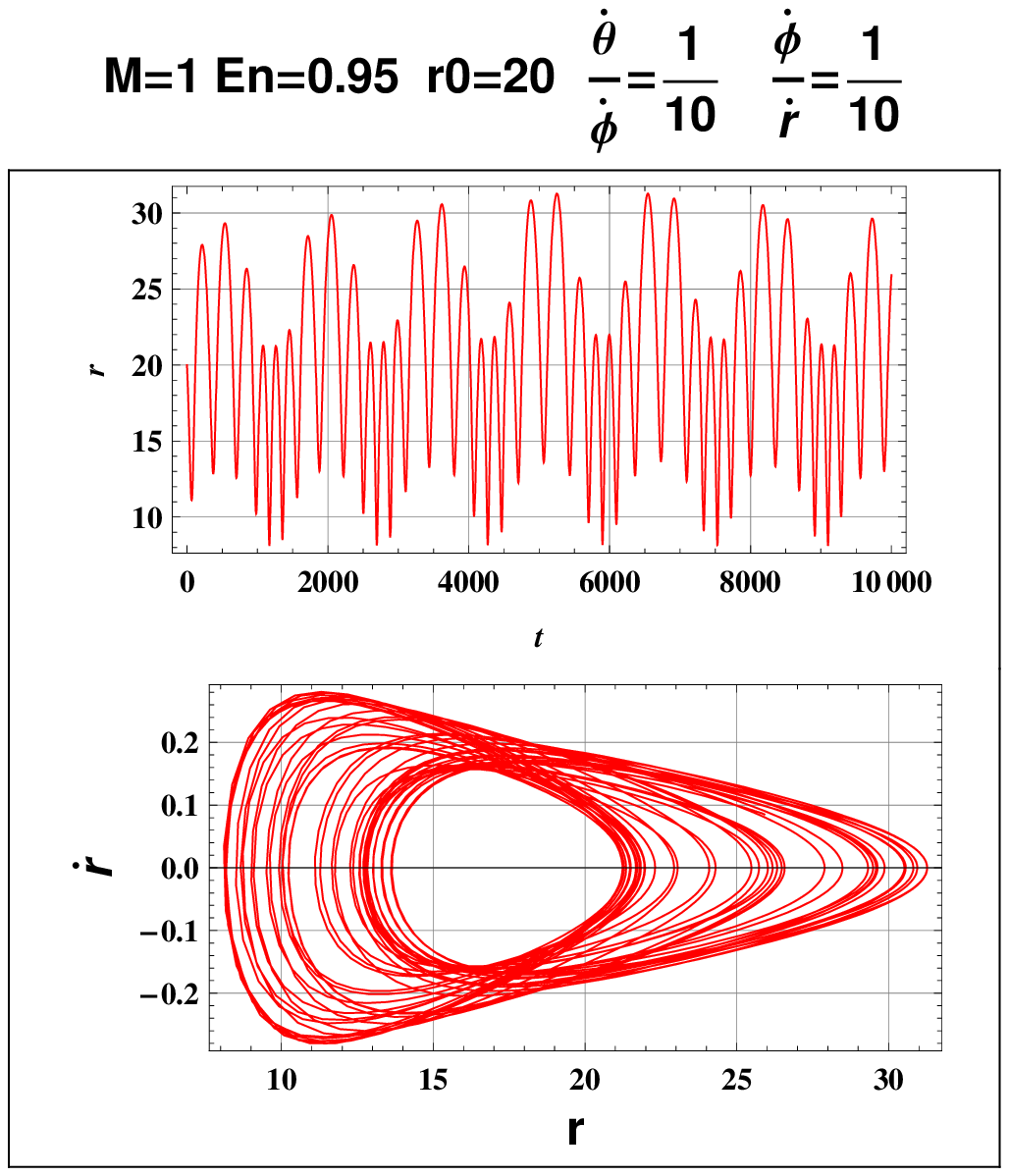}
\includegraphics[scale=0.35]{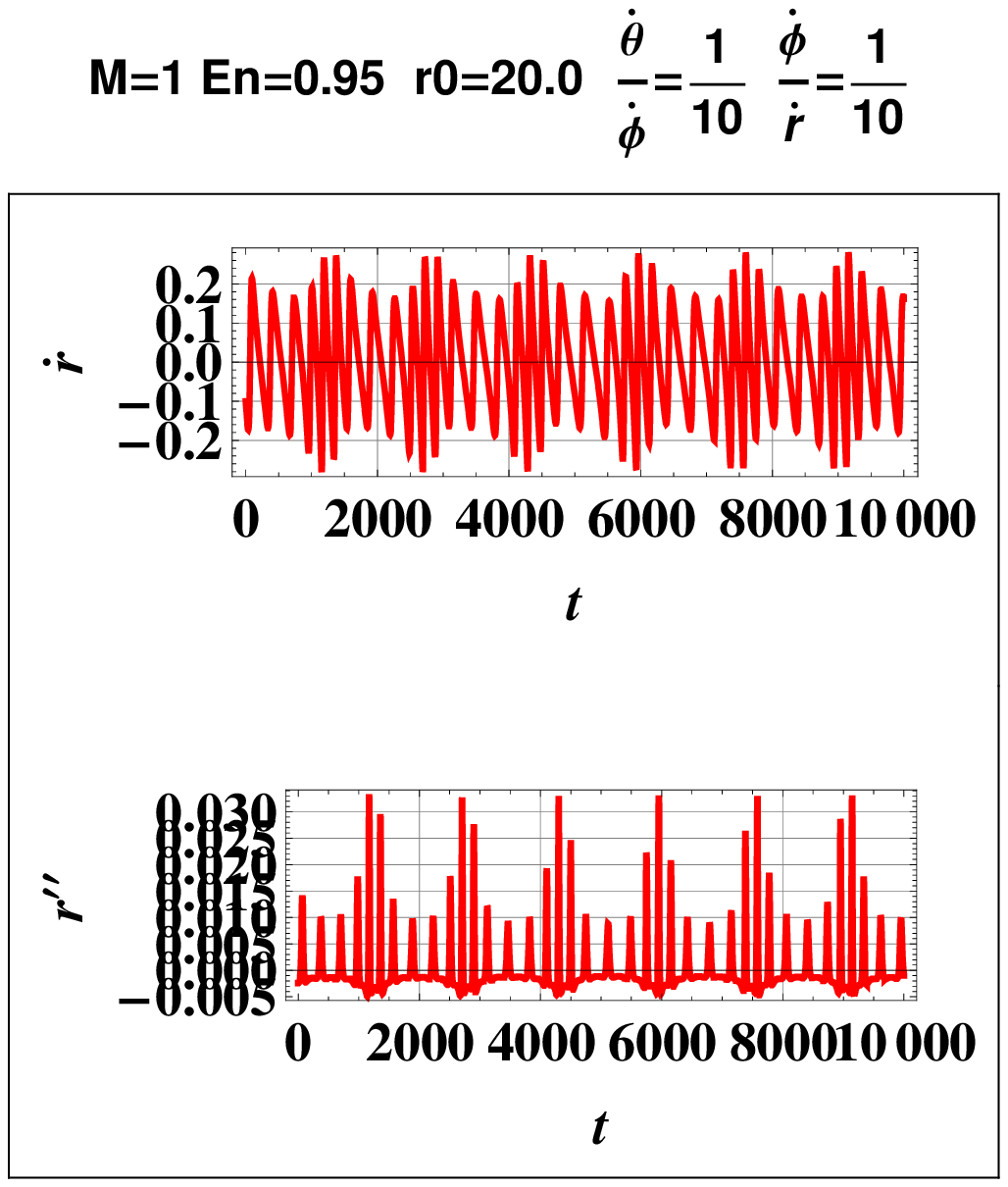}
 \tabularnewline
\hline
\end{tabular}
\caption {Plots along the panel lines of:  $r(t)$  (upper
left),phase portrait of $r(t)$ versus $\dot{r}(t)$ (bottom left),
$\dot{r}(t)$ (upper right) and $\ddot{r}(t)$ (bottom right) for a
star  of $1 M_{\odot}$.The examples we are showing were obtained
solving the system for the following parameters and initial
conditions: $\mu\approx 1 M_{\odot}$, $E=0.95$,$\phi_{0}=0$,
$\theta_{0}=\frac{\pi}{2}$,$\dot{\theta_{0}}=\frac{1}{10}\dot{\phi_{0}}$,$\dot{\phi_{0}}=-\frac{1}{10}\dot{r}_{0}$
and $\dot{r}_{0}=-\frac{1}{100}$ and $r_{0}=20 \mu$. The stiffness
is evident from the trend of  $\dot{r}(t)$ and
$\ddot{r}(t)$}\label{Fig:stiffness}
\end{figure}

\begin{figure}[!ht]
\includegraphics[height=0.30\textheight]{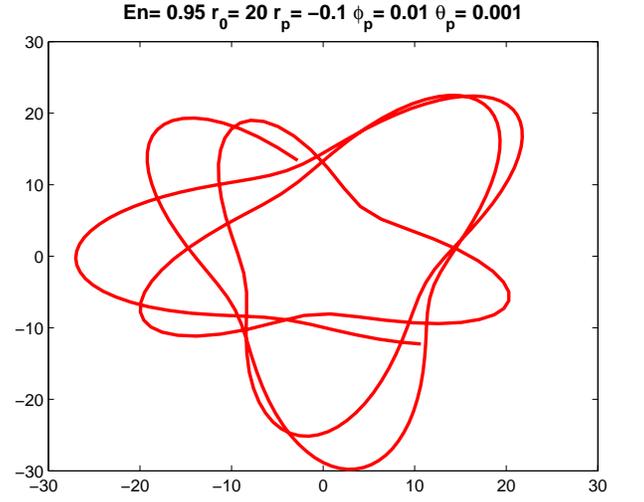}
\caption {Plots of basic orbits (left)
The initial values are:
$M=1M_{\bigodot}$; $E_n=0.95$ in mass units;
 $r_0=20$ in Schwarzschild radii; $\dot\phi=-\frac{\dot
r}{10}$; $\dot\theta=\frac{\dot\phi}{10}$. } \label{fig:orbit}
\end{figure}

\begin{figure}[!ht]
\includegraphics[height=0.30\textheight]{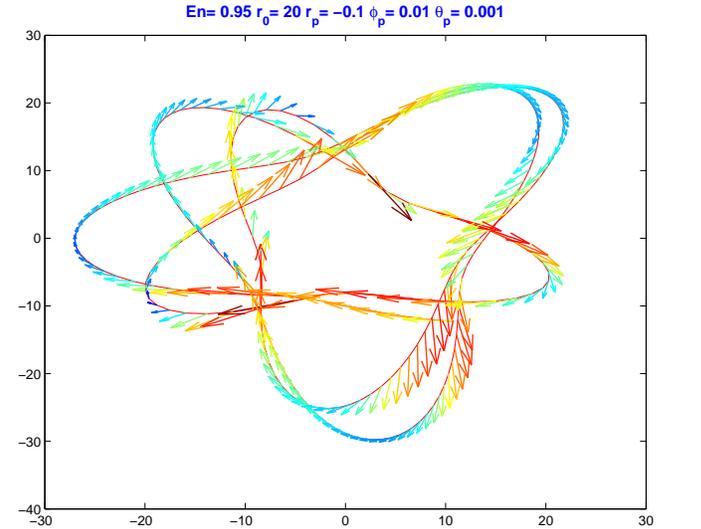}\tabularnewline
\caption {Plots of basic orbits with the
associated velocity field. The arrows indicate the
instantaneous velocities. The initial values are:
$M=1M_{\bigodot}$; $E_n=0.95$ in mass units;
 $r_0=20$ in Schwarzschild radii; $\dot\phi=-\frac{\dot
r}{10}$; $\dot\theta=\frac{\dot\phi}{10}$. } \label{fig:orbit1}
\end{figure}

\begin{figure}[!ht]
\begin{tabular}{|c|c|}
\hline
\includegraphics[height=0.15\textheight]{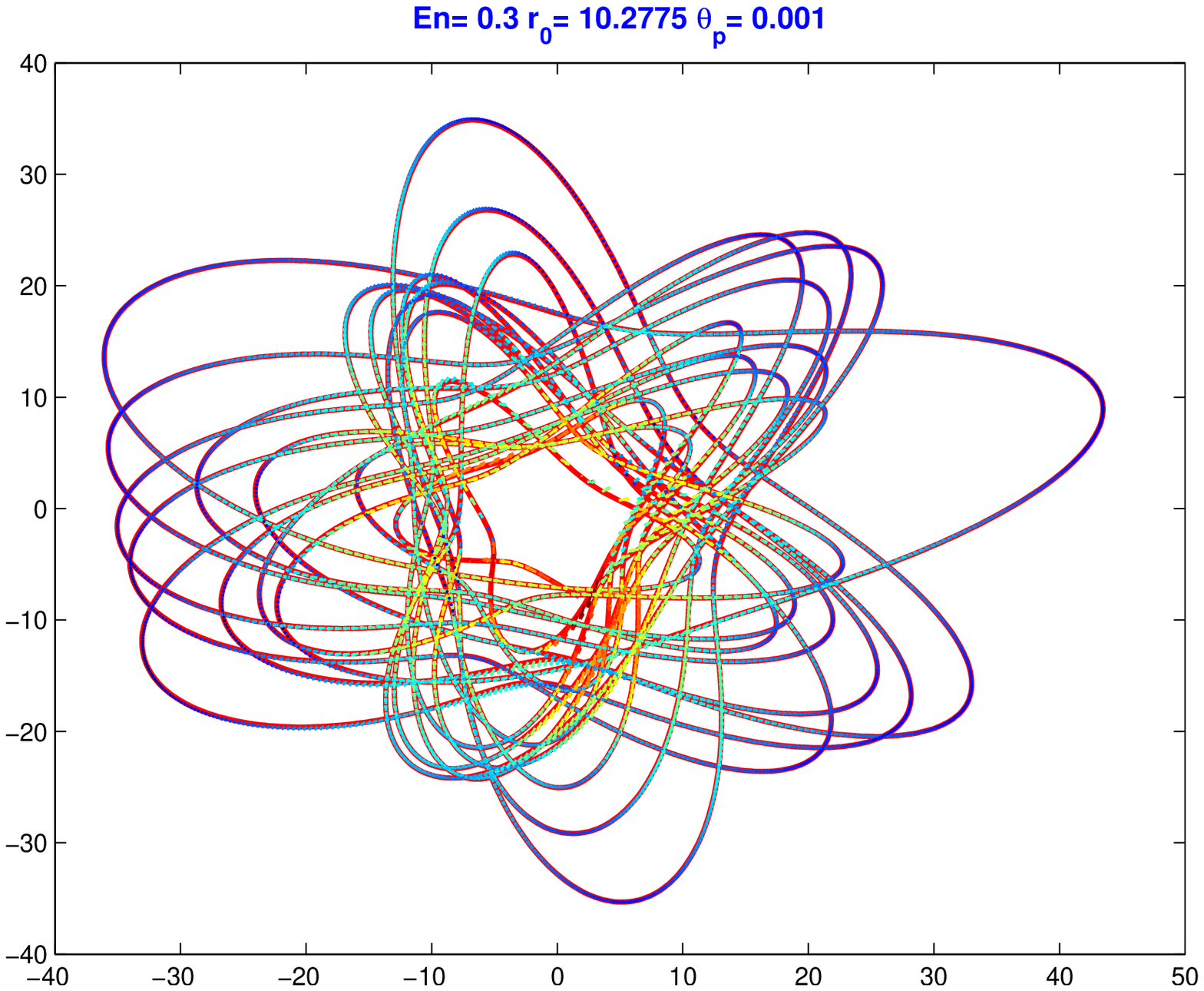}&
\includegraphics[height=0.15\textheight]{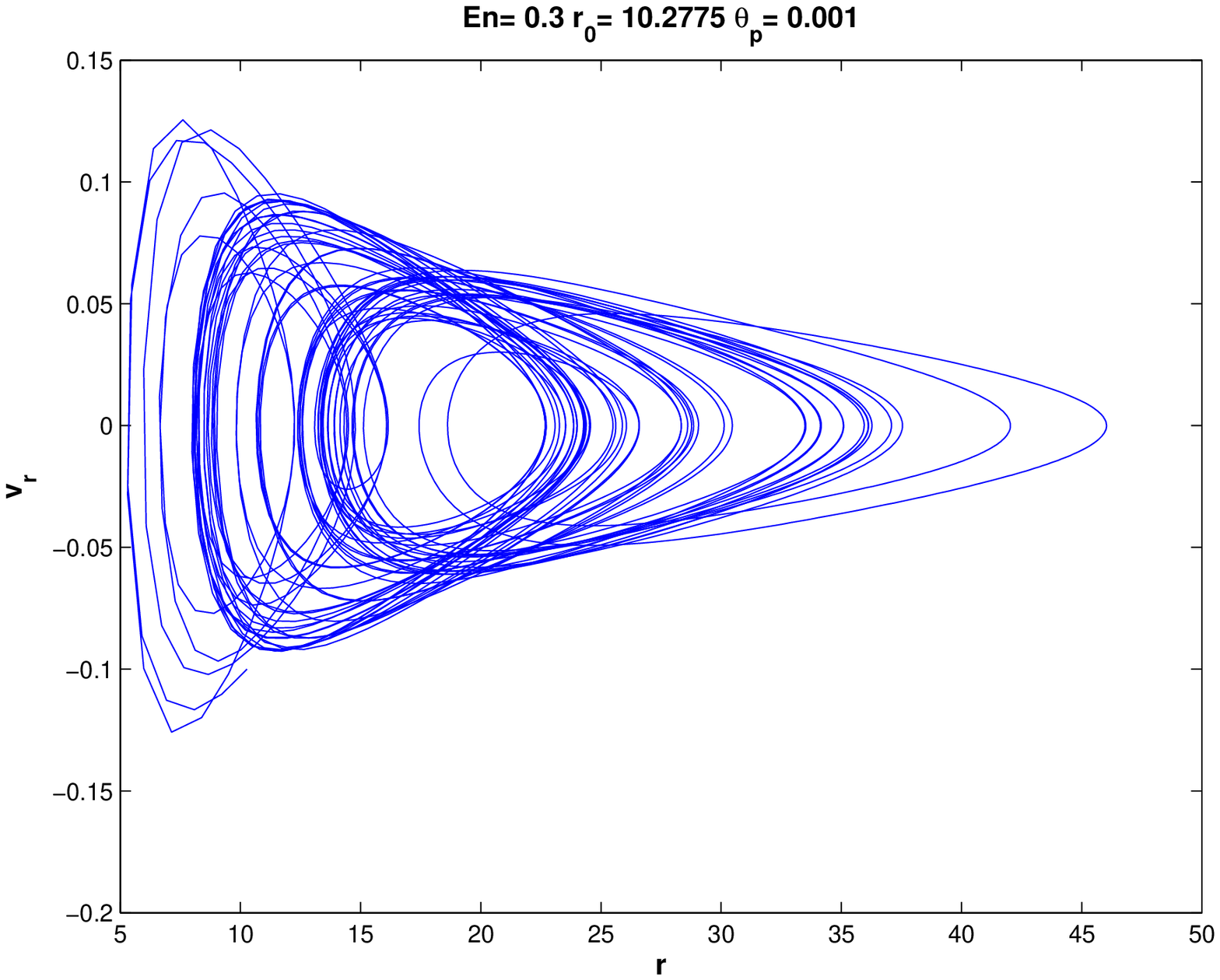}\tabularnewline
\hline
\includegraphics[height=0.15\textheight]{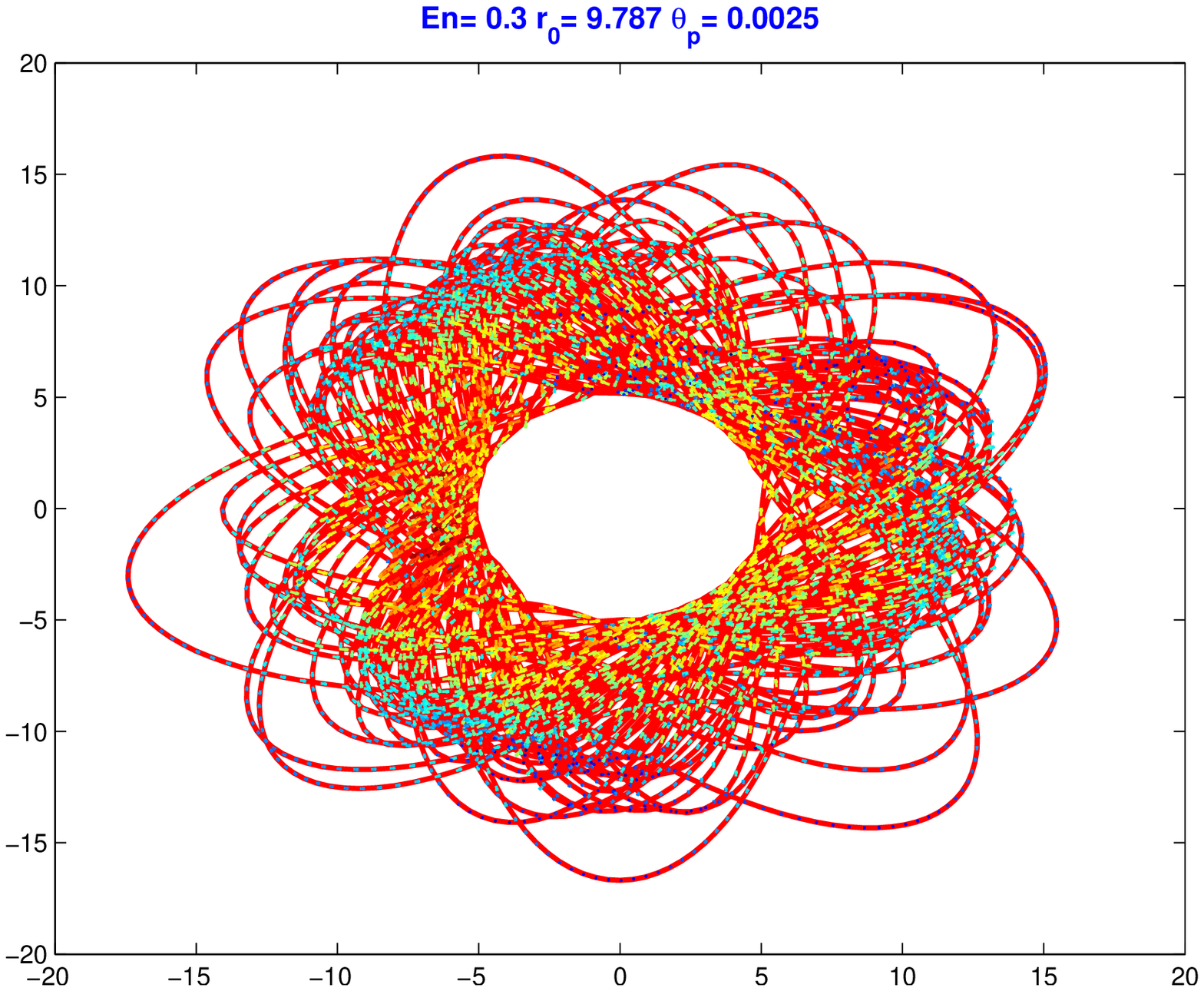}&
\includegraphics[height=0.15\textheight]{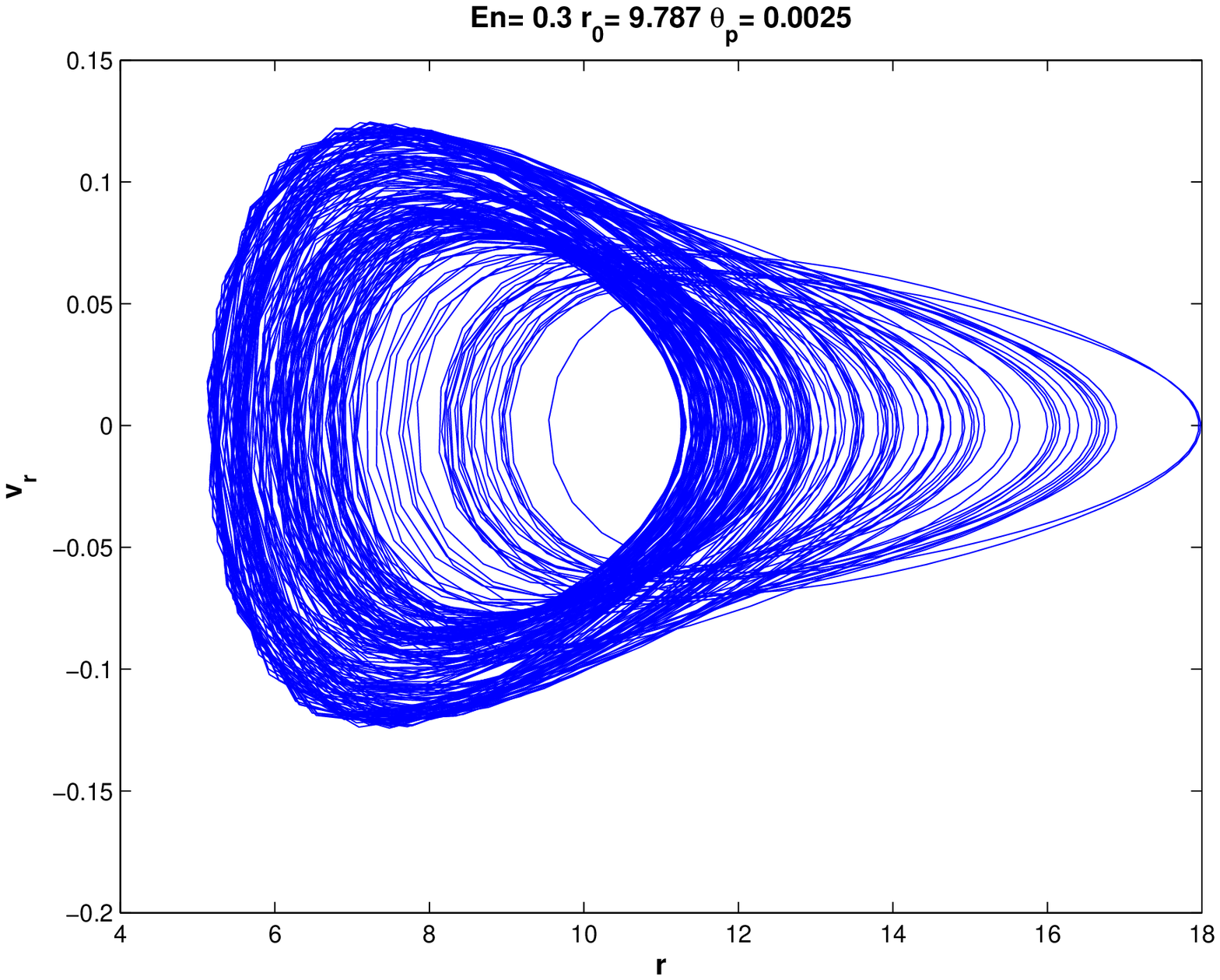}\tabularnewline
\hline
\includegraphics[height=0.15\textheight]{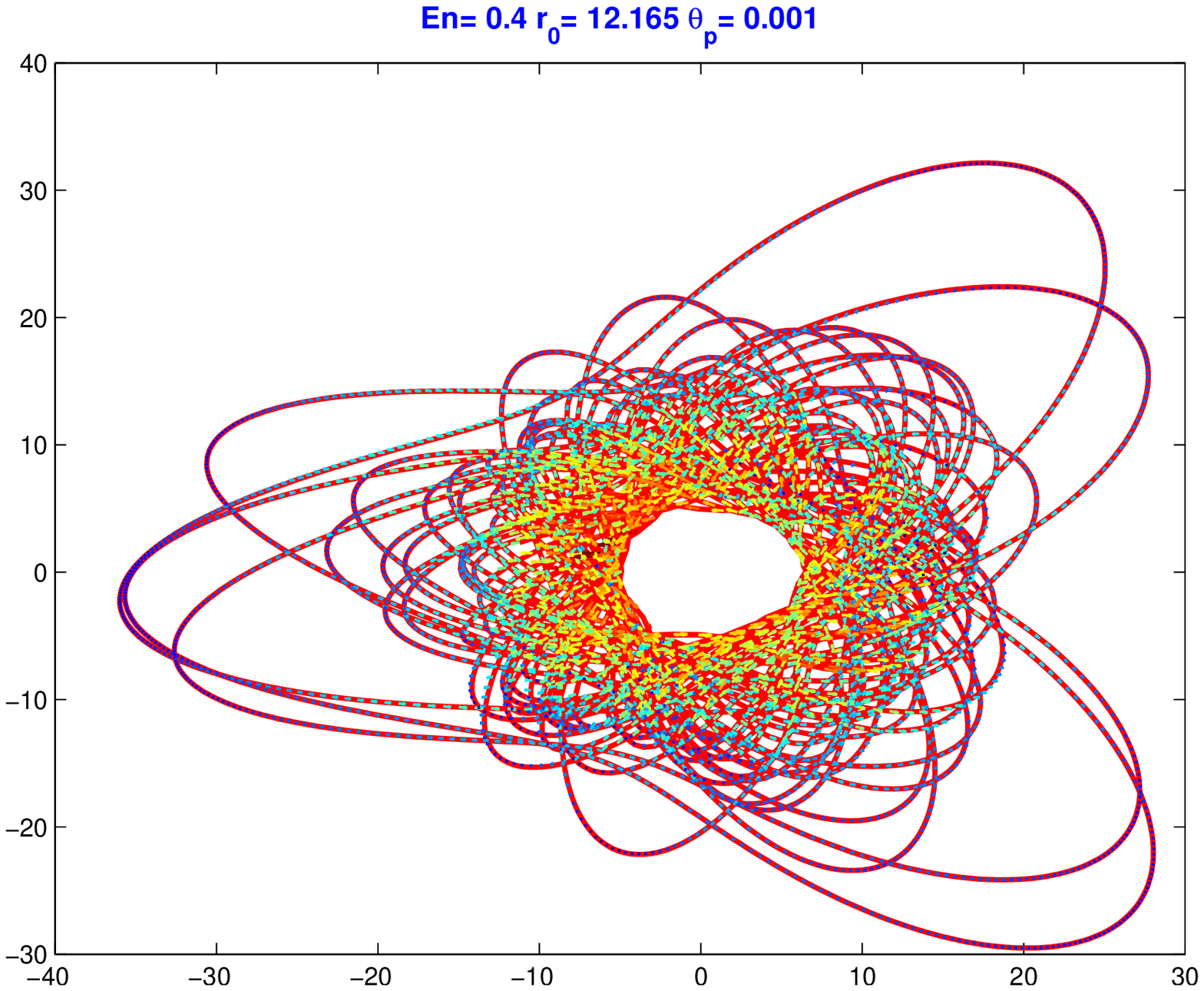}&
\includegraphics[height=0.15\textheight]{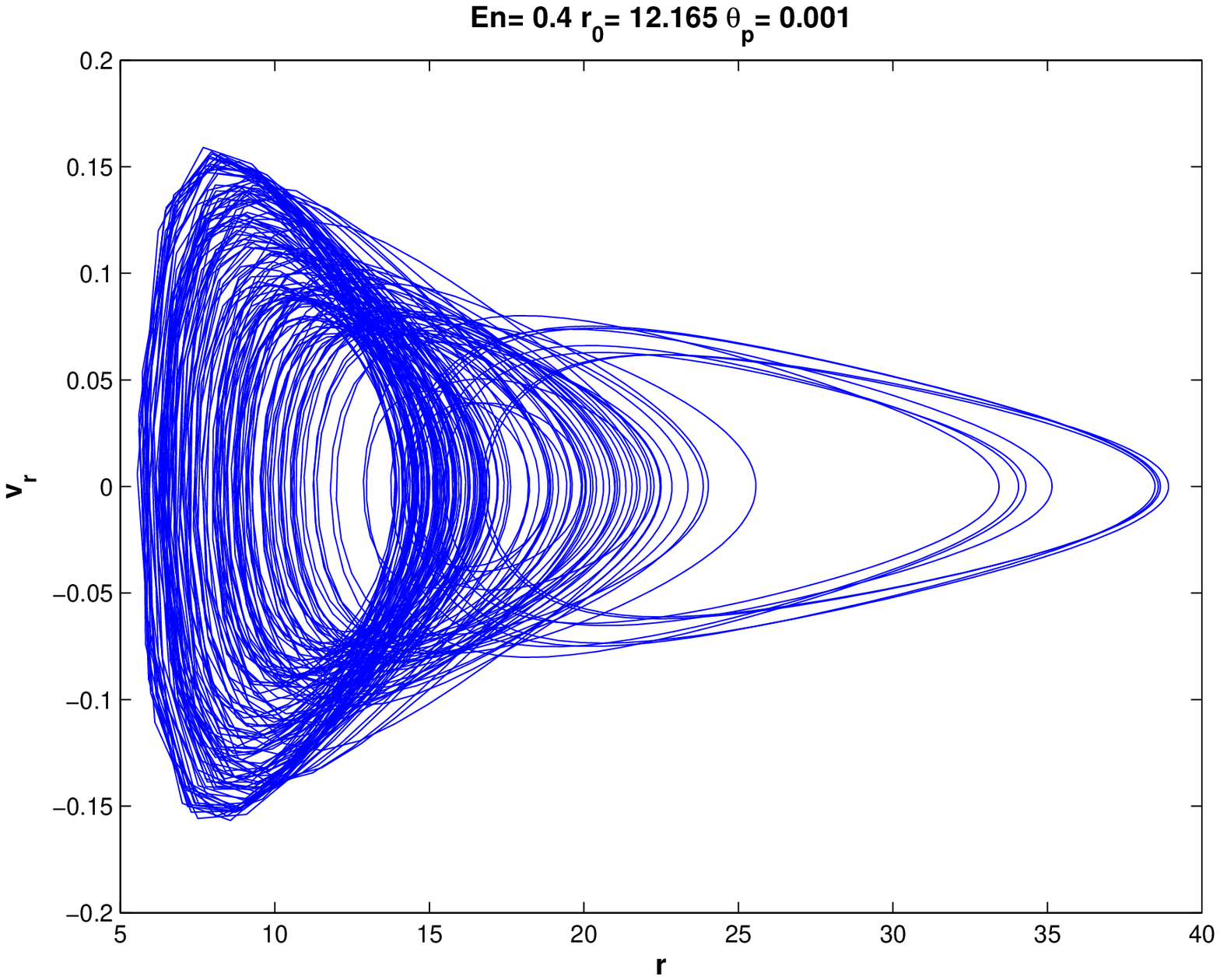}\tabularnewline
\hline
\includegraphics[height=0.15\textheight]{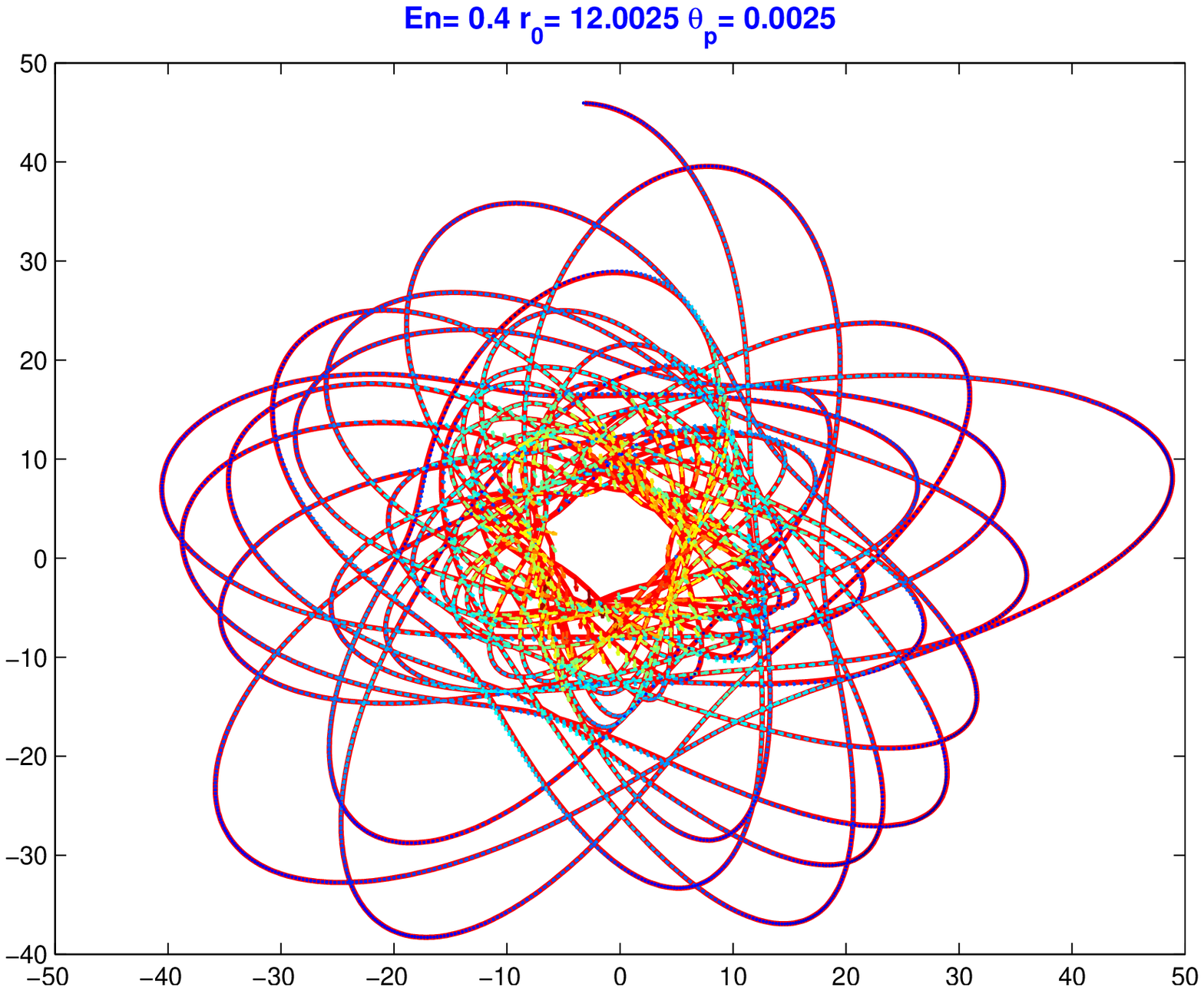}&
\includegraphics[height=0.15\textheight]{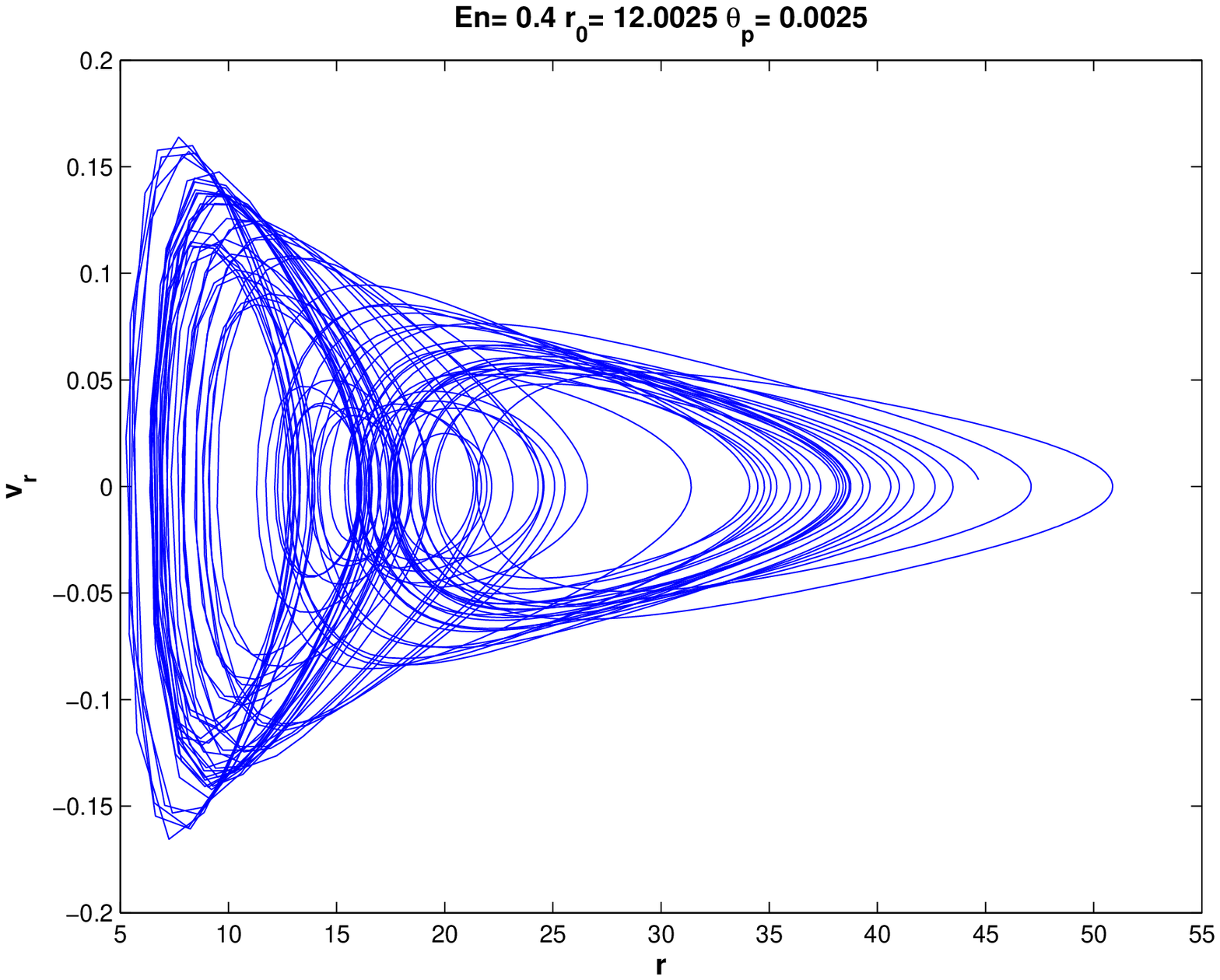}\tabularnewline
\hline
\end{tabular}
\caption {Plots of orbits with  various  energy values. For each
value of  energy, four plots are shown: the first on the left
column is the orbit, with the orbital velocity field in false
colors. The color scale goes from blue to red in increasing
velocity. The second on the left column is the orbit with a
different nutation angular velocity. On the right column, the
phase portraits $\displaystyle{\dot r=\dot r(r(t))}$ are shown. Energy varies
from $0.3$ to $0.4$, in mass units. The stability of the system is
highly sensitive either to very small variation of $r_{0}$ or to
variation on the initial conditions on both precession and
nutation angular velocities: it is sufficient a variation of few
percent on $r_{0}$ to induce system instability} \label{fig:1}
\end{figure}

In Fig.\ref{Fig:stiffness}, the results for a given value of nutation
angular velocity with a time span of $10000$ steps is shown. It is
interesting to see that, by increasing the initial nutation
angular velocity, being fixed all the other initial conditions, we
get curves with decreasing frequencies for $\dot{r}(t)$ and
$\ddot{r}(t)$. This fact is relevant to have an insight on  the
orbital motion stability (see Fig.\ref{fig:1}). The effect of gravitomagnetic terms are taken into
account, in Fig.
\ref{fig:orbit}-\ref{fig:orbit1}, showing the basic orbits  (left) and the orbit
with the associated velocity field in false colors (right). From a
rapid inspection of the right panel, it is clear the sudden
changes of velocity direction induced by the gravitomagnetic
effects.

To show  the orbital velocity field, a rotation
and a projection of the orbits along the axes of maximal energy can be performed.
In other words, by a {\it Singular Value Decomposition} of the
de-trended positions and velocities, the
eigenvectors corresponding to the largest eigenvalues can be selected, and, of
course, those representing the highest energy components (see Fig
\ref{fig:orbit}-\ref{fig:orbit1}).

\begin{figure}[!ht]
\includegraphics[height=0.30\textheight]{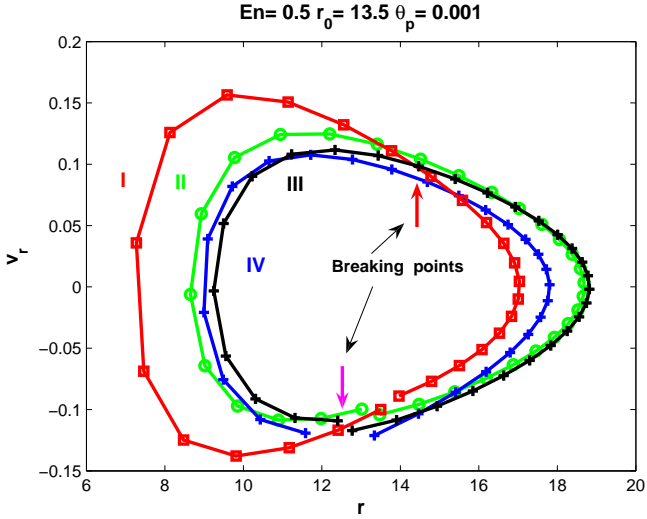}
\caption {Breaking points  examples: on the left  panel, the first
four orbits in the phase plane are shown: the red one is labelled
 I, the green is II, the black is  III and the fourth is
IV. As it is possible to see, the orbits in the phase plane are
not closed and they do not overlap at the orbital closure points;
 we have called this feature {\it breaking points}. In this dynamical
situation, a small perturbation can lead the system to a
transition to the chaos as prescribed by the
Kolmogorov-Arnold-Moser (KAM) theorem  \cite{binney}. On the right
panel, it is shown the initial orbit with the initial (square) and
final (circles) points marked in black.}
\label{fig:Breakstab}
\end{figure}

\begin{figure}[!ht]
\includegraphics[height=0.30\textheight]{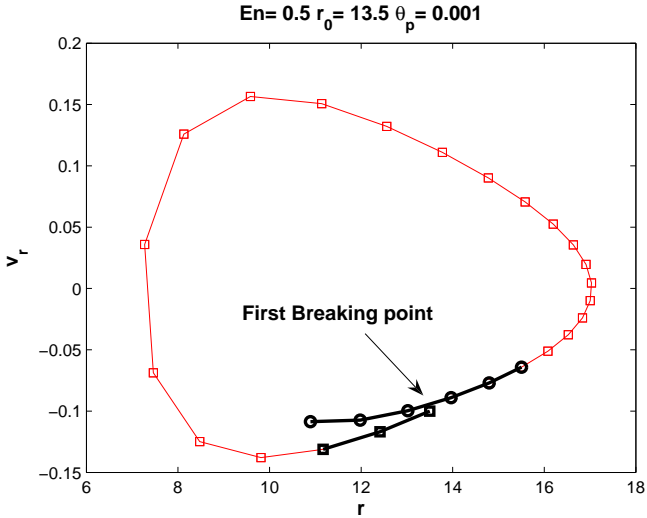}
\caption{In this figure it is shown the initial orbit with
the initial (squares) and final(circles) points marked
in black.}
 \label{fig:Breakstab1}
\end{figure}
The  above differential equations for the parametric orbital
motion are non-linear and with time-varying coefficients. In order
to have a well-posed Cauchy problem, we have to  define:
\begin{itemize}
\item the initial and final boundary condition problems;
\item the stability and the dynamical equilibrium of solutions.
\end{itemize}
We can start by  solving the Cauchy problem, as in the classical
case,  for the initial condition putting $\dot{r}=0$ ,
$\displaystyle{\dot{\phi}=0}$, $\displaystyle{\dot{\theta}=0}$ and $\displaystyle{\theta=\frac{\pi}{2}}$ and
the result we get is that the orbit is not planar being
$\displaystyle{\ddot{\theta}\neq 0}$. In this case,  we are compelled to solve
numerically the system of second order differential equations and
to treat carefully the initial conditions, taking into account the
high non-linearity of the system. A similar discussion, but for
different problems, can be found in \cite{cutler,cutler1}.

A series of numerical trials on the orbital parameters can be done
in order to get an empirical insight on the orbit stability. The
parameters involved in this analysis are the mass, the energy, the
orbital radius, the initial values of $r,\phi,\theta$ and the
angular precession and nutation velocities $\dot{\phi}$ and
$\dot{\theta}$ respectively. We have empirically assumed initial
conditions on $\dot r$, $\dot\phi$ and $\dot\theta$.

The trials can be organized in two series, i.e.
constant mass and energy variation and constant energy and mass
variation.
\begin{itemize}
\item In the first class of trials, we assume the mass equal
to $M=1M_{\bigodot}$  and the energy $E_n$ (in mass units) varying
step by step. The initial orbital radius $r_0$ can  be  changed,
according to the step in energy: this allows  to find out
numerically the dynamical equilibrium of the orbit. We have also
chosen, as varying parameters, the ratios of the precession
angular velocity $\dot\phi$ to the radial angular velocity $\dot
r$ and the ratio of the nutation angular velocity $\dot\theta$ and
the precession angular velocity $\dot\phi$. The initial condition
on $\phi$ has been assumed to be $\phi_0=0$ and the initial
condition on $\theta$ has been $\displaystyle{\theta_0=\frac{\pi}{2}}$. For $M=1$
(in Solar masses) , $\displaystyle{\frac{\dot\theta}{\dot\phi}=\frac{1}{2}}$ and
$\displaystyle{\dot{\phi}=-\frac{\dot{r}}{10}}$,  can be found out two different
empirical linear equations, according to the different values of
$\displaystyle{\dot{\theta},\dot{\phi}}$. One obtains  a rough guess of the initial
distance $r_0=r_0(E_n)$ around which it is possible to find a guess
on the equilibrium of the initial radius, followed by trials and
errors procedure.

\item In the second class of trials, we have assumed the
variation of the initial orbital radius for different values of
mass at a constant energy value equal to $E_n=0.95$ in mass units.
With this conditions, we assume ${\displaystyle
\dot\phi=\frac{\dot{r}}{10}}$ and assume that $\dot{\theta}$ takes
the two values $1/2$ and $1/10$. One can approach the problem also
considering the mass parameterization, at a given fixed energy, to
have an insight of the effect of  mass variation on the initial
conditions. The masses have been varied between 0.5 and 20 Solar
masses and the distances have been found to vary according to the
two 3rd-order polynomial functions, according to the different
values of $\dot{\theta}$ with respect to the mass (for details see [\cite{SMFL}])
\end{itemize}

In summary, the numerical calculations, if optimized, allow to put
in evidence the specific contributions of gravitomagnetic
corrections on orbital motion. In particular, specific
contributions due to nutation and precession emerge when higher
order terms in $v/c$ are considered.

The conclusion of this part of the review is that orbits are
highly characterized by the velocity regime of the moving bodies.
The order of the parameter $v/c$ determines the global shape of
the trajectories. Our task is now to show how the motion of
sources is related to the features of emitted GWs, in particular
to their production and to the  profile of waves.

 \part{\large Production and signature of gravitational waves }
The first derivation of gravitational radiation in GR
is due to Einstein.  His initial calculation \cite{ae1916} was
"marred by an error in calculation" (Einstein's words), and was
corrected in 1918 \cite{ae1918} (albeit with an overall factor of
two error).  Modulo a somewhat convoluted history (discussed in great
detail by Kennefick \cite{dk1997}) owing (largely) to the
difficulties of analyzing radiation in a nonlinear theory, Einstein's
final result stands today as the leading-order "quadrupole formula" for
gravitational wave emission.  This formula plays a role in gravity
theory analogous to the dipole formula for electromagnetic radiation,
showing GWs arise
from accelerated masses exactly as electromagnetic waves arise from
accelerated charges.
The quadrupole formula tells us that GWs are difficult to produce ---
very large masses moving at relativistic speeds are needed.  This
follows from the weakness of the gravitational interaction.  A
consequence of this is that it is {\it extremely} unlikely there will
ever be an interesting laboratory source of GWs.  The only objects
massive and relativistic enough to generate detectable GWs are
astrophysical.  Indeed, experimental confirmation of the existence of
GWs has come from the study of binary neutron star systems --- the
variation of the mass quadrupole in such systems is large enough that
GW emission changes the system's characteristics on a timescale short
enough to be observed.
Intuitively, it is clear that measuring these waves must be difficult
--- the weakness of the gravitational interaction ensures that the
response of any detector to gravitational waves is very small.
Nonetheless, technology has brought us to the point where detectors
are now beginning to set interesting upper limits on GWs from some
sources \cite{ligo1,ligo2,ligo3,ligo4}. The first direct detection could be,hopefully, not too far in the future.\par
\section{Gravitational waves in linearized gravity}
The most natural starting point for any discussion of GWs is {\it
linearized gravity} \cite{gravitation,shapiro,maggiore}.  Linearized gravity is an adequate approximation
to GR  when the spacetime metric, may be
treated as deviating only slightly from a flat metric, $\eta_{\mu\nu}$:
\begin{displaymath}
g_{\mu\nu} = \eta_{\mu\nu} + h_{\mu\nu},\qquad ||h_{\mu\nu}|| \ll 1\;.
\end{displaymath}
Here
$||h_{\mu\nu}||$ means ``the magnitude of a typical non-zero component of
$h_{\mu\nu}$''.  Note that the condition $||h_{\mu\nu}|| \ll 1$ requires both
the gravitational field to be weak \footnote{We will work in geometrized coordinates putting $c=G=1$}, and in addition constrains the
coordinate system to be approximately Cartesian.  We will refer to
$h_{\mu\nu}$ as the metric perturbation; as we will see, it encapsulates
GWs, but contains additional, non-radiative degrees of freedom as
well.  In linearized gravity, the smallness of the perturbation means
that we only keep terms which are linear in $h_{\mu\nu}$ --- higher order
terms are discarded.  As a consequence, indices are raised and lowered
using the flat metric $\eta_{\mu\nu}$.  The metric perturbation $h_{\mu\nu}$
transforms as a tensor under Lorentz transformations, but not under
general coordinate transformations.

We now compute all the quantities which are needed to describe
linearized gravity.  The components of the affine connection
(Christoffel coefficients) are given by
\begin{eqnarray*}
{\Gamma^\mu}_{\nu\rho} &=& \frac{1}{2}\eta^{\mu\sigma}\left(\partial_\rho h_{\rho\nu} +
\partial_\nu h_{\sigma\rho} -\partial_\sigma h_{\nu\rho}\right)
\nonumber\\
&=& \frac{1}{2}\left(\partial_\rho {h^\mu}_\nu + \partial_\nu {h^\mu}_{\rho}
-\partial^\mu h_{\nu\rho}\right)\;.
\label{eq:connection}
\end{eqnarray*}
Here $\partial_\mu$ means the partial derivative $\partial / \partial x^\mu$.
Since we use $\eta_{\mu\nu}$ to raise and lower indices, spatial indices
can be written either in the "up" position or the "down" position
without changing the value of a quantity: $f^x = f_x$.  Raising or
lowering a time index, by contrast, switches sign: $f^t = -f_t$.  The
Riemann tensor we construct in linearized theory is then given by
\begin{eqnarray}
R^\mu_{\nu\rho\sigma} &=& \partial_\rho{\Gamma^\mu}_{\nu\sigma} - \partial_\sigma{\Gamma^\mu}_{\nu\rho}
\nonumber\\
&=& \frac{1}{2}\left(\partial_\rho\partial_\nu {h^\mu}_\sigma +
\partial_\sigma\partial^\mu h_{\nu\rho} - \partial_\rho\partial^\mu h_{\nu\sigma} -
\partial_\sigma\partial_\nu {h^\mu}_\rho\right)\,.\nonumber\\
\label{eq:riemann}
\end{eqnarray}
From this, we construct the Ricci tensor
\begin{displaymath}
R_{\mu\nu} = {R^\rho}_{\mu\rho\nu}
= \frac{1}{2}\left(\partial_\rho\partial_\nu {h^\rho}_\mu + \partial^\rho
\partial_\mu h_{\nu\rho} - \Box h_{\mu\nu} - \partial_\mu\partial_\nu h\right)\;,
\label{eq:ricci}
\end{displaymath}
where $h = {h^\mu}_\mu$ is the trace of the metric perturbation, and $\displaystyle{\Box
= \partial_\rho\partial^\rho = \nabla^2 - \partial_t^2}$ is the wave
operator.  Contracting once more, we find the curvature scalar:
\begin{displaymath}
R = {R^\mu}_\mu = \left(\partial_\rho\partial^\mu {h^\rho}_\mu - \Box h\right)
\label{eq:scalar}
\end{displaymath}
and finally build the Einstein tensor:
\begin{eqnarray*}
G_{\mu\nu} &=& R_{\mu\nu} - \frac{1}{2}\eta_{\mu\nu} R
\nonumber\\
&=& \frac{1}{2}\left(\partial_\rho\partial_\nu {h^\rho}_\mu + \partial^\rho
\partial_\mu h_{\nu\rho} - \Box h_{\mu\nu} - \partial_\mu\partial_\nu h
\right.\nonumber\\
& &\qquad\left.-\eta_{\mu\nu}\partial_\rho\partial^\sigma {h^\rho}_\sigma + \eta_{\mu\nu} \Box
h\right)\;.
\label{eq:einstein_h}
\end{eqnarray*}
This expression is a bit unwieldy.  Somewhat remarkably, it can be
cleaned up significantly by changing notation: rather than working
with the metric perturbation $h_{\mu\nu}$, we use the {\it trace-reversed}
perturbation $\displaystyle{\bar h_{\mu\nu} = h_{\mu\nu} - \frac{1}{2}\eta_{\mu\nu} h}$.  (Notice
that $\displaystyle{\bar {h^\mu}_\mu = -h}$, hence the name ``trace reversed''.)
Replacing $h_{\mu\nu}$ with\\
 $\displaystyle{\bar h_{\mu\nu} + \frac{1}{2}\eta_{\mu\nu} h}$ in Eq.\
(\ref{eq:einstein_h}) and expanding, we find that all terms with the
trace $h$ are canceled.  What remains is
\begin{eqnarray}
G_{\mu\nu} = \frac{1}{2}\left(\partial_\sigma\partial_\nu {{\bar h}^\rho}_{\ \mu} +
\partial^\rho \partial_\mu \bar h_{\mu\nu} - \Box \bar h_{\mu\nu} -\eta_{\mu\nu}
\partial_\rho\partial^\sigma {{\bar h}^\rho}_{\ \sigma}\right)\;.\nonumber\\
\label{eq:einstein_hbar}
\end{eqnarray}
This expression can be simplified further by choosing an appropriate
coordinate system, or {\it gauge}.  Gauge transformations in general
relativity are just coordinate transformations.  A general
infinitesimal coordinate transformation can be written as ${x^a}' =
x^\mu + \xi^\mu$, where $\xi^\mu(x^\nu)$ is an arbitrary infinitesimal vector
field.  This transformation changes the metric via

\begin{equation}
h_{\mu\nu}' = h_{\mu\nu} - 2\partial_{(\mu} \xi_{\nu)}\;,
\label{eq:metric_transform}
\end{equation}
so that the trace-reversed metric becomes
\begin{eqnarray*}
\bar h_{\mu\nu}' &=& h_{\mu\nu}' - \frac{1}{2}\eta_{\mu\nu} h'
\nonumber\\
&=& \bar h_{\mu\nu} - 2\partial_{(\nu}\xi_{\mu)} + \eta_{\mu\nu}\partial^\rho\xi_\rho\;.
\label{eq:barmetric_transform}
\end{eqnarray*}
A class of gauges that are commonly used in studies of radiation are
those satisfying the {\it Lorenz gauge} condition
\begin{equation}
\partial^\mu \bar h_{\mu\nu} = 0.
\label{eq:lorentzgauge}
\end{equation}
(Note the close analogy to Lorentz gauge
\footnote{Fairly recently, it
has become widely recognized that this gauge was in fact invented by
Ludwig Lorenz, rather than by Hendrik Lorentz.  The inclusion of the
"t" seems most likely due to confusion between the similar names;
see Ref.\cite{vanbladel} for detailed discussion.  Following the
practice of Griffiths (\cite{griffiths}, p. 421), we bow to the
weight of historical usage in order to avoid any possible confusion.}

in electromagnetic theory, $\partial^\mu A_\mu = 0$, where $A_\mu$ is the
potential vector.)
Suppose that our metric perturbation is not in Lorentz gauge.  What
properties must $\xi_\mu$ satisfy in order to {\it impose} Lorentz
gauge?  Our goal is to find a new metric $h'_{\mu\nu}$ such that
$\partial^\mu \bar h'_{\mu\nu} = 0$:
\begin{eqnarray*}
\partial^\mu \bar h_{\mu\nu}' &=& \partial^\mu \bar h_{\mu\nu} -
\partial^\mu\partial_\nu\xi_\mu - \Box \xi_\nu + \partial_\nu \partial^\rho\xi_\rho
\nonumber\\
&=& \partial^\mu \bar h_{\mu\nu} - \Box\xi_\nu\;.
\label{eq:ll1}
\end{eqnarray*}
Any metric perturbation $h_{\mu\nu}$ can therefore be put into a Lorentz
gauge by making an infinitesimal coordinate transformation that satisfies
\begin{equation}
\Box \xi_\nu = \partial^\mu \bar h_{\mu\nu}\;.
\label{eq:wave}
\end{equation}
One can always find solutions to the wave equation (\ref{eq:wave}),
thus achieving Lorentz gauge.
The amount  of gauge freedom has now been reduced

The amount of gauge freedom has now been reduced from 4 freely
specifiable functions of 4 variables to 4 functions of 4 variables
that satisfy the homogeneous wave equation $\Box \xi^\nu =0$, or,
equivalently, to 8 freely specifiable functions of 3 variables on an
initial data hypersurface.

Applying the Lorentz gauge condition (\ref{eq:lorentzgauge}) to the
expression (\ref{eq:einstein_hbar}) for the Einstein tensor, we find
that all but one term vanishes:
\begin{displaymath}
G_{\mu\nu} = -\frac{1}{2}\Box \bar h_{\mu\nu}\;.
\label{eq:einstein_lg}
\end{displaymath}
Thus, in Lorentz gauges, the Einstein tensor simply reduces to the
wave operator acting on the trace reversed metric perturbation (up to
a factor $-1/2$).  The linearized Einstein equation is therefore
\begin{equation}
\Box \bar h_{\mu\nu} = - 16 \pi T_{\mu\nu}\;;
\label{eq:elin}
\end{equation}
in vacuum, this reduces to
\begin{equation}
\Box \bar h_{\mu\nu} = 0\;.
\label{eq:elin1}
\end{equation}
Just as in electromagnetism, the equation (\ref{eq:elin}) admits a
class of homogeneous solutions which are superpositions of plane
waves:
\begin{displaymath}
{\bar h}_{\mu\nu}({\bf x},t) = {\rm Re} \int d^3 k \ A_{\mu\nu}({\bf k}) e^{i
  ({\bf k} \cdot {\bf x} - \omega t)}\;.
\label{eq:planewaves}
\end{displaymath}
Here, $\omega = |{\bf k}|$.  The complex coefficients $A_{\mu\nu}({\bf
k})$ depend on the wavevector ${\bf k}$ but are independent of ${\bf
x}$ and $t$.  They are subject to the constraint $k^\mu A_{\mu\nu} = 0$
(which follows from the Lorentz gauge condition), with $k^\mu =
(\omega,{\bf k})$, but are otherwise arbitrary.  These solutions are the
gravitational waves.
    
 \subsection{Transverse traceless (TT) gauge in globally vacuum spacetimes}
\label{sec:TTgauge}
 We now specialize to globally vacuum spacetimes in which $T_{\mu\nu}=0$
everywhere, and which are asymptotically flat (for our purposes,
$h_{\mu\nu} \to 0$ as $r \to \infty$).  Equivalently, we specialize to the
space of homogeneous, asymptotically flat solutions of the linearized
Einstein Eq. (\ref{eq:elin}).  For such spacetimes one can, along
with choosing Lorentz gauge, further specialize the gauge to make the
metric perturbation be purely spatial

\begin{equation}
h_{00} = h_{0i} =0
\label{eq:spatial}
\end{equation}
and traceless
\begin{equation}
h = h_i^{\ i} =0.
\label{eq:traceless}
\end{equation}
The Lorentz gauge condition (\ref{eq:lorentzgauge}) then
implies that the spatial metric perturbation is transverse:
\begin{displaymath}
\partial_i h_{ij} =0.
\label{eq:transverse}
\end{displaymath}
This is called the transverse-traceless gauge, or TT gauge.  A metric
perturbation that has been put into TT gauge will be written
$\displaystyle{h_{\mu\nu}^{\rm TT}}$.  Since it is traceless, there is no distinction
between $\displaystyle{h^{\rm TT}_{\mu\nu}}$ and $\displaystyle{\bar h^{\rm TT}_{\mu\nu}}$.

The conditions (\ref{eq:spatial}) and (\ref{eq:traceless}) comprise 5
constraints on the metric, while the residual gauge freedom in Lorentz
gauge is parameterized by 4 functions that satisfy the wave equation.
It is nevertheless possible to satisfy these conditions,
essentially because the metric perturbation satisfies the linearized
vacuum Einstein equation.  When the TT gauge conditions are satisfied,
the gauge is completely fixed.

One might wonder {\it why} we would choose the TT gauge.  It is certainly
not necessary; however, it is extremely {\it convenient}, since the TT
gauge conditions completely fix all the local gauge freedom.  The
metric perturbation $\displaystyle{h_{\mu\nu}^{\rm TT}}$ therefore contains only
physical, non-gauge information about the radiation.  In the TT gauge
there is a close relation between the metric perturbation and the
linearized Riemann tensor $\displaystyle{R_{\mu\nu\rho\sigma}}$ [which is invariant under the
local gauge transformations (\ref{eq:metric_transform}) by Eq.\
(\ref{eq:riemann})], namely
\begin{displaymath}
R_{i0j0} = - \frac{1}{2} {\ddot h}_{ij}^{\rm TT}.
\label{eq:Rtitj}
\end{displaymath}
In a globally vacuum spacetime, all non-zero components of the Riemann
tensor can be obtained from $R_{i0j0}$ via Riemann's symmetries
and the Bianchi identity.  In a
more general spacetime, there will be components that are not related
to radiation.

\begin{figure}
\begin{center}
\includegraphics[width=0.40\textwidth]{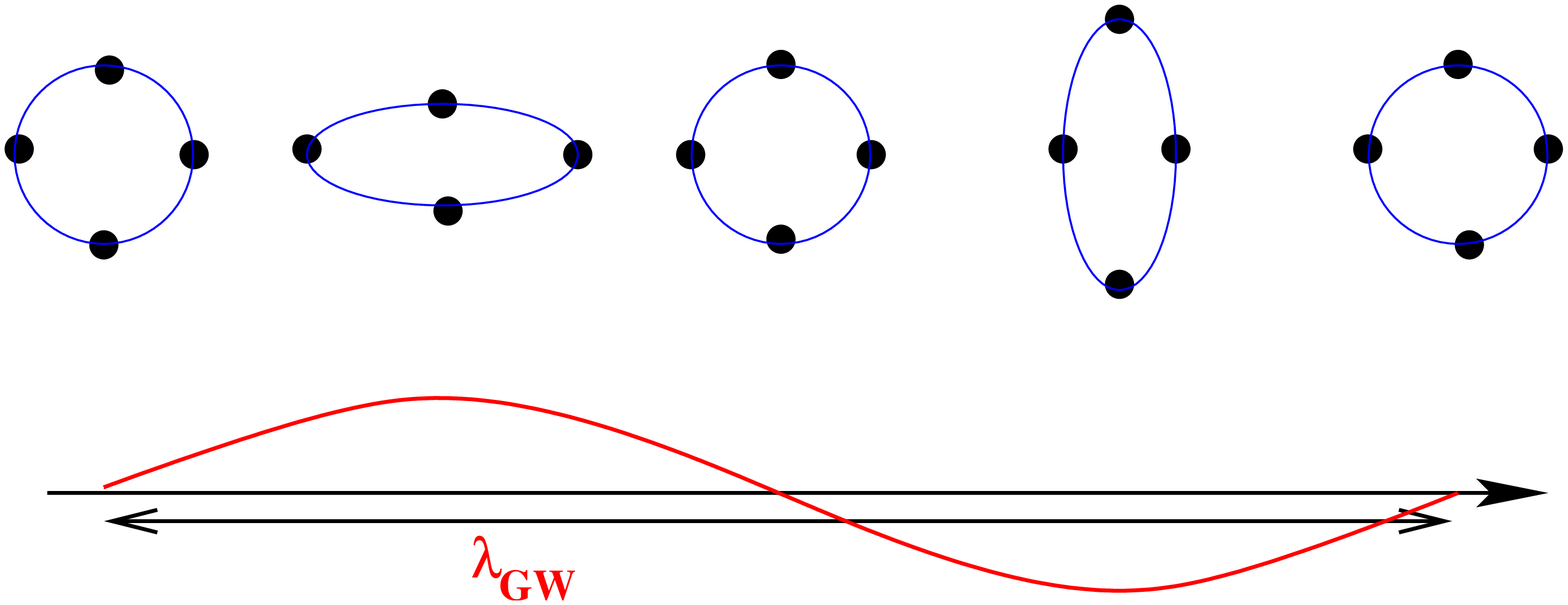} 
\caption{We show how point particles along a ring move as a result of 
the interaction  with a GW propagating in the direction perpendicular to the 
plane of the ring. This figure refers to a wave 
with $+$ polarization. 
\label{fig:2}}
\end{center}
\end{figure}

\begin{figure}
\begin{center}
\includegraphics[width=0.40\textwidth]{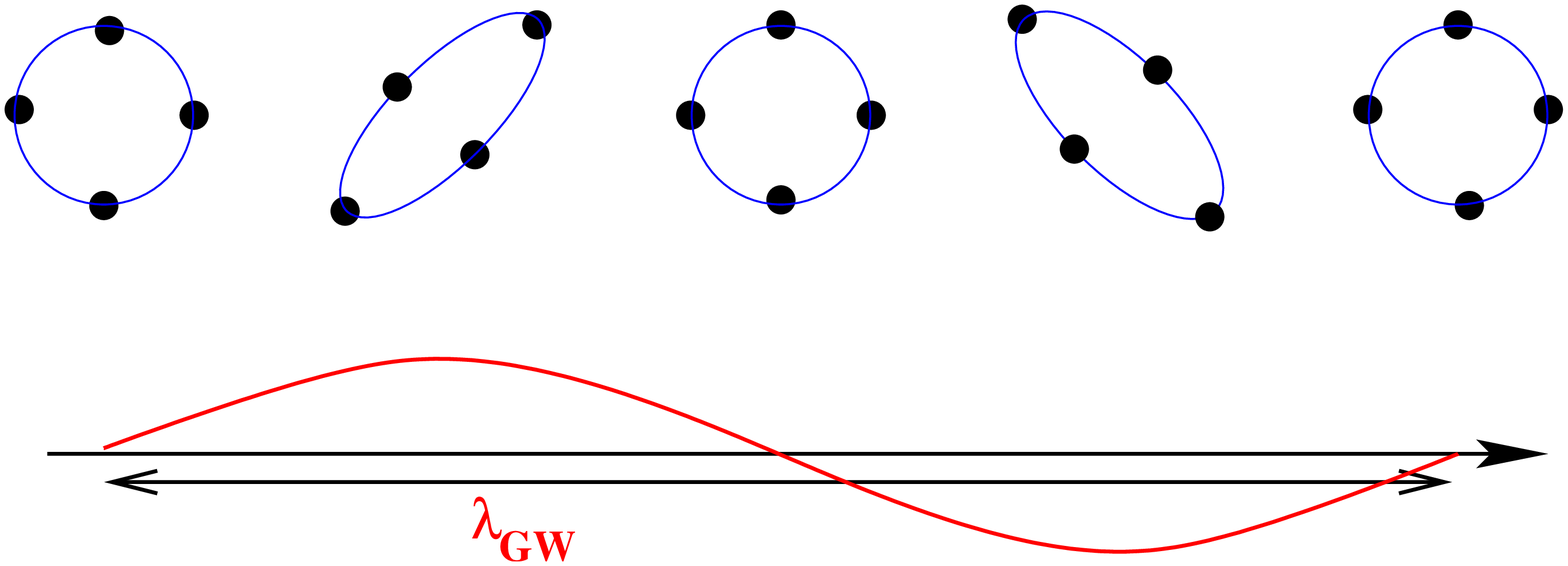}
\caption{We show how point particles along a ring move as a result of 
the interaction  with a GW propagating in the direction perpendicular to the 
plane of the ring. This figure refers to a wave with $\times$ polarization. 
\label{fig:3}}
\end{center}
\end{figure}

Transverse traceless gauge also exhibits the fact that gravitational
waves have two polarization components.  For example, consider a GW
which propagates in the $z$ direction: $h^{\rm TT}_{ij} = h^{\rm
TT}_{ij}(t - z)$ is a valid solution to the wave equation $\Box h^{\rm
TT}_{ij} = 0$.  The Lorentz condition $\partial_z h^{\rm TT}_{zj} = 0$
implies that $h^{\rm TT}_{zj}(t - z) = \mbox{constant}$.  This
constant must be zero to satisfy the condition that $h_{ab} \to 0$ as
$r \to \infty$.  The only non-zero components of $h^{\rm TT}_{ij}$ are
then $h^{\rm TT}_{xx}$, $h^{\rm TT}_{xy}$, $h^{\rm TT}_{yx}$, and
$h^{\rm TT}_{yy}$.  Symmetry and the tracefree condition
(\ref{eq:traceless}) further mandate that only two of these are
independent:
\begin{eqnarray*}
h^{\rm TT}_{xx} &=& -h^{\rm TT}_{yy} \equiv h_+(t-z)\;;
\\
h^{\rm TT}_{xy} &=& h^{\rm TT}_{yx} \equiv h_\times(t-z)\;.
\label{eq:pols0}
\end{eqnarray*}
The quantities $h_+$ and $h_\times$ are the two independent waveforms
of the GW (see Fig.\ \ref{fig:2},\ref{fig:3}) \cite{gravitation,MR}.

To illustrate the effect of GWs on free falling (FF) particles, 
we consider a ring of point particles initially at rest with respect to a 
FF frame attached to the center of the ring, as shown in Fig.~\ref{fig:2},\ref{fig:3}. 
\section{Interaction of gravitational waves with a detector}
\label{sec:detect}

The usual notion of ``gravitational force'' disappears in GR, replaced instead by the idea that freely falling bodies
follow geodesics in spacetime.  Given a spacetime metric $g_{\mu\nu}$ and
a set of spacetime coordinates $x^\mu$, geodesic trajectories are given
by the equation
\begin{eqnarray*}
\frac{d^2 x^\mu}{d\tau^2} +
{\Gamma^\mu}_{\nu\rho}\frac{dx^\nu}{d\tau}\frac{dx^\rho}{d\tau} = 0\;,
\label{eq:geod_eqn}
\end{eqnarray*}
where $\tau$ is the proper time as measured by an observer travelling
along the geodesic.  By writing the derivatives in the above geodesic
equation  in terms of coordinate time $t$ rather
than proper time $\tau$, and by combining the $\mu=t$ (i.e $0$ coordinate) equation with the
spatial, $\mu=j$ (i.e. spatial coordinates) equations, we obtain an equation for the coordinate
acceleration:
\begin{eqnarray}
\frac{d^2 x^i}{dt^2} &=& - (\Gamma^i_{00} + 2 \Gamma^i_{0j} v^j +
\Gamma^i_{jk} v^j v^k) +\nonumber\\&& v^i (\Gamma^0_{00} + 2 \Gamma^0_{0j} v^j +
\Gamma^0_{jk} v^j v^k),
\label{eq:geod_eqn1}
\end{eqnarray}
where $v^i = dx^i/dt$ is the coordinate.

Let us now specialize to linearized theory, with the non-flat part of
our metric dominated by a GW in TT gauge.  Further, let us specialize
to non-relativistic motion for our test body.  This implies that $v^i
\ll 1$, and to a good approximation we can neglect the velocity
dependent terms in Eq.\ (\ref{eq:geod_eqn1}):
\begin{displaymath}
\frac{d^2 x^i}{dt^2} + {\Gamma^i}_{00} = 0\;.
\end{displaymath}
In linearized theory and TT gauge,
\begin{displaymath}
{\Gamma^i}_{00} = \Gamma^{0}_{i0} = \frac{1}{2}\left(2\partial_t h^{\rm
TT}_{j0} - \partial_j h^{\rm TT}_{00}\right) = 0\, ,
\label{eq:linGamma}
\end{displaymath}
since $h^{\rm TT}_{00} = 0$.  Hence, we find that $\displaystyle{d^2x^i/dt^2 = 0}$.

This result could mean that the GW has no effect.This is not true since it
just tells us that, in TT gauge, the {\it coordinate location} of a
slowly moving, freely falling (here in after FF) body is unaffected by the GWs.  In
essence, the coordinates move with the waves.

This result illustrates why, in GR, it is important to
focus upon coordinate-invariant observables (a naive interpretation
of the above result would be that freely falling bodies are not
influenced by GWs).  In fact the GWs cause the {\it proper separation}
between two FF particles to oscillate, even if the {\it
coordinate separation} is constant.  Consider two spatial FF
 particles, located at $z = 0$, and separated on the $x$ axis
by a coordinate distance $L_c$.  Consider a GW in TT gauge that
propagates down the $z$ axis, $h^{\rm TT}_{\mu\nu}(t,z)$.  The proper
distance $L$ between the two particles in the presence of the GW is
given by
\begin{eqnarray}
L &=& \int_0^{L_c} dx\,\sqrt{g_{xx}} = \int_0^{L_c} dx\,\sqrt{1 +
h^{\rm TT}_{xx}(t,z = 0)}
\nonumber\\
&\simeq& \int_0^{L_c} dx\,\left[1 + \frac{1}{2} h^{\rm TT}_{xx}(t,z =
  0)\right]=\nonumber\\&&= L_c\left[1 + \frac{1}{2} h^{\rm TT}_{xx}(t,z =
  0)\right]\;.
\label{eq:waveeffect}
\end{eqnarray}
Notice that we use the fact that the coordinate location of each
particle is fixed in TT gauge.  In a gauge in which the particles move
with respect to the coordinates, the limits of integration would have
to vary.  Eq. (\ref{eq:waveeffect}) tells us that the proper
separation of the two particles oscillates with a fractional length
change $\delta L/L$ given by
\begin{equation}
\frac{\delta L}{L} \simeq \frac{1}{2} h^{\rm TT}_{xx}(t,z = 0)\;.
\label{eq:strainans}
\end{equation}

Although we used TT gauge to perform this calculation, the result is
gauge independent; we will derive it in a different gauge momentarily.
Notice that $h^{\rm TT}_{xx}$ acts as a strain, a fractional length
change.  The magnitude $h$ of a wave is often referred to as the
``wave strain''.  The proper distance we have calculated here is a
particularly important quantity since it directly relates to the
accumulated phase which is measured by laser interferometric GW
observatories .  The
``extra'' phase $\delta \phi$ accumulated by a photon that travels
down and back the arm of a laser interferometer in the presence of a
GW is $\displaystyle{\delta\phi = 4\pi \delta L/\lambda}$, where $\lambda$ is the
photon's wavelength and $\delta L$ is the distance the end mirror
moves relative to the beam splitter\footnote{This description of the
phase shift only holds if $L \ll \lambda$, so that the metric
perturbation does not change value very much during a light travel
time.  This condition will be violated in the high frequency regime
for space-based GW detectors; a more careful analysis of the phase
shift is needed in this case {\cite{lhh00}}.}.  We now give a
different derivation of the fractional length change
(\ref{eq:strainans}) based on the concept of {\it geodesic deviation}.
Consider a geodesic in spacetime given by $x^\mu = z^\mu(\tau)$, where
$\tau$ is the proper time, with four velocity $u^\mu(\tau) =
dz^\mu/d\tau$.  Suppose we have a nearby geodesic $x^\mu(\tau) = z^\mu(\tau)
+ L^\mu(\tau)$, where $L^\mu(\tau)$ is small.  We can regard the
coordinate displacement $L^\mu$ as a vector ${\vec L} = L^\mu \partial_\mu$
on the first geodesic; this is valid to first order in ${\vec L}$.
Without loss of generality, we can make the connecting vector be
purely spatial: $L^\mu u_\mu =0$.  Spacetime curvature causes the
separation vector to change with time, the geodesics will move
further apart or closer together, with an acceleration given by the
geodesic deviation equation

\begin{equation}
u^\nu \nabla_\nu (u^\rho \nabla_\rho L^\mu) = - {R^\mu}_{\nu\rho\sigma}[{\vec z}(\tau)] u^\nu
L^\rho u^\sigma\;;
\label{eq:geod_dev}
\end{equation}

see, e.g., Ref.\cite{jh03}.  This equation is valid to
linear order in $L^\mu$; fractional corrections to this equation will
scale as $L / {\cal L}$, where ${\cal L}$ is the lengthscale over
which the curvature varies.

For application to GW detectors, the shortest lengthscale ${\cal
L}$ is the wavelength $\lambda$ of the GWs.  Thus, the geodesic
deviation equation will have fractional corrections of order $L /
\lambda$.  For ground-based detectors $L \lesssim $ a few km, while
$\lambda \gtrsim 3000 {\rm km}$; thus
the approximation will be valid.  For detectors with $L \gtrsim
\lambda$ (e.g. the space based detector LISA) the analysis here is not
valid and other techniques must be used to analyze the detector.

A convenient coordinate system to analyze the geodesic deviation
equation (\ref{eq:geod_dev}) is the {\it local proper reference frame}
of the observer who travels along the first geodesic.  This coordinate
system is defined by the requirements
\begin{equation}
z^i(\tau) = 0,\ \ \ \ \ g_{\mu\nu}(t,{\bf 0}) = \eta_{\mu\nu}, \ \ \ \
\Gamma^\mu_{\nu\rho}(t,{\bf 0}) =0,
\label{eq:lprf}
\end{equation}
which imply that the metric has the form
\begin{equation}
ds^2 = - dt^2 + d {\bf x}^2 + O\left(\frac{{\bf x}^2}{{\cal R}^2}
\right).
\label{eq:lprfmetric}
\end{equation}
Here ${\cal R}$ is the radius of curvature of spacetime, given by
${\cal R}^{-2} \sim ||R_{\mu\nu\rho\sigma}||$.  It also follows from the gauge
conditions (\ref{eq:lprf}) that proper time $\tau$ and coordinate time
$t$ coincide along the first geodesic, that ${\vec u} = \partial_t$
and that $L^\mu =(0,L^i)$.

Consider now the proper distance between the two geodesics, which are
located at $x^i=0$ and $x^i = L^i(t)$.  From the metric
(\ref{eq:lprfmetric}) we see that this proper distance is just $|{\bf
L}| = \sqrt{L_i L_i}$, up to fractional corrections of order $L^2 /
{\cal R}^2$.  For a GW of amplitude $h$ and wavelength $\lambda$ we
have ${\cal R}^{-2} \sim h / \lambda^2$, so the fractional errors are
$\sim h L^2 / \lambda^2$.  (Notice that ${\cal R} \sim {\cal
L}/\sqrt{h}$  the wave's curvature scale ${\cal R}$ is much larger
than the lengthscale ${\cal L}$ characterizing its variations.)  Since we are
restricting attention to detectors with $L \ll \lambda$, these
fractional errors are much smaller than the fractional distance
changes $\sim h$ caused by the GW.
Therefore, we can simply identify $|{\bf L}|$ as the proper
separation.

We now evaluate the geodesic deviation equation (\ref{eq:geod_dev})
in the local proper reference frame coordinates.  From the conditions
(\ref{eq:lprf}) it follows that we can replace the covariant time
derivative operator $u^\mu \nabla_\mu$ with $\partial / (\partial t)$.
Using ${\vec u} = \partial_t$ and $L^\mu = (0,L^i)$, we get
\begin{equation}
\frac{d^2 L^i(t)}{dt^2} =-  {R}_{i0j0}(t,{\bf 0}) L^j(t) \;.
\label{eq:observable}
\end{equation}
Note that the key quantity entering into the equation, $R_{i0j0}$, is
gauge invariant in linearized theory, so we can use any convenient
coordinate system to evaluate it.  Using the expression
(\ref{eq:Rtitj}) for the Riemann tensor in terms of the TT gauge
metric perturbation $h_{ij}^{\rm TT}$ we find
\begin{displaymath}
\frac{d^2 L^i}{dt^2} = \frac{1}{2}\frac{d^2h^{\rm TT}_{ij}}{dt^2}L^j\;.
\label{eq:geod_dev_detector}
\end{displaymath}
Integrating this equation using $L^i(t) = L^i_0 + \delta L^i(t)$ with
$|\delta {\bf L}| \ll |{\bf L}_0|$ gives
\begin{equation}
\delta L^i(t) = \frac{1}{2}h^{\rm TT}_{ij}(t) L_0^j\;.
\label{eq:response}
\end{equation}

\begin{figure}
\begin{center}
\begin{tabular}{cc}
\includegraphics[width=0.20\textwidth]{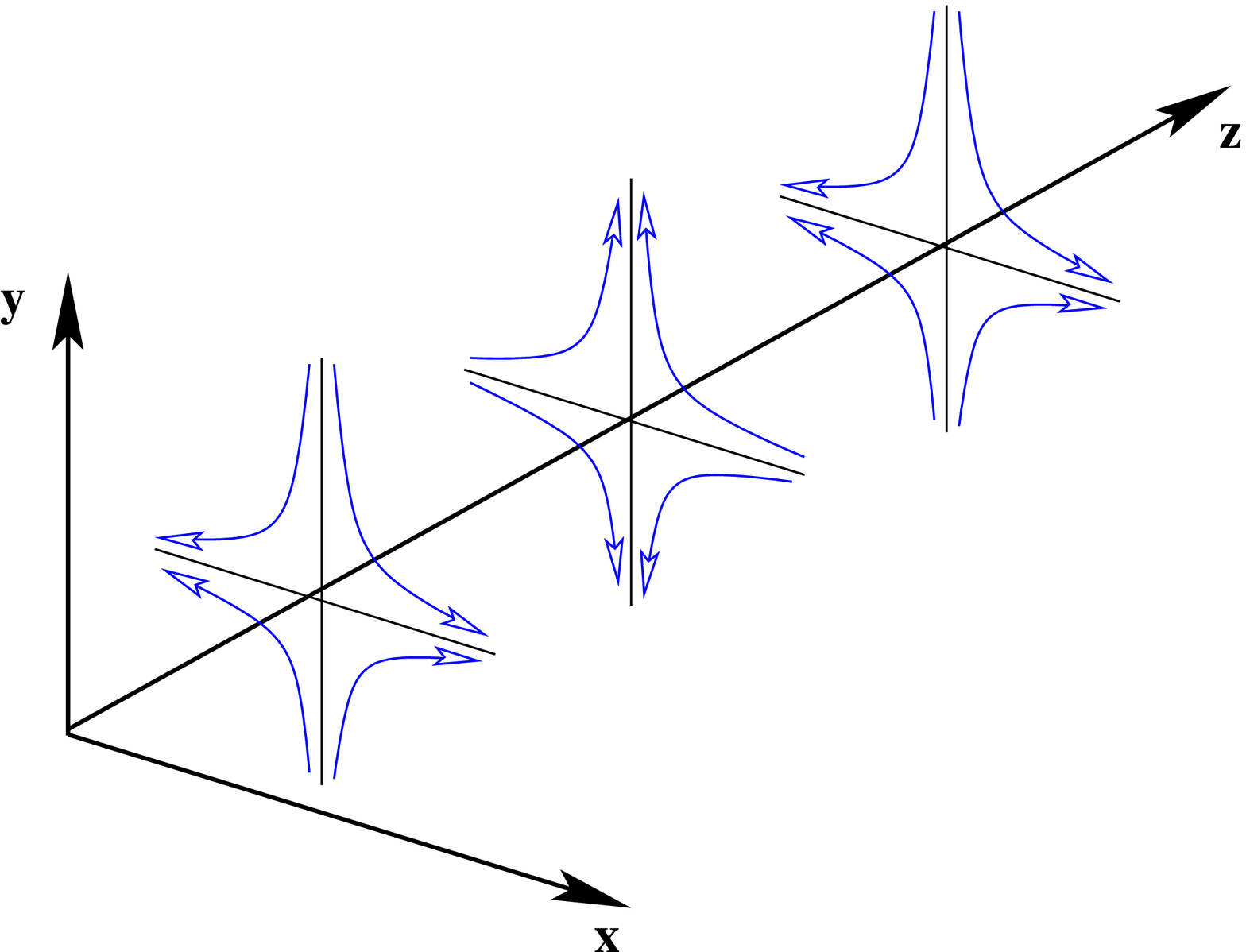} &
\includegraphics[width=0.20\textwidth]{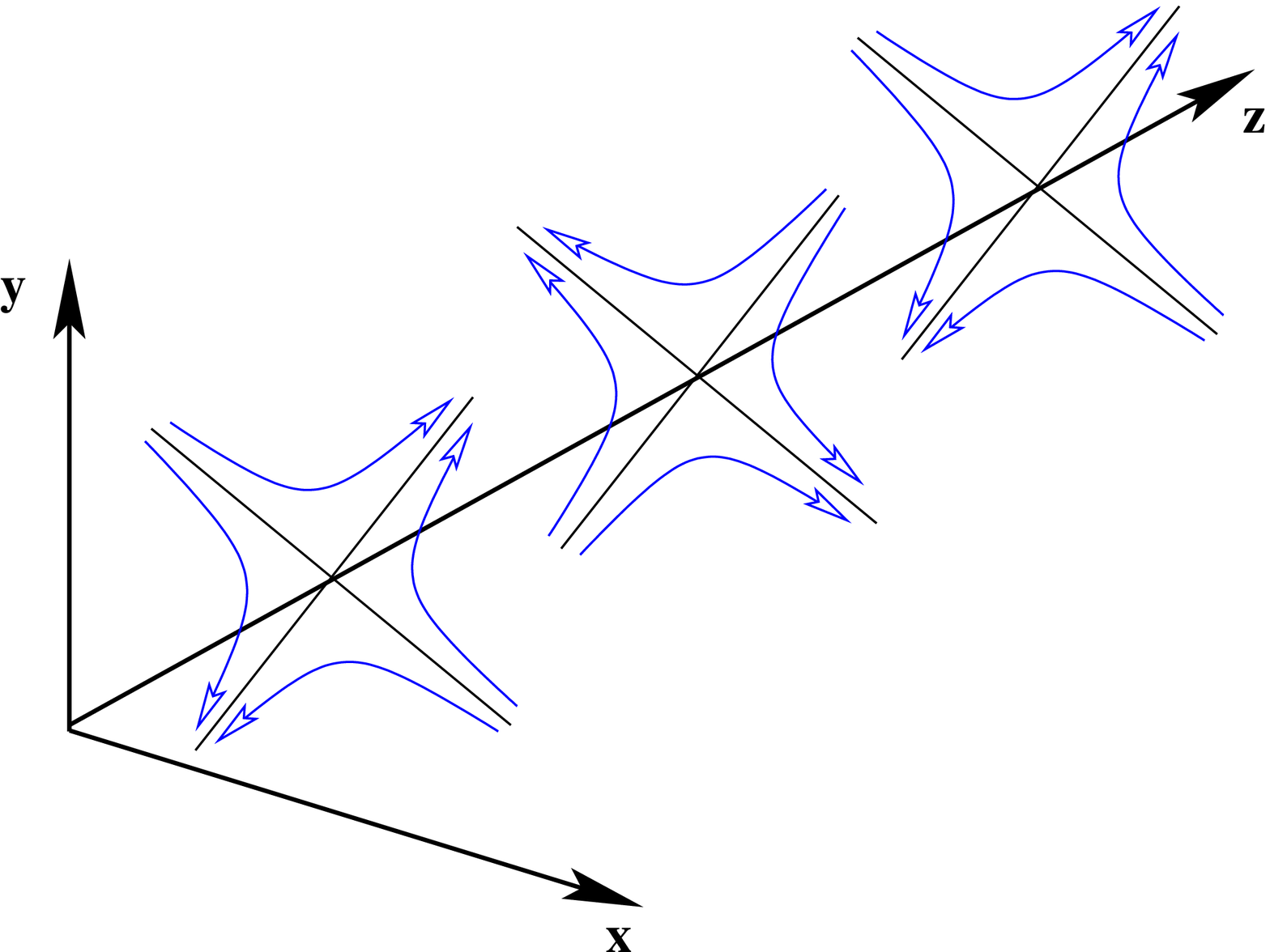} \\
\end{tabular}
\caption{Lines of force associated to the $+$ (left panel) and $\times$ (right panel) polarizations.\label{fig:forcelines}}
\end{center}
\end{figure}

This equation is ideal to analyze an interferometric GW detector.
We choose Cartesian coordinates such that the interferometer's two
arms lie along the $x$ and $y$ axes, with the beam splitter at the
origin.  For concreteness, let us imagine that the GW propagates down
the $z$-axis.  Then, as discussed in Sec.\ \ref{sec:TTgauge}, the only
non-zero components of the metric perturbation are $h^{\rm TT}_{xx} =
-h^{\rm TT}_{yy} = h_+$ and $h^{\rm TT}_{xy} = h^{\rm TT}_{yx} =
h_\times$, where $h_+(t-z)$ and $h_\times(t-z)$ are the two
polarization components.  We take the ends of one of the
interferometer's two arms as defining the two nearby geodesics; the
first geodesic is defined by the beam splitter at ${\bf x}=0$, the
second by the end-mirror.  From Eq.\ (\ref{eq:response}), we then find
that the distances $L = | {\bf L}|$ of the arms end from the beam
splitter vary with time as
\begin{eqnarray*}
\frac{\delta L_x}{L} &=& \frac{1}{2} h_+\;,
\nonumber\\
\frac{\delta L_y}{L} &=& -\frac{1}{2} h_+\;.
\end{eqnarray*}
(Here the subscripts $x$ and $y$ denote the two different arms, not
the components of a vector).  These distance variations  are then measured
via laser interferometry.  Notice that the GW (which is typically a
sinusoidally varying function) acts tidally, squeezing along one axis
and stretching along the other.  In this configuration, the detector is
sensitive only to the $+$ polarization of the GW.  The $\times$
polarization acts similarly, except that it squeezes and stretches
along a set of axes that are rotated with respect to the $x$ and $y$
axes by $45^\circ$.  The force lines corresponding to the two
different polarizations are illustrated in Fig.\
{\ref{fig:forcelines}}.

\
Of course, we don't expect nature to provide GWs that so perfectly
align with our detectors.  In general, we will need to account for the
detector's {\it antenna pattern}, meaning that we will be sensitive to
some weighted combination of the two polarizations, with the weights
depending upon the location of a source on the sky, and the relative
orientation of the source and the detector \cite{KT87}. 

Finally, in our analysis so far of detection we have assumed that the
only contribution to the metric perturbation is the GW contribution.
However, in reality time-varying near zone gravitational fields
produced by sources in the vicinity of the detector will also be
present.  From Eq. (\ref{eq:observable}) we see that the quantity
that is actually measured by interferometric detectors is the
space-time-space-time or electric-type piece $R_{i0j0}$ of the Riemann
tensor (or more precisely the time-varying piece of this within the
frequency band of the detector).  From the general expression
of Riemann tensor (see \cite{gravitation}), we see that $R_{i0j0}$
contains contributions from both $h_{ij}^{\rm TT}$ describing GWs, and
also additional terms describing the time-varying near zone
gravitational fields.  There is no way for the detector to separate
these two contributions, and the time-varying near zone gravitational
fields produced by motions of bedrock, air, human bodies, and
tumbleweeds can all contribute to the output of the detector and act
as sources of noise \cite{ht1998,tw1999,tcreighton}.

\subsection{The generation of gravitational waves}
\label{sec:lin_with_source}

GWs are generated by the matter source term on the
right hand side of the linearized Einstein equation
\begin{equation}
\Box {\bar h}_{\mu\nu} = -16\pi T_{\mu\nu}\;,
\label{eq:LinEinstein_TT}
\end{equation}
cf.\ Eq.\ (\ref{eq:elin}) (presented here in Lorentz gauge).  In this
section we will compute the leading order contribution to the spatial
components of the metric perturbation for a source whose internal
motions are slow compared to the speed of light (``slow-motion
sources'').  We will then compute the TT piece of the metric
perturbation to obtain the standard quadrupole formula for the emitted
radiation.

Eq. (\ref{eq:LinEinstein_TT}) can be solved by using a Green's
function.  A wave equation with source generically takes the form
\begin{displaymath}
\Box f(t, {\bf x}) = s(t, {\bf x})\;,
\label{eq:wave_source}
\end{displaymath}
where $f(t,{\bf x})$ is the radiative field, depending on time $t$ and
position ${\bf x}$, and $s(t,{\bf x})$ is a source function. 
Green's function $G(t,{\bf x};t',{\bf x}')$ is the field which arises
due to a delta function source; it tells how much field is generated
at the ``field point'' $(t,{\bf x})$ per unit source at the ``source
point'' $(t',{\bf x}')$:
\begin{displaymath}
\Box G(t, {\bf x}; t',{\bf x}') = \delta(t - t')\delta({\bf x} - {\bf x}')\;.
\label{eq:wave_greens}
\end{displaymath}
The field which arises from our actual source is then given by
integrating Green's function against $s(t,{\bf x})$:
\begin{displaymath}
f(t, {\bf x}) = \int dt'd^3 x'\,G(t, {\bf x}; t',{\bf x}')\,s(t',{\bf
x}')\;\;.
\end{displaymath}
The Green's function associated with the wave operator $\Box$ is very
well known (see, e.g., {\cite{jackson}}):
\begin{displaymath}
G(t, {\bf x}; t', {\bf x}') = -\frac{\delta(t' - [t - |{\bf x} - {\bf
x}'|/c])}{4\pi|{\bf x} - {\bf x}'|}\;.
\label{eq:BoxGreen}
\end{displaymath}
The quantity $t - |{\bf x} - {\bf x}'|/c$ is the {\it retarded time};
it takes into account the lag associated with the propagation of
information from events at ${\bf x}$ to position ${\bf x}'$.  The
speed of light $c$ has been restored here to emphasize the causal
nature of this Green's function; we set it back to unity in what
follows.

Applying this result to Eq.\ ({\ref{eq:LinEinstein_TT}}), we find
\begin{displaymath}
{\bar h}_{ab}(t,{\bf x}) = 4\int d^3x' \frac{T_{\mu\nu}(t - |{\bf x} - {\bf x}'|,
{\bf x}')}{|{\bf x} - {\bf x}'|}\;.
\label{eq:naivesolution}
\end{displaymath}
Projected transverse and
traceless, as already mentioned, the radiative degrees of freedom are contained
entirely in the spatial part of the metric. First, consider the spatial part of the metric:
\begin{displaymath}
{\bar h}_{ij}(t,{\bf x}) = 4\int d^3x' \frac{T^{ij}(t - |{\bf x} - {\bf
x}'|,{\bf x}')} {|{\bf x} - {\bf x}'|}\;.
\label{eq:spatialsolution}
\end{displaymath}
We have raised indices on the right-hand side, using the rule that the
position of spatial indices in linearized theory is irrelevant.

We now evaluate this quantity at large distances from the source.
This allows us to replace the factor $|{\bf x} - {\bf x}'|$ in the
denominator with $r = |{\bf x}|$.  The corresponding fractional errors
scale as $\sim L / r$, where $L$ is the size of the source; these
errors can be neglected.  We also make the same replacement in the
time argument of $T_{ij}$:
\begin{displaymath}
T_{ij}(t-|{\bf x} - {\bf x}'|,{\bf x}') \approx T_{ij}(t-r,{\bf x}').
\label{eq:slowmotion}
\end{displaymath}
Using the formula $|{\bf x} - {\bf x}'| = r - n^i x^{'\,i} + O(1/r)$
where $n^i = x^i/r$, we see that the fractional errors generated by
the replacement (\ref{eq:slowmotion}) scale as $L/\tau$, where $\tau$
is the timescale over which the system is changing.  This quantity is
just the velocity of internal motions of the source (in units with
$c=1$), and is therefore small compared to one by our assumption.
These replacements give
\begin{equation}
{\bar h}_{ij}(t,{\bf x}) = \frac{4}{r}\int d^3x'\,T^{ij}(t - r,{\bf x}')\;,
\label{eq:almost_quadrupole}
\end{equation}
which is the first term in a multipolar expansion of the radiation field.

Eq. (\ref{eq:almost_quadrupole}) almost gives us the quadrupole
formula that describes GW emission (at leading order).  To get the
remaining part  there, we need to manipulate this equation a bit.  The
stress-energy tensor must be conserved, which means $\partial_\mu T^{\mu\nu}
= 0$ in linearized theory.  Breaking this up into time and space
components, we have the conditions
\begin{displaymath}
\partial_t T^{00} + \partial_i T^{0i} = 0\,,
\end{displaymath}
\begin{displaymath}
\partial_t T^{0i} + \partial_j T^{ij} = 0\,.
\label{eq:cons_of_stress1}
\end{displaymath}
From this, it follows that
\begin{equation}
\partial^2_t T^{00} = \partial_k\partial_l T^{kl}\;.
\label{eq:cons_of_stress2}
\end{equation}
Multiplying both sides of this equation by $x^i x^j$, we first
manipulate the left-hand side:
\begin{displaymath}
\partial^2_t T^{00} x^i x^j = \partial_t^2\left(T^{00} x^i x^j\right)\;.
\end{displaymath}
Next, manipulate the right-hand side of Eq.\ (\ref{eq:cons_of_stress2}); multiplying by $x^i x^j$, we obtain:
\begin{displaymath}
\partial_k\partial_l T^{kl} x^i x^j =
\partial_k\partial_l \left(T^{kl} x^i x^j\right) -
2\partial_k\left(T^{ik} x^j + T^{kj} x^i\right) + 2 T^{ij}\;.
\end{displaymath}
This identity is easily verified by expanding the derivatives and
applying the identity $\partial_i x^j = {\delta_i}^j$.  We thus have
\begin{displaymath}
\partial_t^2\left(T^{00} x^i x^j\right) =
\partial_k\partial_l \left(T^{kl} x^i x^j\right) -
2\partial_k\left(T^{ik} x^j + T^{kj} x^i\right) + 2 T^{ij}\;.
\end{displaymath}
This yields
\begin{eqnarray*}
\frac{4}{r} \int d^3 x'\,T_{ij} &=&
\frac{4}{r}\int d^3x'
\left[\frac{1}{2}\partial_t^2\left(T^{00} x^{\prime i} x^{\prime
j}\right) +\right.\nonumber\\
& &\left. +\partial_k\left(T^{ik} x^{\prime j} + T^{kj} x^{\prime
i}\right) - \frac{1}{2} \partial_k\partial_l \left(T^{kl}
x^{\prime i}x^{\prime j}\right)\right]
\nonumber\\
&=& \frac{2}{r}\int d^3x'\,\partial_t^2\left(T^{00} x^{\prime i}
x^{\prime j}\right)
\nonumber\\
&=& \frac{2}{r}\frac{\partial^2}{\partial t^2}\int d^3x'\,T^{00}
x^{\prime i} x^{\prime j}
\nonumber\\
&=& \frac{2}{r}\frac{\partial^2}{\partial t^2}\int
d^3x'\,\rho\,x^{\prime i} x^{\prime j}\;.
\label{eq:oh_so_close}
\end{eqnarray*}
In going from the first to the second line, we used the fact that the
second and third terms under the integral are divergences.  Using
Gauss's theorem, they can thus be recast as surface integrals; taking
the surface outside the source, their contribution is zero.  In going
from the second to the third line, we used the fact that the
integration domain is not time dependent, so we can take the
derivatives out of the integral.  Finally, we used the fact that
$T^{00}$ is the mass density $\rho$.  Defining the second moment
$Q_{ij}$ of the mass distribution via
\begin{equation}
Q_{ij}(t) = \int d^3x'\,\rho(t,{\bf x}') \,x^{\prime i}x^{\prime j}\;,
\label{eq:quad_moment1}
\end{equation}
and combining Eqs.\ (\ref{eq:almost_quadrupole}) and
(\ref{eq:oh_so_close}) we get
\begin{equation}
{\bar h}_{ij}(t,{\bf x}) = \frac{2}{r}\frac{d^2Q_{ij}(t-r)}{dt^2}\;.
\label{eq:h_from_I}
\end{equation}
When we subtract the trace from $Q_{ij}$, we obtain the {\it
quadrupole momentum} tensor:
\begin{displaymath}
{\cal Q}_{ij} = Q_{ij} - \frac{1}{3}\delta_{ij} Q, \qquad
Q = Q_{ii}\;.
\label{eq:quad_moment}
\end{displaymath}

To complete the derivation, we must project out the non-TT pieces of
the right-hand side of Eq.\ (\ref{eq:h_from_I}).
Since we are working to leading order in $1/r$, at each field point
${\bf x}$, this operation reduces to algebraically projecting the
tensor perpendicularly to the local direction of propagation ${\bf n}
= {\bf x} / r$, and subtracting off the trace.
It is useful to introduce the projection tensor,
\begin{displaymath}
P_{ij} = \delta_{ij} - n_i n_j\;.
\end{displaymath}
This tensor eliminates vector components parallel to ${\bf n}$,
leaving only transverse components.  Thus,
\begin{displaymath}
{\bar h}^T_{ij} = {\bar h}_{kl}P_{ik}P_{jl}
\end{displaymath}
is a transverse tensor.  Finally, we remove the trace; what remains is
\begin{equation}
h^{\rm TT}_{ij} = {\bar h}_{kl} P_{ik} P_{jl} -
\frac{1}{2}P_{ij}P_{kl}{\bar h}_{kl}\;.
\label{eq:quad_formula1}
\end{equation}
Substituting Eq.\ (\ref{eq:h_from_I}) into (\ref{eq:quad_formula1}),
we obtain our final quadrupole formula:
\begin{displaymath}
h^{\rm TT}_{ij}(t,{\bf x}) = \frac{2}{r}\frac{d^2{\cal
Q}_{kl}(t-r)}{dt^2}\left[P_{ik}({\bf n})P_{jl}({\bf n}) -
\frac{1}{2}P_{kl}({\bf n})P_{ij}({\bf n})\right] \, ,
\label{eq:quad_formula}
\end{displaymath}
or 
\begin{equation}
h^{\rm TT}_{ij}(t,{\bf x}) = \frac{2G}{rc^4} \ddot{{\cal Q}_{kl}}(t-r)P_{ijkl} \, .
\label{eq:quad_formula}
\end{equation}

One can now search for wave solutions of (\ref{eq:LinEinstein_TT}) from a
system of masses undergoing arbitrary motions, and then obtain the
power radiated. The result, assuming the source dimensions very
small with respect to the wavelengths, (quadrupole approximation
\cite{landau}), is that the power ${\displaystyle
\frac{dE}{d\Omega}}$ radiated in a solid angle $\Omega$ is

\begin{equation}
\frac{dE}{d\Omega}=\frac{G}{8\pi c^{5}}\left(\frac{d^{3}Q_{ij}}{dt^{3}}\right)^{2}\label{eq:P}\end{equation}

 If one sums (\ref{eq:P}) over the two allowed
polarizations, one obtains
\begin{eqnarray}
&&\sum_{pol}\frac{dE}{d\Omega}= \frac{G}{8\pi c^{5}}\left[\frac{d^{3}Q_{ij}}{dt^{3}}\frac{d^{3}Q_{ij}}{dt^{3}}-2n_{i}\frac{d^{3}Q_{ij}}{dt^{3}}n_{k}\frac{d^{3}Q_{kj}}{dt^{3}}+\right.\nonumber \\
 &  & \left.-\frac{1}{2}\left(\frac{d^{3}Q_{ii}}{dt^{3}}\right)^{2}+\frac{1}{2}\left(n_{i}n_{j}\frac{d^{3}Q_{ij}}{dt^{3}}\right)^{2}+\frac{d^{3}Q_{ii}}{dt^{3}}n_{j}n_{k}\frac{d^{3}Q_{jk}}{dt^{3}}\right]\nonumber\\\label{eq:sommatoria}\end{eqnarray}
where $\hat{n}$ is the unit vector in the  radiation direction.
The total radiation rate is obtained by integrating
(\ref{eq:sommatoria}) over all directions of emission; the result
is

\begin{equation}
{\cal F}^{GW}=\frac{dE}{dt}=-\frac{G\left\langle Q_{ij}^{(3)}Q^{(3)ij}\right\rangle }{45c^{5}}\label{eq:dEdt}\end{equation}

where the index (3) represents the number of differentiations with respect to
time, the symbol $<>$ indicates that the quantity is averaged over
several wavelengths. 

\subsection{Extension to sources with non-negligible self gravity}
\label{sec:quad1}

Concerning our derivation of the quadrupole formula (\ref{eq:quad_formula}) we assumed
the validity of the linearized Einstein equations.  In particular, the
derivation is not applicable to systems with weak (Newtonian) gravity
whose dynamics are dominated by self-gravity, such as binary star
systems\footnote{Stress energy conservation in linearized gravity,
$\partial^\mu T_{\mu\nu} =0$, forces all bodies to move on geodesics of
the Minkowski metric.}.  This shortcoming of the above
linearized-gravity derivation of the quadrupole formula was first
pointed out by Eddington.  However, it is very straightforward to
extend the derivation to encompass systems with non-negligible self gravity.

In full GR, we define the quantity ${\bar h}^{\mu\nu}$ via
\begin{displaymath}
\sqrt{-g} g^{\mu\nu} = \eta^{\mu\nu} - {\bar h}^{\mu\nu},
\end{displaymath}
where $\eta^{\mu\nu} \equiv {\rm diag}(-1,1,1,1)$.
When gravity is weak, this definition coincides with
our previous definition of ${\bar h}^{\mu\nu}$ as a trace-reversed metric
perturbation.  We impose the harmonic gauge condition
\begin{equation}
\partial_\mu (\sqrt{-g} g^{\mu\nu}) = \partial_\mu {\bar h}^{\mu\nu} =0.
\label{eq:harmonic}
\end{equation}
In this gauge, the Einstein equation can be written as
\begin{equation}
\Box_{\rm flat} {\bar h}^{\mu\nu} = - 16 \pi ( T^{\mu\nu} + t^{\mu\nu} ),
\label{eq:enonlin}
\end{equation}
where $\Box_{\rm flat} \equiv \eta^{\mu\nu} \partial_\mu \partial_\nu$ is the
flat-spacetime wave operator, and $t^{\mu\nu}$ is a pseudotensor that is
constructed from ${\bar h}^{\mu\nu}$. Taking a coordinate divergence of
this equation and using the gauge condition (\ref{eq:harmonic}), stress-energy conservation can be written
\begin{equation}
\partial_\mu (T^{\mu\nu} + t^{\mu\nu}) =0.
\label{eq:nonlincons}
\end{equation}

Eqs. (\ref{eq:harmonic})- (\ref{eq:enonlin}) and
(\ref{eq:nonlincons}) are precisely the same equations as are used in
the linearized-gravity derivation of the quadrupole formula, except
for the fact that the stress energy tensor $T^{\mu\nu}$ is replaced by
$T^{\mu\nu} + t^{\mu\nu}$.  Therefore the derivation of the last subsection
carries over, with the modification that the formula
(\ref{eq:quad_moment1}) for $Q_{ij}$ is replaced by
\begin{displaymath}
Q_{ij}(t) = \int d^3 x' \left[T^{00}(t,{\bf x}') + t^{00}(t,{\bf
    x}')\right] x^{\prime i}x^{\prime j}.
\end{displaymath}
In this equation the term $t^{00}$ describes gravitational binding
energy, roughly speaking. For systems with weak gravity, this term is
negligible in comparison with the term $T^{00}$ describing the
rest-masses of the bodies.  Therefore the quadrupole formula
(\ref{eq:quad_formula}) and the original definition
(\ref{eq:quad_moment1}) of $Q_{ij}$ continue to apply to the more
general situation considered here.

\subsection{Dimensional analysis}

The rough form of the leading GW field that we just derived, Eq.\
(\ref{eq:quad_formula}), can be deduced using simple physical
arguments.  First, we define some moments of the mass distribution.
The zeroth moment is just the mass itself:
\begin{displaymath}
M_0 \equiv \int \rho\,d^3x = M\;.
\end{displaymath}
More accurately, this is the total mass-energy of the source.
Next, we define the dipole moment:
\begin{displaymath}
M_1 \equiv \int \rho\,x_i\,d^3x = ML_i\;.
\end{displaymath}
$L_i$ is a vector with the dimension of length; it describes the
displacement of the center of mass from the origin we chose.  As such,
$M_1$ is clearly not a very meaningful quantity --- we can change its
value simply by choosing a different origin.

If our mass distribution exhibits internal motion, then moment of the
{\it mass current}, $j_i = \rho v_i$, are also important.  The first
momentum is the spin angular momentum:
\begin{displaymath}
S_1 \equiv \int \rho v_j\,x_k\,\epsilon_{ijk}\,d^3x = S_i\;.
\end{displaymath}
Finally, we look at the second momentum of the mass distribution:
\begin{displaymath}
M_2 \equiv \int \rho\,x_i\,x_j\,d^3x = M L_{ij}
\end{displaymath}
where $L_{ij}$ is a tensor with the dimension length squared.

Using dimensional analysis and simple physical arguments, it is simple
to see that the first moment that can contribute to GW emission is
$M_2$.  Consider first $M_0$.  We want to combine $M_0$ with the
distance to our source, $r$, in such a way as to produce a
dimensionless wavestrain $h$.  The only way to do this (bearing in
mind that the strain should fall off as $1/r$, and restoring factors
of $G$ and $c$) is to put
\begin{displaymath}
h \sim \frac{G}{c^2}\frac{M_0}{r}\;.
\label{eq:hnewton}
\end{displaymath}
 Conservation
of mass-energy tells us that $M_0$ for an isolated source cannot vary
dynamically.  This $h$ cannot be radiative; it corresponds to a
Newtonian potential, rather than a GW.

Let us consider now the momentum $M_1$.  In order to get the right dimensions,we
must take one time derivative:
\begin{displaymath}
h \sim \frac{G}{c^3}\frac{d}{dt}\frac{M_1}{r}\;.
\end{displaymath}
The extra factor of $c$ converts the dimension of the time derivative
to space, so that the whole expression is dimensionless. Think
carefully about the derivative of $M_1$:
\begin{displaymath}
\frac{dM_1}{dt} = \frac{d}{dt}\int\rho\,x_i\,d^3x =
\int\rho\,v_i\,d^3x = P_i\;.
\end{displaymath}
This is the total momentum of our source.  Our guess for the form of a
wave corresponding to $M_1$ becomes
\begin{equation}
h \sim \frac{G}{c^3}\frac{P}{r}\;.
\label{eq:hboost}
\end{equation}
Also this formula cannot describe a GW. The momentum of an
isolated source must be conserved.  By boosting into a different
Lorentz frame, we can always set $P = 0$.  Terms like this can only be
gauge artifacts; they do not correspond to radiation. Indeed, terms
like (\ref{eq:hboost}) appear in the metric of a moving BH,
and correspond to the relative velocity of the BH and the observer, \cite{membrane}.

  Dimensional analysis tells us that radiation from
$S_1$ must take the form
\begin{displaymath}
h \sim \frac{G}{c^4}\frac{d}{dt}\frac{S_1}{r}.
\end{displaymath}
Conservation of angular momentum tells us that the total spin of an
isolated system cannot change, so we must reject also this term. 
Finally, we examine $M_2$:
\begin{displaymath}
h \sim \frac{G}{c^4}\frac{d^2}{dt^2}\frac{M_2}{r}\;.
\end{displaymath}
There is {\it no} conservation principle that allows us to reject this
term.  This is
the quadrupole formula we derived earlier, up to numerical factors.

In ``normal'' units, the prefactor of this formula turns out to be
$G/c^4$ --- a small number divided by a {\it very} big number.  In
order to generate interesting amounts of GWs, the variation quadrupole momentum must be enormous.  The only interesting sources of GWs will
be those which have very large masses undergoing extremely rapid
variation; even in this case, the strain we expect from typical sources
is tiny.  The smallness of GWs reflects the fact that gravity is the
weakest of the fundamental interactions.

\subsection{Numerical estimates}
\label{subsec:numerstrain}

Consider a binary star system, with stars of mass $m_1$ and $m_2$ in a
circular orbit with separation $R$.  The quadrupole moment is given by
\begin{equation}
{\cal Q}_{ij} = \mu\left(x_i x_j - \frac{1}{3}R^2\delta_{ij}\right)\;,
\end{equation}
where 
${\bf x}$ is the relative displacement, with $|{\bf x}| = R$.  We
use the center-of-mass reference frame, and 
choose the coordinate axes so that the binary lies in the $xy$ plane,
so $x = x_1 = R\cos\Omega t$, $y = x_2 = R\sin\Omega t$, $z = x_3 =
0$.  Let us further choose to evaluate the field on the $z$ axis, so
that ${\bf n}$ points in the $z$-direction.  The projection operators
in Eq.\ (\ref{eq:quad_formula}) then simply serve to remove the $zj$
components of the tensor.  Bearing this in mind, the quadrupole
formula (\ref{eq:quad_formula}) yields
\begin{displaymath}
h^{\rm TT}_{ij} = \frac{2 \ddot {\cal Q}_{ij}}{r}\;.
\end{displaymath}
The quadrupole moment tensor is
\begin{displaymath}
{\cal Q}_{ij} = \mu R^2\left[
\begin{array}{ccc}
\cos^{2}\Omega t-\frac{1}{3}  & \cos\Omega t \sin\Omega t & 0\\
\cos\Omega t\sin\Omega t & \cos^2\Omega t - \frac{1}{3} & 0 \\
0 & 0 & -\frac{1}{3} \\
\end{array}
\right]\;;
\end{displaymath}
its second derivative is
\begin{displaymath}
\ddot{\cal Q}_{ij} = -2\Omega^2\mu R^2
\left[\begin{array}{ccc}
 \cos2\Omega t & \sin2\Omega t & 0 \\
-\sin2\Omega t &-\cos2\Omega t & 0 \\
0 & 0 & 0\\\end{array}\right]\;.
\end{displaymath}
The magnitude $h$ of a typical non-zero component of $h^{\rm TT}_{ij}$
is
\begin{displaymath}
h = \frac{4\mu\Omega^2 R^2}{r} = \frac{4\mu M^{2/3}\Omega^{2/3}}{r}\;.
\end{displaymath}
We used Kepler's 3rd law\footnote{In units with $G = 1$, and for
circular orbits of radius $R$, $R^3\Omega^2 = M$.} to replace $R$ with
powers of the orbital frequency $\Omega$ and the total mass $M = m_1 +
m_2$.
The combination of masses here, $\mu M^{2/3}$,
appears quite often in studies of GW emission from binaries; it
motivates the definition of the {\it chirp mass}:
\begin{equation}
{\cal M} = \mu^{3/5}M^{2/5}\;.
\end{equation}
For the purpose of numerical estimate, we will take the members of
the binary to have equal masses, so that $\mu = M/4$:
\begin{displaymath}
h = \frac{M^{5/3}\Omega^{2/3}}{r}\;.
\end{displaymath}
Finally, we insert numbers corresponding to plausible sources:
\begin{eqnarray*}
h &\simeq& 10^{-21}\left(\frac{M}{2\,M_\odot}\right)^{5/3}
\left(\frac{1\,\mbox{hour}}{P}\right)^{2/3}
\left(\frac{1\,\mbox{kiloparsec}}{r}\right)
\nonumber\\
&\simeq& 10^{-22}\left(\frac{M}{2.8\,M_\odot}\right)^{5/3}
\left(\frac{0.01\,\mbox{sec}}{P}\right)^{2/3}
\left(\frac{100\,\mbox{Megaparsecs}}{r}\right)\;.
\end{eqnarray*}
The first line corresponds roughly to the mass, distance and orbital
period ($P = 2\pi/\Omega$) expected for the many close binary white
dwarf systems in our G
alaxy.  Such binaries are so common that they
are likely to be a confusion limited source of GWs for space-based
detectors, acting in some cases as an effective source of noise.  The
second line contains typical parameter values for binary neutron stars
that are on the verge of spiralling together and merging.  Such waves
are targets for the ground-based detectors that have recently begun
operations.  The {\it tiny} magnitude of these waves illustrates why
detecting GWs is so difficult.
The emission of GWs costs energy and to compensate for the loss of energy, 
the radial separation $R$ between the two bodies must decrease. We shall 
now derive how the orbital frequency and GW frequency change in time, 
using Newtonian dynamics and the balance equation  

\begin{equation}
\frac{d E_{\rm orbit}}{d t} = - P\,.
\label{eq:84}
\end{equation}

At Newtonian order, $E_{\rm orbit} = - m_1\,m_2/(2R)$. Thus, $\dot{R} = -2/3\,(R\,\Omega)\,
(\dot{\Omega}/\Omega^2)$. As long as $\dot{\Omega}/\Omega^2 \ll 1$, the radial 
velocity is smaller than the tangential velocity and the binary's motion  
is well approximated by an adiabatic sequence of quasi-circular orbits.
Eq. (\ref{eq:84}) implies that the orbital frequency varies as 
\begin{equation}
\frac{\dot{\Omega}}{\Omega^2}=\frac{96}{5}\,\nu\,\left (\frac{G M\Omega}{c^3} \right )^{5/3}\,,
\label{eq}
\end{equation}
and the GW frequency $f_{\rm GW} = 2 \omega$,  
\begin{equation}
\dot{f}_{\rm GW} = \frac{96}{5}\pi^{8/3}\,\left (\frac{{\cal M}}{c^3} \right )^{5/3}\,f_{\rm GW}^{11/3}\,.
\label{eq:85}
\end{equation}

 Introducing the time 
to coalescence $\tau = t_{\rm coal} -t$, and integrating Eq.~(\ref{eq:85}), we get
\begin{equation}
f_{\rm GW} \simeq 130 \left (\frac{1.21 M_\odot}{\cal M}\right )^{5/8}\, 
\left (\frac{1 {\rm sec}}{\tau} \right )^{3/8}\,{\rm Hz}\,,
\label{eq:86}
\end{equation}
where $1.21 M_\odot$ is the chirp mass of a NS-NS binary. 
Eq.(\ref{eq:86})  predicts, { \it e.g.} coalescence times of 
$ \sim 17 {\rm min}, 2 {\rm sec}, 1 {\rm msec}$, for $f_{\rm GW} \sim 
10, 100 ,10^3$ Hz. Using the above equations, 
it is straightforward  to compute the relation between the radial 
separation and the GW frequency. We find
\begin{equation}
R \simeq 300 \left (\frac{M}{2.8 M_\odot} \right )^{1/3}\,
\left (\frac{100\, {\rm Hz}}{f_{\rm GW}} \right )^{2/3}\, {\rm km}\,.
\label{eq:87}
\end{equation}
Finally, a useful quantity is the number of GW cycles, defined by 
\begin{equation} 
{\cal N}_{\rm GW} = \frac{1}{\pi} \int_{t_{\rm in}}^{t_{\rm fin}} 
\Omega(t)\,dt = \frac{1}{\pi} \int_{\Omega_{\rm in}}^{\Omega_{\rm fin}} 
\frac{\Omega}{\dot{\Omega}}\,d\Omega\,.
\label{cycles}
\end{equation}
Assuming $\Omega_{\rm fin} \gg \Omega_{\rm in}$, we get 
\begin{equation}
{\cal N}_{\rm GW} \simeq 10^4\,\left (\frac{{\cal M}}{1.21 M_\odot} \right )^{-5/3}\,
\left (\frac{f_{\rm in}}{10 {\rm Hz}} \right )^{-5/3}\,.
\label{eq:88}
\end{equation}

\section{Gravitational waves from sources in  Newtonian motion}

We have now all the ingredient to characterize gravitational radiation with respect to the motion of the sources, {\it i.e.} with respect to different types of stellar encounters. Let us start with the Newtonian cases.
With the above formalism,  it is possible to estimate the amount of
energy emitted in the form of GWs from a system of massive objects
interacting among them \cite{pm1,pm2}. Considering the quadrupole components for two bodies interacting in a Newtonian gravitational field, we have:
\begin{equation}
\begin{array}{lll}
Q_{xx}=\mu r^2(3\cos{^2\phi}\sin{^2\theta}-1)~,\\ \\
Q_{yy}=\mu r^2(3\sin{^2\phi}\sin{^2\theta}-1)~,\\ \\
Q_{zz}=\frac{1}{2} r^2 \mu  (3 \cos2 \theta+1) ~,\\ \\
Q_{xz}=Q_{zx}=r^2 \mu  (\frac{3}{2} \cos\phi \sin2\theta)~,\\ \\
Q_{yz}=Q_{zy}=r^2 \mu  (\frac{3}{2} \sin 2\theta \sin \phi)~,\\ \\
Q_{xy}=Q_{yx}=r^2 \mu  \left(\frac{3}{2} \sin ^2\theta \sin2\phi\right)~,
\end{array}\label{eq:quadrupoli}
\end{equation}
where the masses $m_{1}$ and $m_{2}$ have  polar coordinates
$\{r_{i}\cos\theta\cos\phi,\; r_{i}\cos\theta\sin\phi,\:
r_{i}\sin\theta\}$ with $i=1,2$ . We will work in the equatorial plane
($\theta=\pi/2$). The origin of the motions is
taken at the center of mass. Such components can be differentiated
with respect to time, as in Eq.(\ref{eq:dEdt}), in order to derive the amount of gravitational radiation in the various Newtonian orbital motions.

\subsection{Gravitational wave luminosity from circular and elliptical orbits}

The first case we are going to consider is that of closed circular and elliptical orbits.
Using  Eq.(\ref{eq:traiettoria}), let us derive  the angular
velocity equation
\begin{displaymath}
\dot{\phi}=\frac{\sqrt{G l (m_{1}+m_{2})} (\epsilon  \cos\phi+1)^2}{l^2}
\label{eq:angularvelo}
\end{displaymath}

and then, from Eqs.(\ref{eq:quadrupoli}), the third derivatives of quadrupolar components
for the elliptical orbits are:

\begin{eqnarray*}
\frac{d^{3}Q_{xx}}{dt^{3}}&=&\beta(24 \cos\phi+\epsilon  (9 \cos2 \phi)+11)) \sin \phi \\
\frac{d^{3}Q_{yy}}{dt^{3}}&=&-\beta(24 \cos\phi+\epsilon(13+9 \cos2 \phi)) \sin\phi)\\
\frac{d^{3}Q_{zz}}{dt^{3}}&=&-2\beta \epsilon \sin\phi \\
\frac{d^{3}Q_{xy}}{dt^{3}}&=&\beta(24 \cos\phi+\epsilon  (11+9 \cos2 \phi)) \sin\phi) \\
\end{eqnarray*}

where
\begin{displaymath}
\beta=\frac{G l (m_{1}+m_{2}))^{3/2} \mu  (\epsilon
\cos\phi+1)^2}{l^4}\,.
\end{displaymath}
Being
\begin{eqnarray*}
 Q_{ij}^{(3)}Q^{(3)ij}=\frac{G^3}{l^5}\left[ (m_{1}+m_{2})^3 \mu ^2 (1+\epsilon  \cos\phi)^4\right.\\
\left(415 \epsilon ^2+3 (8 \cos\phi+3 \epsilon  \cos2 \phi) \right.\\
\left.(72 \cos\phi+\epsilon  (70+27 \cos2 \phi)))
\sin\phi^{2}\right]
 \end{eqnarray*}
the total power radiated is  given by
\begin{displaymath}
\frac{dE}{dt}=\frac{G^3}{45c^5l^5}f(\phi),\end{displaymath}
where
\begin{eqnarray*}
f(\phi)&=&(m_{1}+m_{2})^3 \mu ^2 (1+\epsilon  \cos\phi)^4\times\\ &&
(415 \epsilon ^2+3 (8 \cos\phi+3 \epsilon  \cos2 \phi)\times \\ &&
(72 \cos\phi+\epsilon  (70+27 \cos2 \phi)))
\sin\phi^{2}.
\end{eqnarray*}
The total energy emitted in the form of gravitational radiation,
during the interaction, is :
\begin{displaymath}
\Delta E=\int^{\infty}_0 \left|\frac{dE}{dt}\right| dt~.
\end{displaymath}
From Eq.(\ref{eq:momang1}), we can adopt the angle $\phi$ as a
suitable integration variable. In this case, the energy emitted
for $\phi_1<\phi<\phi_2$ is
\begin{displaymath}
\Delta E(\phi_1,\phi_2)
=\frac{G^3}{45c^5l^5}\int^{\phi_2}_{\phi_1}f(\phi)~d\phi~,\label{eq:integraleenergia}
\end{displaymath}
and the total energy can be determined from the previous relation
in the limits $\phi_1\rightarrow 0$ and $\phi_2\rightarrow
\pi$. Thus, one has
\begin{displaymath}
\Delta E=\frac{G^4 \pi  (m_{1}+m_{2})^3 \mu ^2}{l^5c^5}F(\epsilon)~
\end{displaymath}
where $F(\epsilon)$ depends on the initial conditions only and it is
given by
\begin{displaymath}
F(\epsilon)=\frac{ \left(13824+102448 \epsilon ^2+59412 \epsilon ^4+2549 \epsilon ^6\right)}{2880}.
\end{displaymath}
In other words, the gravitational wave luminosity strictly depends
on the configuration and kinematics of the binary system.

\subsection{Gravitational wave luminosity from parabolic and hyperbolic orbits}

In this case, we use   Eq.(\ref{eq:traie})  and Eq.
(\ref{eq:quadrupoli}) to calculate the quadrupolar formula for
parabolic and hyperbolic orbits. The angular velocity is

\begin{displaymath}
\dot{\phi}=l^2 L (\epsilon  \cos\phi+1)^2,
\label{eq:angularvelo1}
\end{displaymath}
and the quadrupolar derivatives are

\begin{eqnarray*}
\frac{d^{3}Q_{xx}}{dt^{3}}&=&\rho(24 \cos\phi+\epsilon(9 \cos 2 \phi+11)) \sin \phi, \\
\frac{d^{3}Q_{yy}}{dt^{3}}&=&-\rho(24 \cos\phi+\epsilon  (13+9 \cos2 \phi)) \sin\phi),\\
\frac{d^{3}Q_{zz}}{dt^{3}}&=&-2\rho\epsilon\sin\phi, \\
\frac{d^{3}Q_{xy}}{dt^{3}}&=&-\frac{3}{2}\rho(\epsilon  \cos \phi+1)^2 (5 \epsilon\cos \phi+8 \cos 2 \phi+3 \epsilon  \cos3 \phi),\\
\end{eqnarray*}
where
\begin{displaymath}
\rho=l^4 L^3 \mu  (\epsilon  \cos\phi+1)^2\,.
\end{displaymath}
The  radiated power is given by
\begin{eqnarray*}
\frac{dE}{dt}&=&-\frac{G \rho^2}{120 c^5} \times \\  &&[ 314
   \epsilon ^2+(1152 \cos (\phi+187 \epsilon  \cos 2 \phi \\ &&-3
   (80 \cos 3 \phi+30 \epsilon  \cos4 \phi+48 \cos 5 \phi+9 \epsilon  \cos6 \phi)) \epsilon
   \\ &&-192 \cos4 \phi+576],
\end{eqnarray*}

then
\begin{displaymath}
\frac{dE}{dt}=-\frac{G l^8 L^6 \mu ^2 }{120 c^5 }f(\phi),
\end{displaymath}

where $f(\phi)$, in this case, is
\begin{eqnarray*}
f(\phi)&=& (314
   \epsilon ^2+(1152 \cos (\phi+187 \epsilon  \cos 2 \phi-3
   (80 \cos 3 \phi\\ &&+30 \epsilon  \cos4 \phi+48 \cos 5 \phi+9 \epsilon  \cos6 \phi)) \epsilon+  \phi\\ &&-192 \cos4 \phi+576).
\end{eqnarray*}

Then using  Eq. (\ref{eq:dEdt}), the total energy emitted in the
form of gravitational radiation, during the interaction as a
function of $\phi$, is given by
\begin{eqnarray*}
&& \Delta E(\phi_1,\phi_2)
=-\frac{1}{480 c^5}~d\phi~\times
\nonumber\\ && (G l^8 L^6 \pi  \left(1271 \epsilon ^6+24276 \epsilon ^4+34768
   \epsilon ^2+4608\right) \mu ^2)
,\label{eq:integraleenergia1}
\end{eqnarray*}
and the total energy can be determined from the previous relation
in the limits $\phi_1\rightarrow -\pi$ and $\phi_2\rightarrow\pi$ in the parabolic case. Thus, one has
\begin{eqnarray*}
\Delta E= -\frac{(G l^8 L^6 \pi \mu^2}{480 c^5}F(\epsilon)~,
\end{eqnarray*}
where $F(\epsilon)$ depends on the initial conditions only and it is
given by
\begin{displaymath}
F(\epsilon)= \left(1271 \epsilon ^6+24276 \epsilon ^4+34768
   \epsilon ^2+4608\right)~.
 \end{displaymath}
 In the hyperbolic case, we have that the total energy is determined in the limits
  ${\displaystyle \phi_1\rightarrow \frac{-3\pi}{4}}$ and ${\displaystyle \phi_2\rightarrow\frac{-3\pi}{4}}$, i.e.
 \begin{displaymath}
\Delta E=--\frac{G l^8 L^6\mu^2}{201600 c^5}F(\epsilon)~,
\end{displaymath}
where $F(\epsilon)$ depends on the initial conditions only and is
given by
\begin{eqnarray*}
F(\epsilon)&=& [315 \pi (1271 \epsilon ^6+24276
\epsilon ^4+34768 \epsilon
   ^2+4608)+\\ & & +16 \epsilon[\epsilon[\epsilon(926704 \sqrt{2}-7
   \epsilon  (3319 \epsilon ^2-32632 \sqrt{2} \epsilon
   \\ &&+55200))-383460]+352128 \sqrt{2}]]~.
\end{eqnarray*}

As above, the gravitational wave luminosity strictly depends on
the configuration and kinematics of the binary system.

\subsection{Gravitational wave amplitude from elliptical orbits}

\begin{figure}
\includegraphics[scale=0.5]{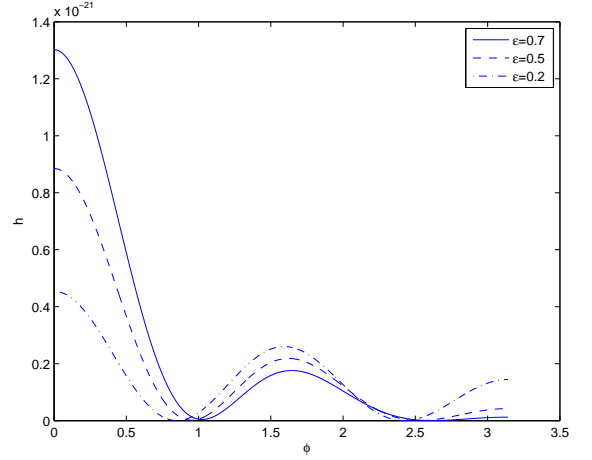}
\caption{The gravitational wave-forms from elliptical orbits shown
as  function of the polar angle $\phi$.  We have fixed
$m_{1}=m_{2}=1.4M_{\odot}$. $m_{2}$ is considered at rest while
$m_{1}$ is moving with initial velocity $v_{0}=200$ Km$s^{-1}$ and
an impact parameter $b=1$ AU. The distance of the GW source is
assumed to be $R=8$ kpc and the eccentricity is  $\epsilon=
0.2,0.5, 0.7.$ } \label{fig:ellisse800pc}
\end{figure}

Beside luminosity, we can characterize also the GW amplitude starting from the motion of  sources. In the case of a binary system and a single amplitude component 
, it is straightforward to show that

\begin{displaymath}
\begin{array}{llllllll}
h^{11}=-\frac{2G}{Rc^4}\frac{G (m_{1}+m_{2}) \mu  (13 \epsilon  \cos \phi+12 \cos2 \phi+\epsilon  (4 \epsilon +3 \cos3 \phi))}{2 l}~,
\\ \\
h^{22}=\frac{2G}{Rc^4}\frac{G (m_{1}+m_{2})  \mu  (17 \epsilon  \cos\phi+12 \cos2 \phi+\epsilon  (8 \epsilon +3 \cos3 \phi))}{2 l}
 ~,
\\ \\
h^{12}=h^{21}=-\frac{2G}{Rc^4}\frac{G (m_{1}+m_{2})  \mu  (13 \epsilon  \cos\phi+12 \cos2 \phi+\epsilon  (4 \epsilon +3 \cos3 \phi))}{2
l}
~,
\end{array}
\end{displaymath}
so that the expected strain amplitude
$h\simeq(h_{11}^2+h_{22}^2+2h_{12}^2)^{1/2}$ turns out to be

\begin{eqnarray*}
h&=&\frac{G^3 (m_{1}+m_{2}) \mu^2}{c^4 Rl^2}\times\\ &&
  (3 (13 \epsilon  \cos\phi+12 \cos2 \phi+\epsilon  (4 \epsilon
+3
\cos3 \phi))^2\\ &&+(17 \epsilon  \cos\phi+
12 \cos2 \phi+\epsilon  (8 \epsilon +3 \cos3 \phi))^2)^{\frac{1}{2}}
~,
\end{eqnarray*}
which, as before, strictly depends on the initial conditions of
the stellar encounter. A remark is in order at this point. A
monochromatic gravitational wave has, at most, two independent
degrees of freedom. In fact, in the TT gauge, we have  $h_{+} =
h_{11} + h_{22}$ and $h_{\times} = h_{12} + h_{21}$ (see e.g.
\cite{bla}). As an example, the amplitude of gravitational wave is
sketched in Fig. \ref{fig:ellisse800pc} for a stellar encounter , in Newtonian motion,
close to the Galactic Center. The adopted initial parameters are
typical of a close impact and are assumed to be $b=1$ AU for the impact factor and
$v_{0}=200$ Km$s^{-1}$ for the initial velocity, respectively. Here, we have fixed
$m_{1}=m_{2}=1.4M_{\odot}$. The impact parameter is defined as
$L=bv$ where $L$ is the angular momentum and $v$ the incoming
velocity. We have chosen  a typical velocity of a star in the
galaxy and we are considering, essentially,  compact objects with
masses comparable to the Chandrasekhar limit $(\sim
1.4M_{\odot})$. This choice is motivated by the fact that
ground-based experiments like VIRGO or LIGO expect to detect
typical GW emissions from the dynamics of these objects or from
binary systems composed by them (see e.g. \cite{maggiore}).

\subsection{Gravitational wave amplitude from parabolic and hyperbolic orbits}

The single components of amplitude for a
parabolic and hyperbolic orbits are
\begin{displaymath}
\begin{array}{llllllll}
h^{11}=-\frac{G l^2 L^2\mu}{Rc^4}(13 \epsilon  \cos \phi+12 \cos2 \phi+\epsilon  (4 \epsilon +3 \cos3 \phi))
~,
\\ \\
h^{22}=\frac{Gl^2 L^2\mu}{Rc^4}(17 \epsilon  \cos\phi+12 \cos2 \phi+\epsilon  (8 \epsilon +3 \cos3 \phi))
 ~,
\\ \\
h^{12}=h^{21}=-\frac{3Gl^2 L^2\mu}{Rc^4} (4 \cos \phi+\epsilon  (\cos2 \phi+3)) \sin\phi~,
\end{array}
\end{displaymath}
and then the expected strain amplitude is

\begin{eqnarray*}
h&=&\frac{2 l^4 L^4 \mu ^2 }{c^4 R}\times
 (10 \epsilon ^4+9  \epsilon ^3\cos 3 \phi+59 \epsilon ^2
   \cos2 \phi \\ &&+59 \epsilon ^2+\left(47 \epsilon
   ^2+108\right)  \epsilon\cos\phi +36)^{\frac{1}{2}}
~,
\end{eqnarray*}
which, as before, strictly depends on the initial conditions of
the stellar encounter. We note that the gravitational wave
amplitude  has the same analytical expression for both cases and
differs only for the value of $\epsilon$ which is $\epsilon=1$ if
the motion is  parabolic and the  polar angle range  is
$\phi\in(-\pi,\pi)$, while it is $\epsilon>1$ and
$\phi\in(-\pi,\pi)$ for hyperbolic orbits. In these cases, we have
non-returning objects.

\begin{figure}
\includegraphics[scale=0.5]{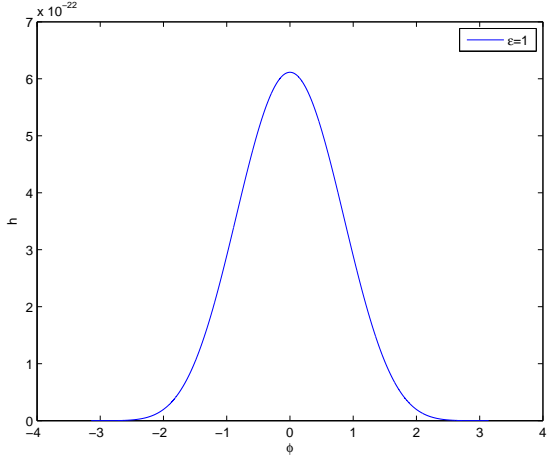}
\caption{The gravitational wave-forms for a parabolic encounter as
a function of the polar angle $\phi$.   As above,
$m_{1}=m_{2}=1.4M_{\odot}$ and $m_{2}$ is considered at rest.
$m_{1}$ is moving with initial velocity $v_{0}=200$ Km$s^{-1}$
with an impact parameter $b=1$ AU. The distance of the GW source
is assumed at $R=8$ kpc. The eccentricity is $\epsilon=1$. }
\label{fig:parabola}
\end{figure}

\begin{figure}
\includegraphics[scale=0.5]{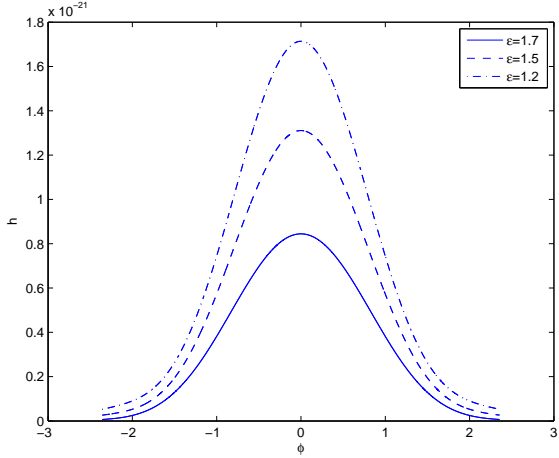}
\caption{The gravitational wave-forms for hyperbolic encounters as
function of the polar angle $\phi$.  As above, we have fixed
$m_{1}=m_{2}=1.4M_{\odot}$. $m_{2}$ is considered at rest while
$m_{1}$ is moving with initial velocity $v_{0}=200$ Km$s^{-1}$ and
an impact parameter $b=1$ AU. The distance of the source is
assumed at $R=8$ kpc. The eccentricity is assumed with the values
$\epsilon=1.2,1.5,1.7$ .} \label{fig:iperbole8}
\end{figure}

The amplitude of the gravitational wave is sketched in Figs.
\ref{fig:parabola} and \ref{fig:iperbole8} for  stellar encounters
close to the Galactic Center. As above, we consider  a close
impact and  assume  $b=1$ AU cm, $v_{0}=200$ Km$s^{-1}$ and
$m_{1}=m_{2}=1.4M_{\odot}$. In summary, we can say that also in the case of Newtonian motion of sources, the orbital features characterize GW luminosities and amplitudes.

 \section{Gravitational waves from sources in relativistic motion}
It is straighforward to extend the above considerations to orbital motions containing Post$-$Newtonian corrections. It is clear that GW luminosity and amplitude are strictly dependent on the parameter ($\frac{v}{c}$) considered at various order of approximation and, as discussed above, the global feature of orbits fully characterize the gravitational emission. Now we study how the waveforms depend on the dynamics of binary and colliding systems and how relativistic corrections modulate the features of gravitational radiation.

 \subsection{Inspiralling  waveform including post-Newtonian corrections}
\label{sec6.2}

As we have shown in the above section the PN method involves an expansion around the Newtonian
limit keeping terms of higher order in the small
parameter~\cite{TD87,PNreview,buonanno}

\begin{equation}
\epsilon \sim \frac{v^2}{c^2} \sim \left |h_{\mu \nu} \right |
\sim \left |\frac{\partial_0 h}{\partial_i h} \right |^2 \sim
\left |\frac{T^{0i}}{T^{00}}\right | \sim \left |\frac{T^{ij}}{T^{00}}
\right |\,.
\end{equation}

In order to be able to determine the dynamics of binary systems with a precision
acceptable for detection, it has been necessary to compute the force determining
the motion of the two bodies and the amplitude of the gravitational radiation
with a precision going beyond the quadrupole formula.
For nonspinning BHs, the two-body equations of motion and the GW
flux are currently known through 3.5PN order~\cite{PNnospin}.
Specifically if we restrict
the discussion to circular orbits, as Eq.~(\ref{eq}) shows,
there exists a natural {\it adiabatic} parameter $\dot{\Omega}/\Omega^2 \cong {\cal O}[(v/c)^5]$.
Higher-order PN corrections to Eq.~(\ref{eq}) have been
computed~\cite{PNnospin,PNspin}, yielding the general equation:
\begin{equation}
\frac{\dot{\Omega}}{\Omega^2} = \frac{96}{5}\,\nu\,v_\Omega^{5/3}\,
\sum_{k=0}^7 {\Omega}_{(k/2)\mathrm{PN}}\,v_\Omega^{k/3}\,
\label{omegadot}
\end{equation}
where $G=1=c$ and where we define $v_\omega \equiv (M\,\omega)^{1/3}$ . The PN-order is given by $\Omega_{(k/2)PN}$ which is, up to $k=7$,
\begin{equation}
{\Omega}_{0\mathrm{PN}} = 1\,,
\label{omegadotSTpn0}
\end{equation}
\begin{equation}
{\Omega}_{0.5\mathrm{PN}} = 0\,,\\
\label{omegadotSTpn05}
\end{equation}
\begin{equation}
{\Omega}_{1\mathrm{PN}} = -\frac{743}{336} -\frac{11}{4}\,\nu\,, \\
\label{omegadotSTpn1}
\end{equation}
\begin{equation}
{\Omega}_{1.5\mathrm{PN}} = 4\pi + \left[-\frac{47}{3}\frac{S_L}{M^2}
-\frac{25}{4}\frac{\delta m}{M}\frac{\Sigma_L}{M^2}\right]\,,\\
\label{omegadotSTpn15}
\end{equation}
\begin{eqnarray}
{\Omega}_{2\mathrm{PN}} &=&
\frac{34\,103}{18\,144}+\frac{13\,661}{2\,016}\,\nu+\frac{59}{18}\,\nu^2 - \nonumber \\
&& \frac{1}{48}\, \nu\,\chi_1\chi_2\left[247\,(\hat {S}_1\cdot\hat{ S}_2)- 721\,
(\boldsymbol{\hat{L}}\cdot\hat{S}_1)(\boldsymbol{\hat{L}}\cdot\hat{S}_2)\right]\,,\nonumber\\
\label{omegadotSTpn2}
\end{eqnarray}
\begin{eqnarray}
{\Omega}_{2.5\mathrm{PN}} &=& -\frac{1}{672}\,(4\,159 +15\,876\,\nu)\,\pi +
 \left[\left(-\frac{31811}{1008}+\right.\right.\nonumber\\ && \left.\left.\frac{5039}{84}\nu\right)\frac{S_L}{M^2}+   \left(-\frac{473}{84}+\frac{1231}{56}\nu\right)\frac{\delta
m}{M}\frac{\Sigma_L}{M^2}\right]\,, \nonumber\\
\label{omegadotSTpn25}
\end{eqnarray}
\begin{eqnarray}
{\Omega}_{3\mathrm{PN}} &=&
\left(\frac{16\,447\,322\,263}{139\,708\,800}-\frac{1\,712}{105}\,\gamma_E+\frac{16}{3}\pi^2\right)+\nonumber\\ &&
\left(-\frac{56\,198\,689}{217\,728}+ \frac{451}{48}\pi^2 \right)\nu
+\nonumber\\ &&\frac{541}{896}\,\nu^2-\frac{5\,605}{2\,592}\,\nu^3
-\frac{856}{105}\log\left[16v^{2}\right]\,,\nonumber\\
\label{omegadotSTpn3}
\end{eqnarray}
\begin{eqnarray}
{\Omega}_{3.5\mathrm{PN}} &=& \left (
-\frac{4\,415}{4\,032}+\frac{358\,675}{6\,048}\,\nu+\frac{91\,495}{1\,512}\,\nu^2
\right )\,\pi\,.\nonumber\\
\label{omegadotSTpn35}
\end{eqnarray}
We denote $\boldsymbol{L} = \mu \,\bf{X} \times \bf{V}$ the
Newtonian angular momentum (with $\bf{X}$ and $\bf{V}$, as above, the two-body center-of-mass
radial separation and relative velocity), and $\boldsymbol{\hat{L}} = \boldsymbol{L} /
|\boldsymbol{L}|$; $\bf{S}_1 =\chi_1\,m_1^2\,\hat{S}_1$ and $\bf{S}_2
=\chi_2\,m_2^2\,\hat{S}_2$ are the spins of the two bodies (with
$\hat{S}_{1,2}$ unit vectors, and $0 < \chi_{1,2} < 1$ for BHs) and

\begin{equation}
\label{spins}
\mathbf{S} \equiv \mathbf{S}_1 + \mathbf{S}_2\,, \quad \mathbf{\Sigma} \equiv M\left[\frac{\mathbf{S}_2}{m_2} -
\frac{\mathbf{S}_1}{m_1}\right]\,.
\end{equation}
Finally, $\delta m = m_1-m_2$ and $\gamma_E=0.577\ldots$ is Euler's constant.

\begin{table*}[t]
\caption{Post-Newtonian contributions to the number of GW
  cycles accumulated from $\Omega_\mathrm{in} =
  \pi\times 10\,\mathrm{Hz}$ to $\Omega_\mathrm{fin} =
  \Omega^\mathrm{ISCO}=1/(6^{3/2}\,M)$ for binaries detectable by
  LIGO and VIRGO. We denote $\kappa_{i} = \hat{S}_i \cdot \boldsymbol{\hat{L}}$ and
   $\xi = \hat{\mathbf{S}}_1\cdot
  \hat{\mathbf{S}}_2$.
\label{tab:1}}
\begin{center}
{\scriptsize
\begin{tabular}{|l|c|c|}\hline
& \multicolumn{1}{c|}{$(10+10)M_\odot$} &
\multicolumn{1}{c|}{$(1.4+1.4)M_\odot$} \\ \hline\hline
Newtonian & $601$ & $16034$ \\
1PN & $+59.3$ & $+441$\\
1.5PN & $-51.4 + 16.0\, \kappa_1\,\chi_1 + 16.0\, \kappa_2\,\chi_2$
& $ -211 + 65.7\,\kappa_1\,\chi_1 + 65.7\, \kappa_2\,\chi_2$ \\
2PN & $+4.1 - 3.3\, \kappa_1\,\kappa_2\,\chi_1\,\chi_2 + 1.1\, \xi\,\chi_1\,\chi_2$
& $+ 9.9 - 8.0\, \kappa_1\,\kappa_2\,\chi_1\,\chi_2 + 2.8
\,\xi\,\chi_1\,\chi_2$ \\
2.5PN & $-7.1 + 5.5\, \kappa_1\,\chi_1 + 5.5\,
\kappa_2\,\chi_2$ & $-11.7 + 9.0\, \kappa_1\,\chi_1 + 9.0\,
\kappa_2\,\chi_2$ \\
3PN & $+2.2$ & $+2.6$ \\
3.5PN & $-0.8$ & $-0.9$ \\ \hline
\end{tabular}}\end{center}
\end{table*}
It is instructive to compute the relative contribution of the
PN terms to the total number of GW cycles accumulating
in the frequency band of LIGO/VIRGO. In Table~\ref{tab:1},
we list the figures obtained by plugging Eq.~(\ref{omegadot})
into Eq.~(\ref{cycles}) \cite{buonanno}. As final frequency, we use the
innermost stable circular orbit (ISCO) of a point particle in
Schwarzschild BH [$f_{\rm GW}^{\rm ISCO} \simeq 4400/(M/M_\odot)$ Hz].

\subsection{The full waveform: inspiral, merger and ring-down}
\label{sec6.3}
\begin{figure}
\begin{center}
\begin{tabular}{cc}
\includegraphics[width=0.5\textwidth]{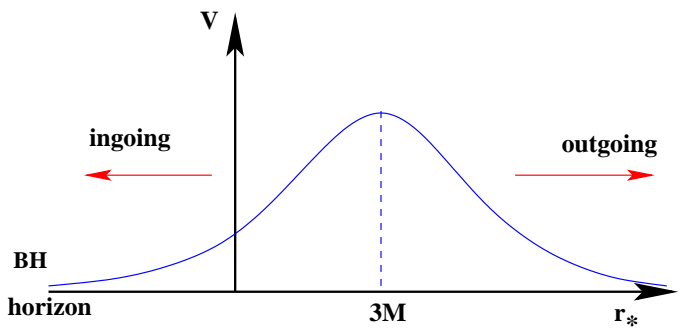} 
\end{tabular}
\caption{We sketch the curvature potential
as function of the tortoise coordinate
$r^*$ associated to metric perturbations of
a Schwarzschild BH.  \label{fig:7}}
\end{center}
\end{figure}

\begin{figure}
\begin{center}
\begin{tabular}{cc}
\includegraphics[width=0.3\textwidth]{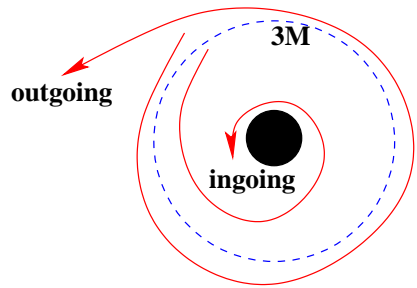}\\
\end{tabular}
\caption{The potential peaks at the last
unstable orbit for a massless particle (the light ring).
Ingoing modes propagate toward the BH horizon, whereas
outgoing modes propagate away from the source. \label{fig:8}}
\end{center}
\end{figure}

After the two BHs merge, the system settles down to a Kerr BH
and emits quasi-normal modes (QNMs),~\cite{qnm,Press}. This phase is commonly known as
the ring-down (RD) phase. Since the QNMs have complex frequencies totally
determined by the BH's mass and spin, the RD waveform is a
superposition of damped sinusoidals. The inspiral and
RD waveforms can be computed analytically. What about the
merger? Since the nonlinearities dominate,
the merger would be described at {\it best} and {\it utterly}
through numerical simulations of Einstein equations. However,
before numerical relativity (NR) results became available, some
analytic approaches were proposed.
In the test-mass limit, $\nu \ll 1$, Refs.~\cite{Davis,Press} realized a long time
ago that the basic physical reason underlying
the presence of a universal merger signal was that
when a test particle falls below $ 3 M$ (the
unstable light storage ring of Schwarzschild), the GW that
it generates is strongly filtered by the curvature potential
barrier centered around it (see Fig.~\ref{fig:7}).
For the equal-mass case $\nu = 1/4$, Price and Pullin~\cite{CLA}
proposed the so-called close-limit approximation,
which consists in switching from the two-body description to the
one-body description (perturbed-BH) close to the light-ring
location. Based on these observations,
the effective-one-body (EOB) resummation scheme~\cite{EOB}
provided a first {\it example} of full
waveform by (i) resumming the PN Hamiltonian, (ii)
modeling the merger as a very short (instantaneous) phase
and (iii) matching the end of the plunge (around the light-ring)
with the RD phase (see Ref.~\cite{lazarus} where similar
ideas were developed also in NR).
The matching was initially done using {\it only} the least
damped QNM whose mass and spin were determined by the binary BH
energy and angular momentum at the end of the plunge.
An example of full waveform is given in Fig.~\ref{fig:7},\ref{fig:6}.

\begin{figure}
\begin{center}
\begin{tabular}{cc}
\includegraphics[width=0.35\textwidth]{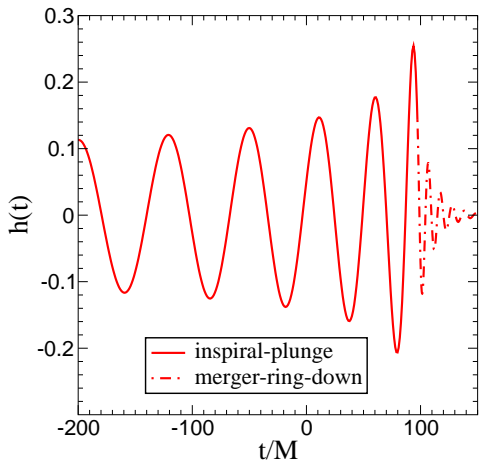} 
\end{tabular}
\caption{GW signal from an equal-mass
nonspinning BH binary as predicted at 2.5PN order by Buonanno and Damour (2000)
in Ref.~\cite{EOB}.
The merger is assumed almost instantaneous and one QNM is included.
\label{fig:6}}
\end{center}
\end{figure}

Today, with the results in NR, we are in the position
of assessing the closeness of analytic to numerical waveforms for
inspiral, merger and RD phases. In Fig.~\ref{fig:6}, we show some
first-order comparisons between the EOB-analytic and NR
waveforms~\cite{BCP} (see also Ref.~\cite{Goddshort}). Similar
results for the inspiral phase but using PN
theory~\cite{PNnospin,PNspin} (without resummation) at 3.5PN order
are given in Refs.~\cite{BCP,Goddshort}. So far, the agreement is
qualitatively good, but more accurate simulations, starting with
the BHs farther apart, are needed to draw robust conclusions.

Those comparisons are suggesting that it should be possible to
design purely analytic templates with the full numerics used to guide the
patching together of the inspiral and RD waveforms.
This is an important avenue to template construction as
eventually hundreds of thousands of waveform
templates may be needed to extract the signal from
the noise, an impossible demand for NR alone.

\begin{figure}
\begin{center}
\begin{tabular}{cc}\includegraphics[width=0.35\textwidth]{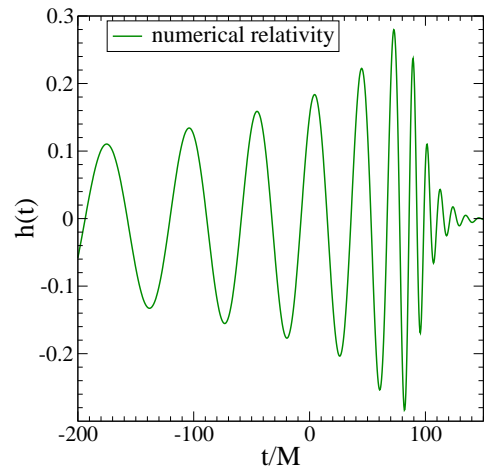}
\end{tabular}
\caption{ GW signal from an equal-mass
BH binary with a small spin $\chi_1=\chi_2 = 0.06$ obtained in
full GR by Pretorius~\cite{BCP}.
\label{fig:6}}
\end{center}
\end{figure}

\section{Gravitational waves with gravitomagnetic corrections}

In this section, we are going to study the evolution of  compact
binary systems,  formed through the  capture of a moving (stellar)
mass $m$ by the gravitational field, whose source is a massive MBH
of mass $M$ where $m \ll M$.  One expects that small compact
objects ($1\div 20 M_{\odot}$) from the surrounding stellar
population will be captured by these black holes following
many-body scattering interactions at a relatively high rate
\cite{Sigurdsson,Sigurdsson2}. It is well known that the  capture
of stellar-mass compact objects by massive  MBHs could constitute,
potentially, a very important target for LISA
\cite{Danzmann,freitag}. However, dynamics has to be carefully
discussed in order to consider and select all effects coming from
standard stellar mass objects inspiralling  over MBHs.

In the first part of this review, we have shown that, in the
relativistic weak field approximation, when considering higher
order corrections to the equations of motion, gravitomagnetic
effects in the theory of orbits,  can be particularly significant,
leading also to chaotic behaviors in the transient regime dividing
stable from unstable trajectories. Generally, such contributions
are discarded since they are considered too small. However, in a
more accurate analysis, this is not true and gravitomagnetic
corrections could give peculiar characterization of dynamics  \cite{SMFL}.

According to these effects, orbits remain rather eccentric  until
the final plunge, and display both extreme  relativistic
perihelion precession and Lense-Thirring
\cite{Thirring1,Thirring,iorio} precession of the orbital plane
due to the spin of MBH, as well as orbital decay. In \cite{Ryan},
it is illustrated how the measured GW-waveforms can  effectively
map out the spacetime geometry close to the MBH. In
\cite{CDDIN,SF},   the classical orbital motion (without
relativistic corrections in the motion of the binary system) has
been studied  in the extreme mass ratio limit $m\ll M$, assuming
the stellar system density and richness as fundamental parameters.
The conclusions have been  that 
\begin{itemize}
\item  the GW-waveforms have been
characterized by the orbital motion (in particular, closed or open
orbits give rise to very different GW-production and waveform
shapes);
\item   in rich and dense stellar clusters, a large
production of GWs can be expected, so that these systems could be
very interesting for the above mentioned ground-based and space
detectors;
\item  the amplitudes of the strongest GW signals are
expected to be roughly an order of magnitude smaller than LISA's
instrumental noise.
\end{itemize}

We investigate the GW emission by binary systems,
in the extreme  mass ratio limit, by the quadrupole approximation,
considering  orbits affected by both nutation and precession
effects,  taking into account also  gravitomagnetic terms in the
weak field approximation of the metric. We will see that
gravitational waves are emitted with a "peculiar" signature
related to the orbital features: such a signature may be a "burst"
wave-form with a maximum in correspondence to the periastron
distance or a modulated waveform, according to the orbit
stability. Here we face this problem discussing in detail the
dynamics of such a phenomenon which could greatly improve the
statistics of possible GW sources.

Besides, we give estimates of the distributions of these sources
and their parameters. It is worth noticing  that the captures
occur when  objects, in the dense stellar cusp surrounding a
galactic MBH, undergo a close encounter, so that the trajectory
becomes tight enough that orbital decay through emission of GWs
dominates the subsequent evolution. According to Refs.
\cite{cutler,cutler1}), for a typical capture, the initial
orbital eccentricity is extremely large (typically $1-e\sim
10^{-6}{-}10^{-3}$) and the initial pericenter distance very small
($r_{\rm p}\sim 8-100 M$, where $M$ is the MBH mass
\cite{FreitagApJ}. The subsequent orbital evolution may (very
roughly) be divided into three stages. In the first and longest
stage the orbit is extremely eccentric, and GWs are emitted in
short ``pulses'' during pericenter passages. These GW pulses
slowly remove energy and angular momentum from the system, and the
orbit gradually shrinks and circularizes. After $\sim 10^3-10^8$
years (depending on the two masses and the initial eccentricity)
the evolution enters its second stage, where the orbit is
sufficiently circular:  the emission can be viewed as continuous.
Finally, as the object reaches the last stable orbit, the
adiabatic inspiral transits to a direct plunge, and the GW signal
cuts off. Radiation reaction quickly circularizes the orbit over
the inspiral phase; however, initial eccentricities are large
enough that a substantial fraction of captures will maintain high
eccentricity  until the final plunge. It has been estimated
\cite{cutler1} that about half of the captures will plunge with
eccentricity $e\gtrsim 0.2$. While individually-resolvable
captures will mostly be detectable during the last $\sim 1-100$
yrs of the second stage (depending on the stellar mass $m$ and the
MBH mass), radiation emitted during the first stage will
contribute significantly to the confusion background. As we shall
see,  the above scenario is heavily modified since the
gravitomagnetic effects play a crucial role in modifying the orbital
shapes that are far from being simply circular or elliptic and no
longer closed.

\subsection{Gravitational waves amplitude considering orbits with gravitomagnetic corrections}

Direct signatures of gravitational radiation are given by
GW-amplitudes and  waveforms. In other words, the identification
of a GW signal is strictly related to the accurate selection of
the waveform shape by interferometers or any possible detection
tool. Such an achievement could give information on the nature of
the GW source, on the propagating medium, and, in principle, on
the gravitational theory producing such a radiation \cite{Dela}.

Considering the formulas of previous Section,  the GW-amplitude
can be evaluated by
\begin{equation}
h^{jk}(t,R)=\frac{2G}{Rc^4}\ddot{Q}^{jk}~, \label{ampli1}
\end{equation}
$R$ being the distance between the source and the observer and,
due to the above polarizations, $\{j,k\}=1,2$.

\begin{figure*}[t!]
\begin{tabular}{|c|c|}
\hline
\tabularnewline
\includegraphics[scale=0.4]{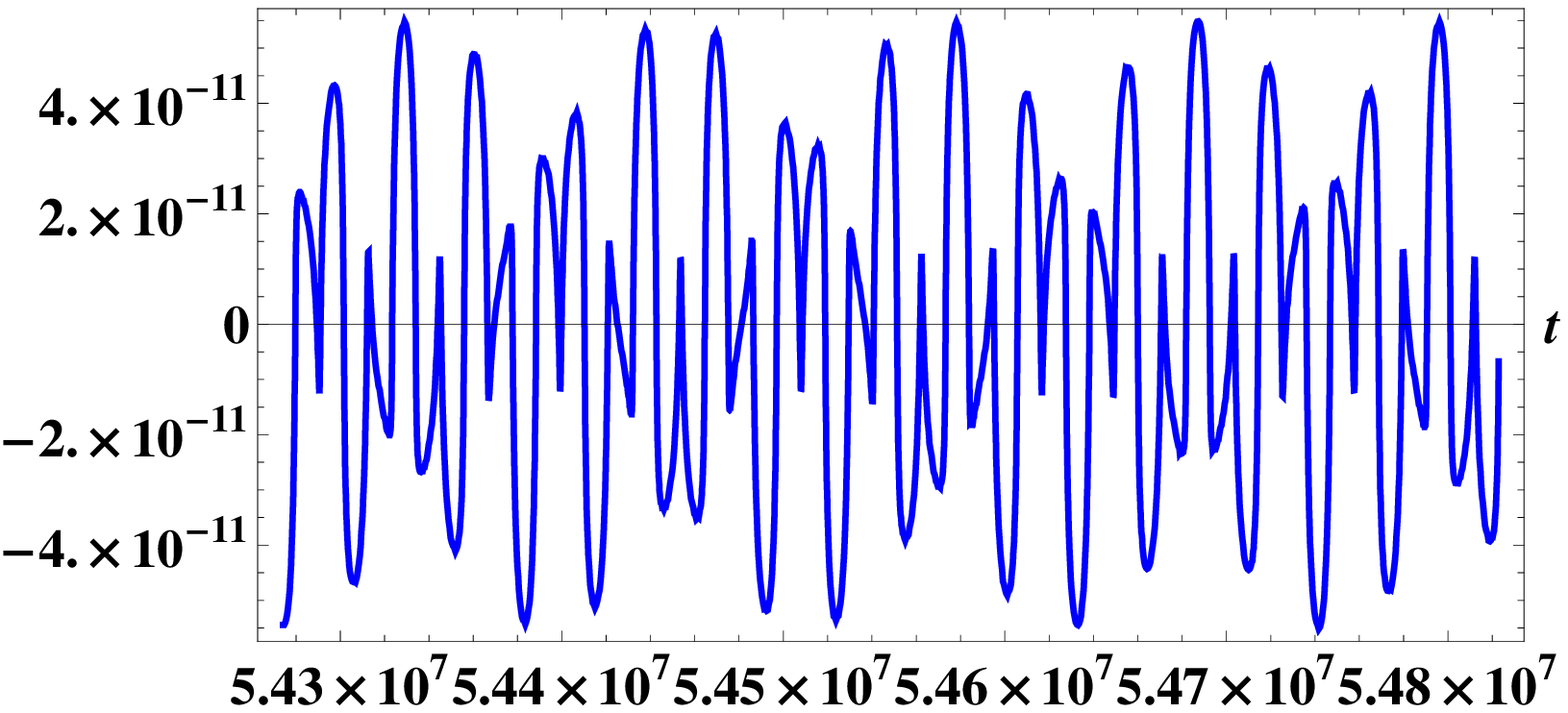}
\includegraphics[scale=0.4]{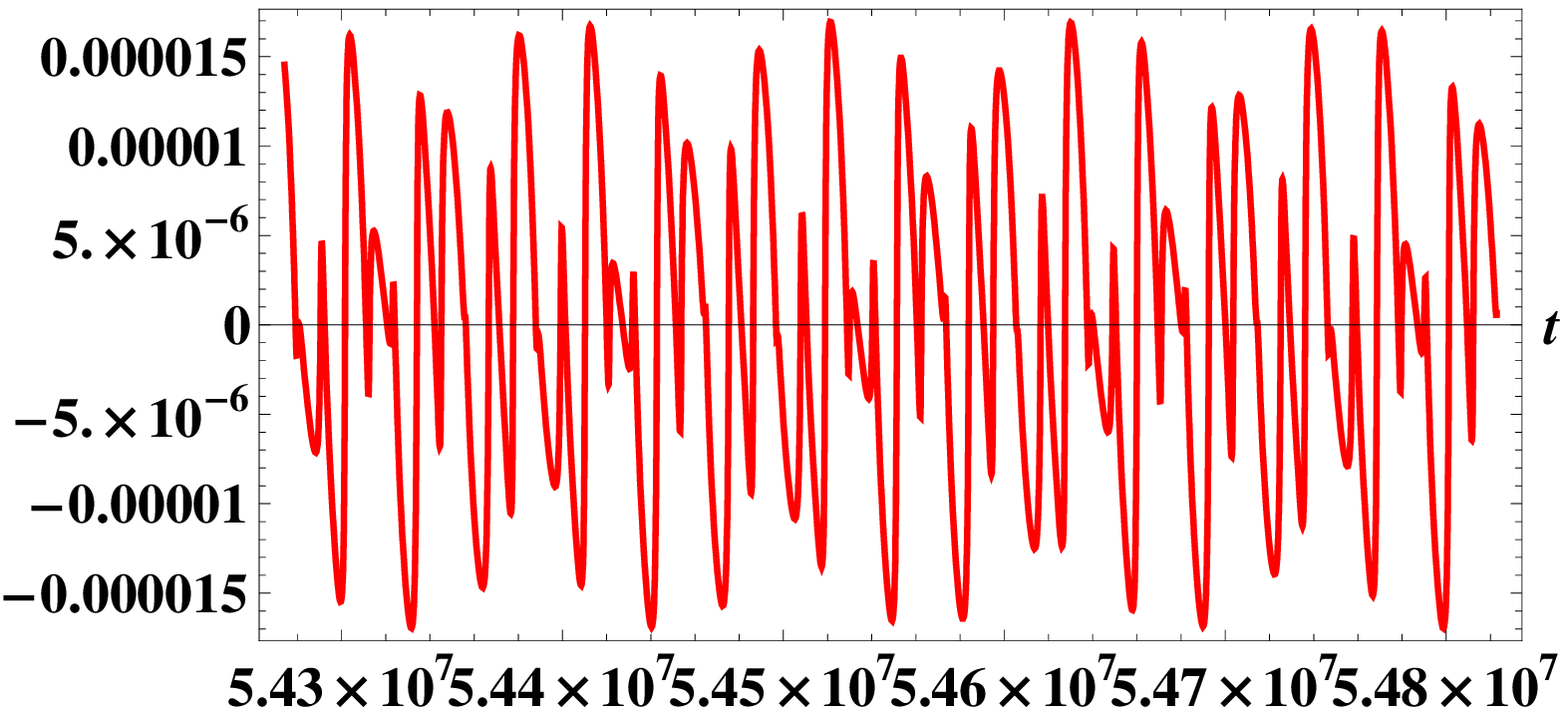}
\tabularnewline
\hline
\includegraphics[scale=0.4]{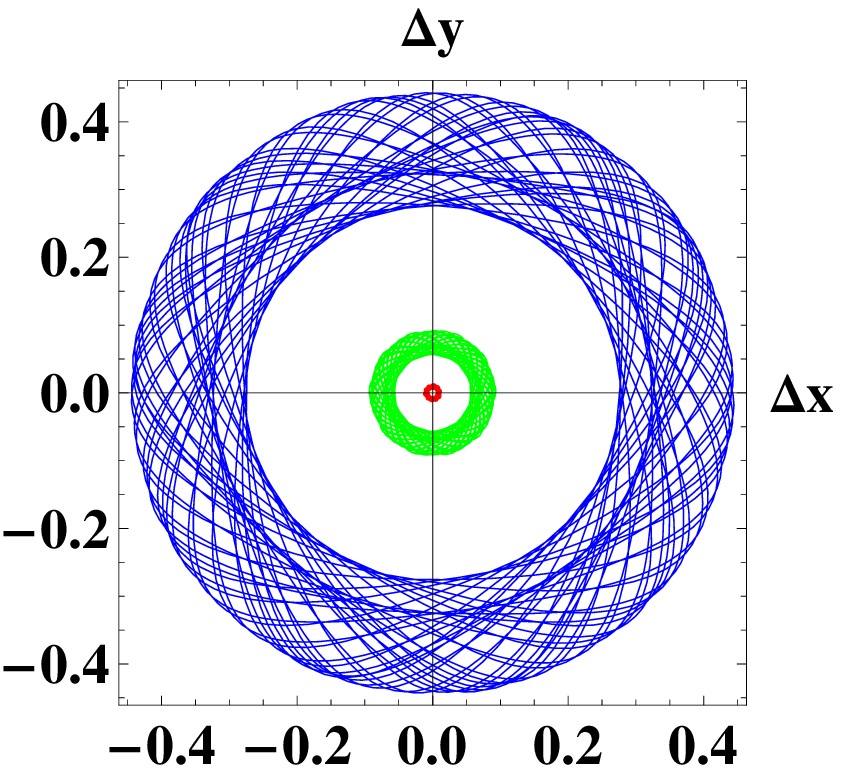}
\includegraphics[scale=0.4]{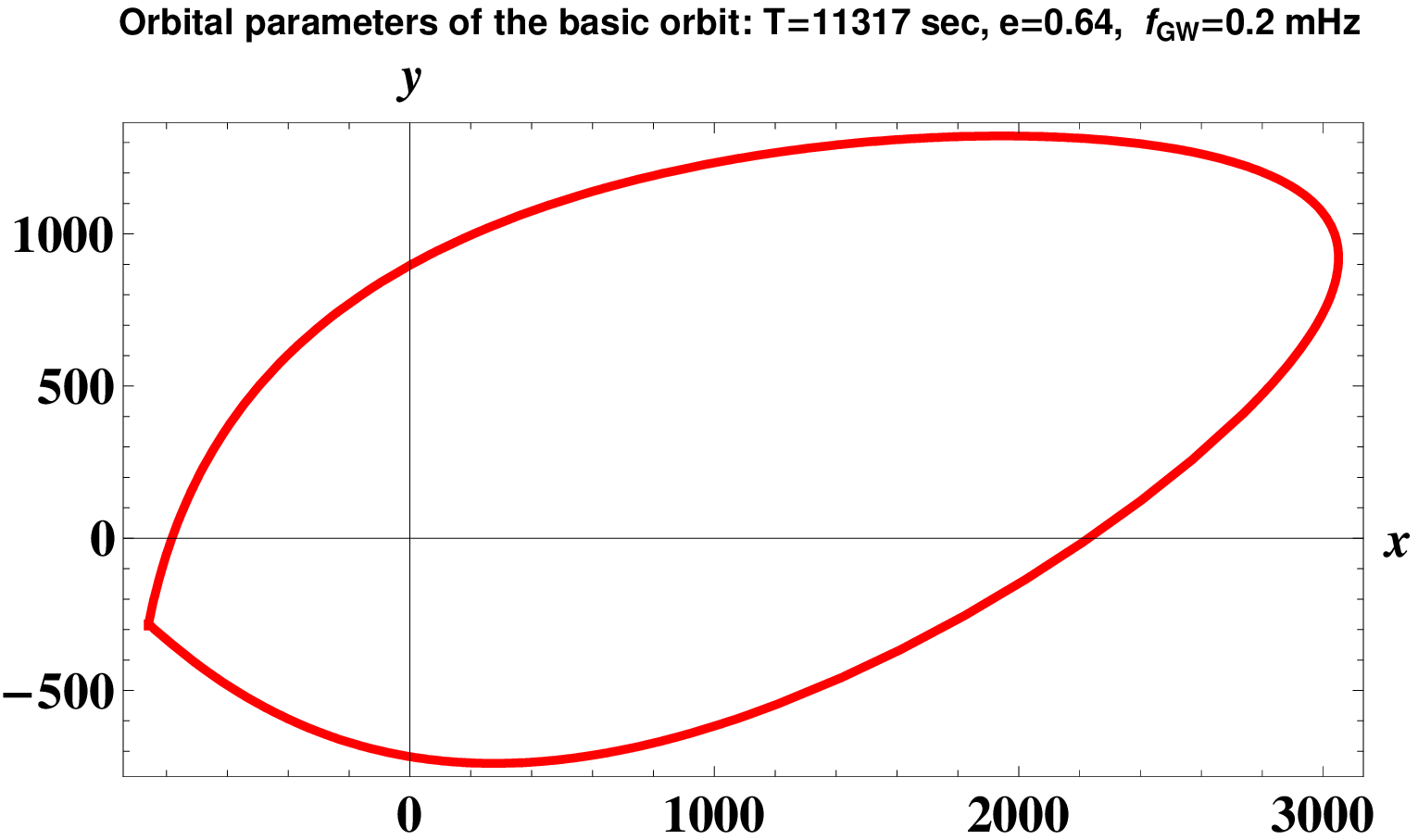}
\tabularnewline
\hline
\end{tabular}
\caption {\footnotesize{Plots of $z_{NO}(t)$ (left upper panel) and
$z_{Grav}(t)$ (right upper panel). It is interesting to see the
differences of about five orders of magnitude between the two
plots. At the beginning, the effect is very small but, orbit by
orbit, it grows and, for a suitable interval of coordinated time,
the effect cannot be neglected (see the left bottom panel in which
the differences in $x$ and $y$, starting from the initial orbits
up to the last ones, by steps of about 1500 orbits, are reported).
The internal red circle represents the beginning, the middle one
is the intermediate situation (green) and the blue one is the
final result of the correlation between $\Delta x$ versus $\Delta
y$, being $\Delta x=x_{Grav}-x_{NO}$ and $\Delta
y=y_{Grav}-y_{NO}$. On the bottom right, it is shown the basic
orbit.}}\label{Fig:03}
\end{figure*}

\begin{figure}[!ht]
\begin{tabular}{|c|c|c|}
\hline
\tabularnewline
\includegraphics [scale=0.60]{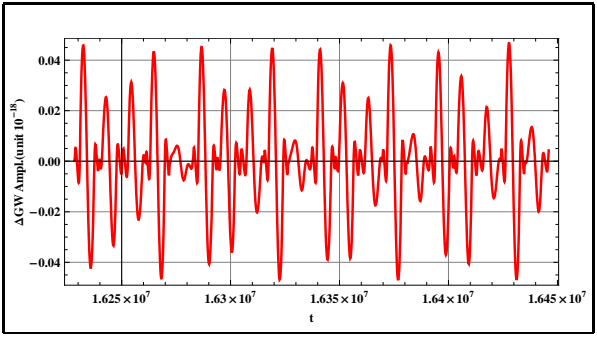}
 \tabularnewline
\hline
\end{tabular}
\caption {Plot of the differences of total gravitational waveform
$h$, with and without the gravitomagnetic orbital correction  for
a neutron star of $1.4 M_{\odot}$ orbiting around a MBH . The
waveform has been  computed at the Earth-distance from SgrA$^*$
(the central Galactic Black Hole). The example we are showing has
been obtained solving the systems for the following parameters and
initial conditions: $\mu\approx1.4 M_{\odot}$, $r_{0}$, $E=0.95$,
$\phi_{0}=0$, $\theta_{0}=\frac{\pi}{2}$, $\dot{\theta}_{0}=0$,
$\dot{\phi_{0}}=-\frac{1}{10}\dot{r}_{0}$ and
$\dot{r}_{0}=-\frac{1}{100}.$ It is worth noticing that frequency
modulation gives cumulative effects after suitable long times.
}\label{Fig:04}
\end{figure}

\begin{figure*}[!ht]
\begin{tabular}{|c|c|c|}
\hline
\tabularnewline
\includegraphics[scale=0.25]{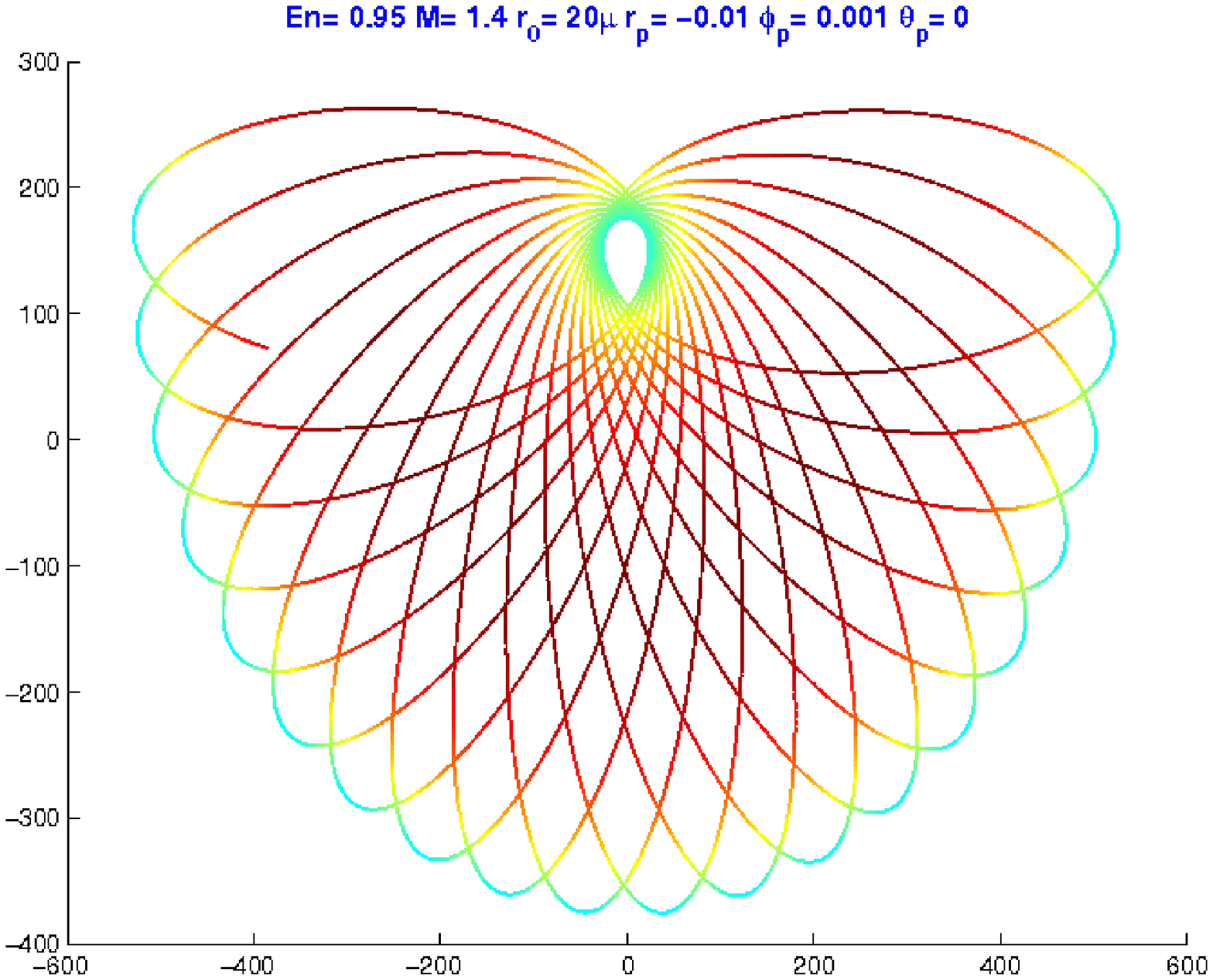}
\includegraphics[scale=0.25]{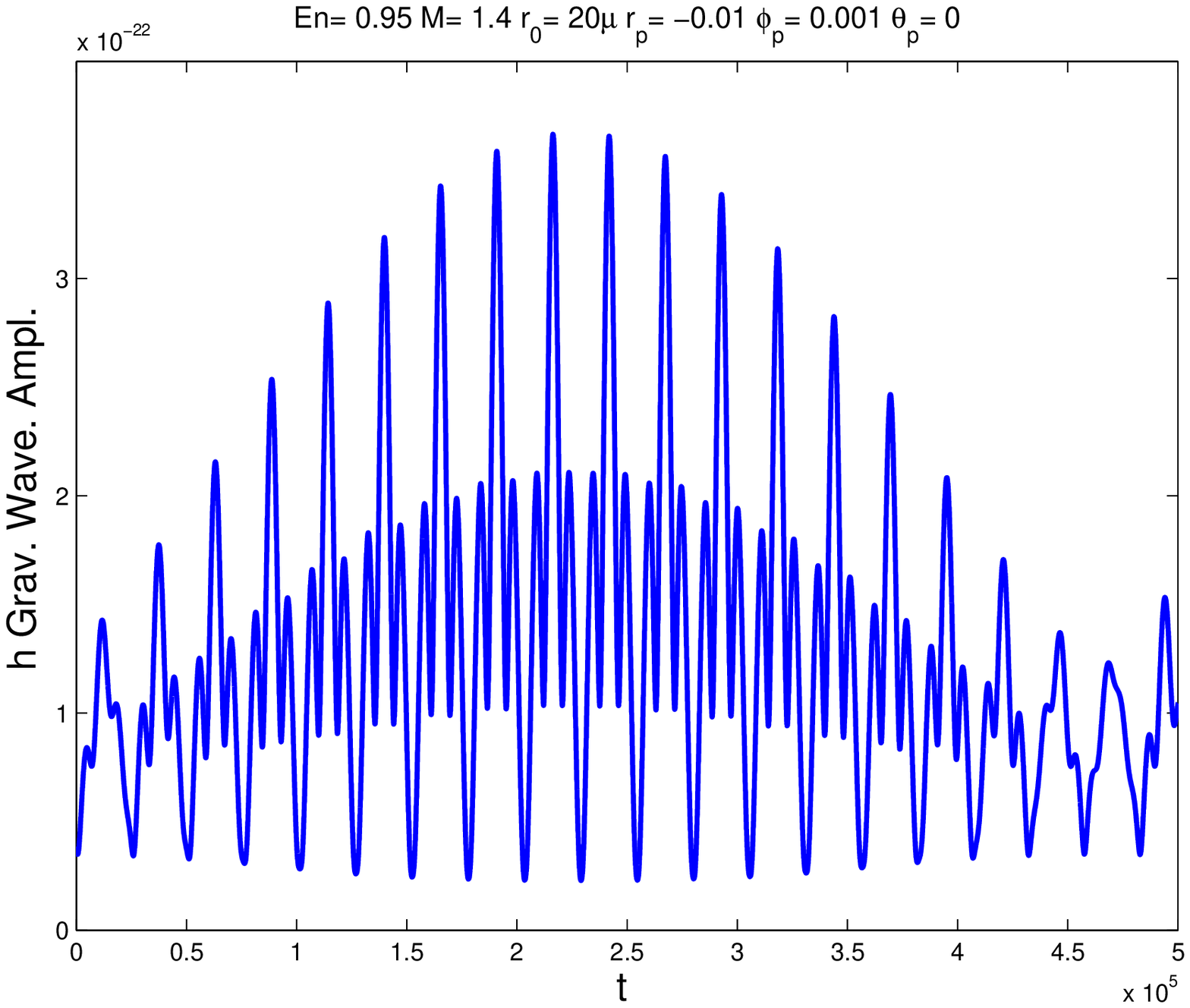}
\includegraphics[scale=0.25]{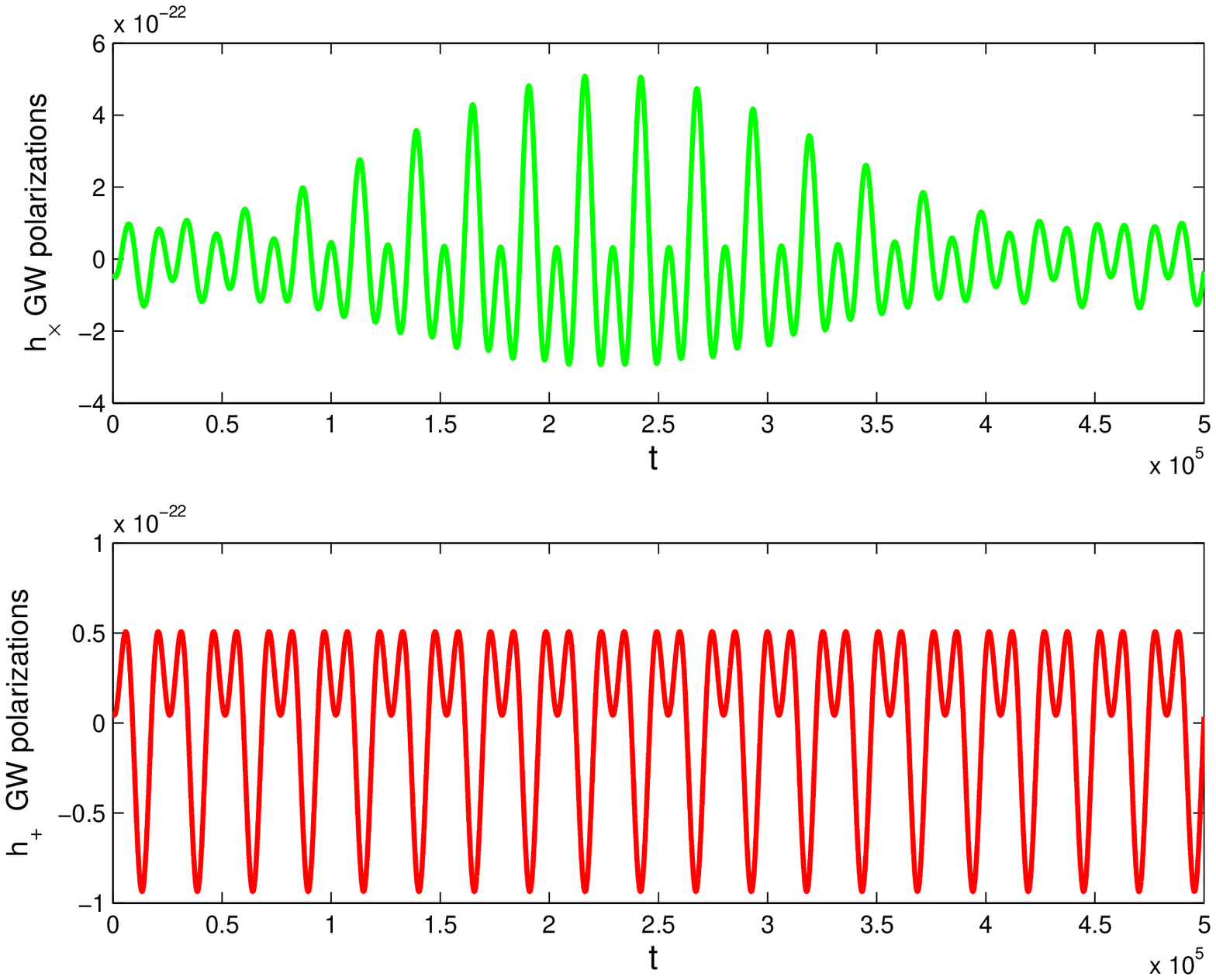}
 \tabularnewline
\hline
 \tabularnewline
\includegraphics[scale=0.25]{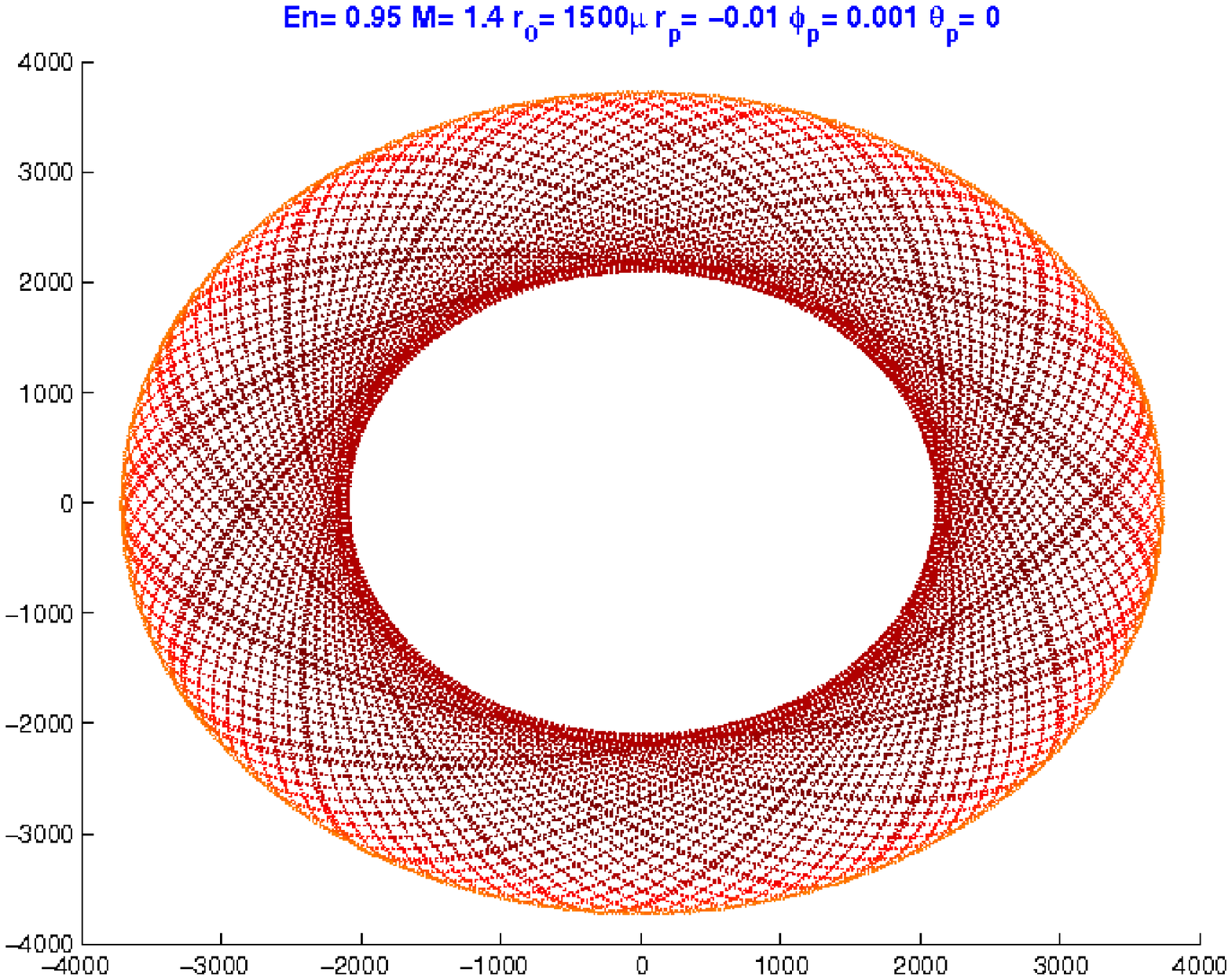}
\includegraphics[scale=0.25]{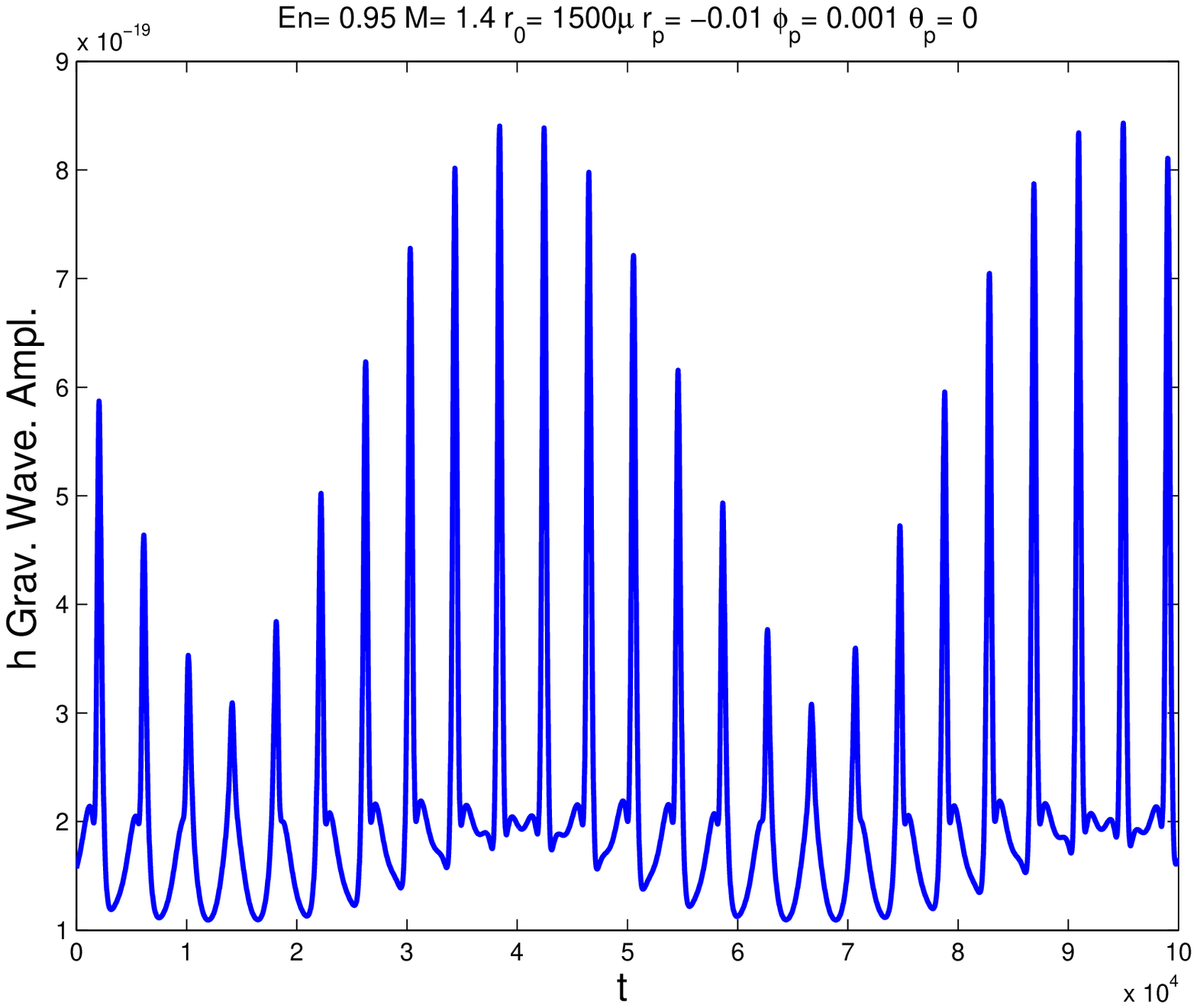}
\includegraphics[scale=0.25]{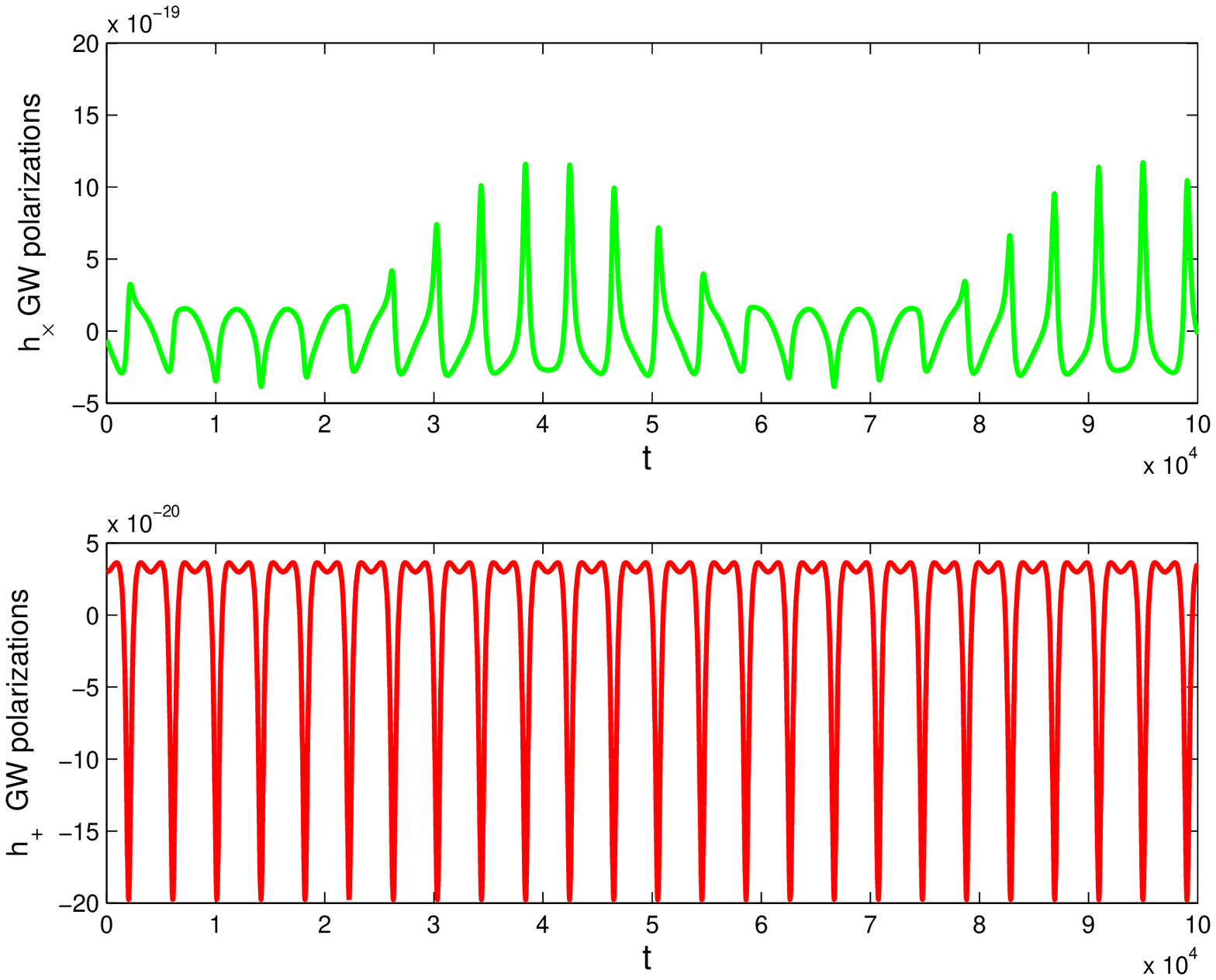}
\tabularnewline
\hline
\tabularnewline
\includegraphics[scale=0.25]{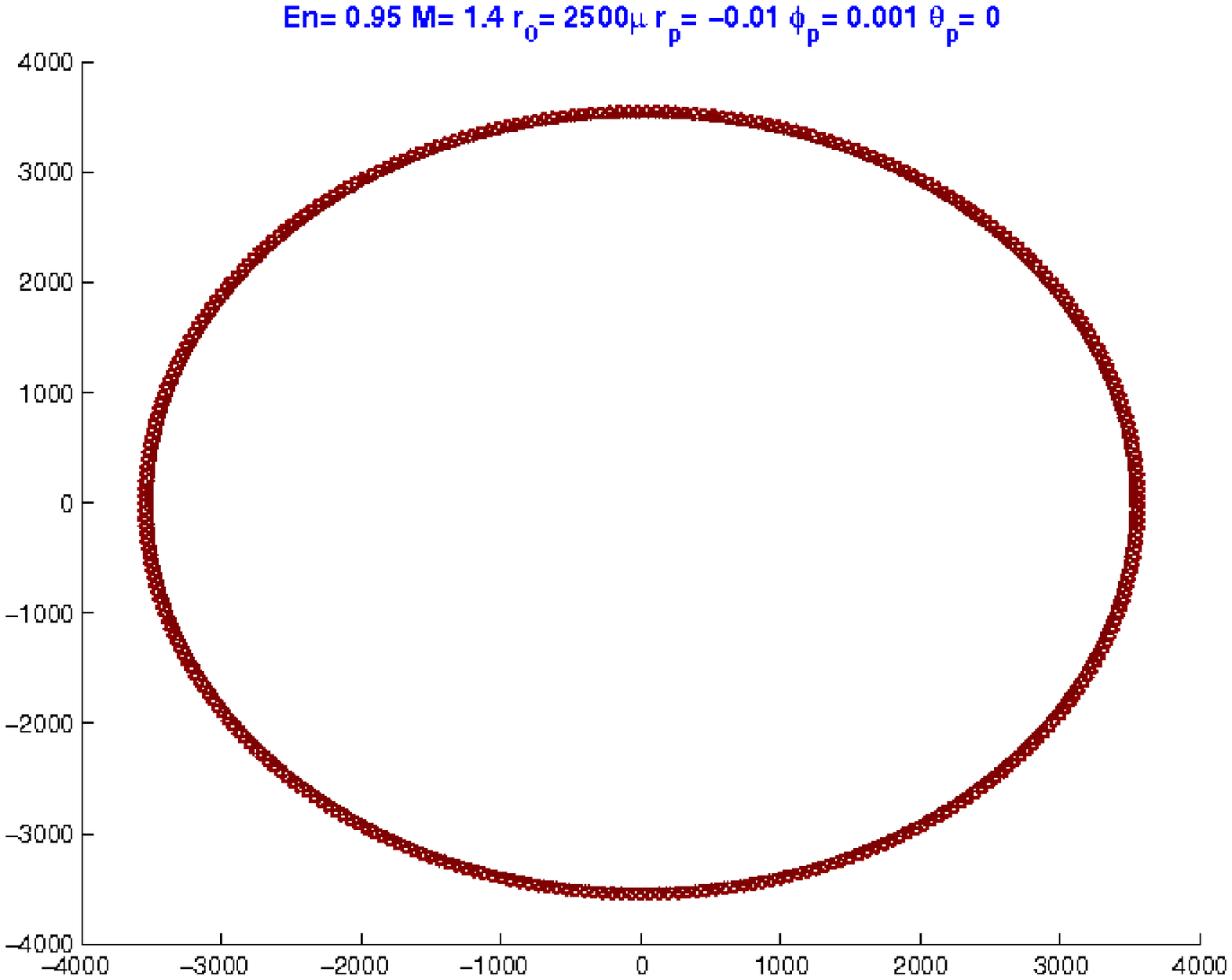}
\includegraphics[scale=0.25]{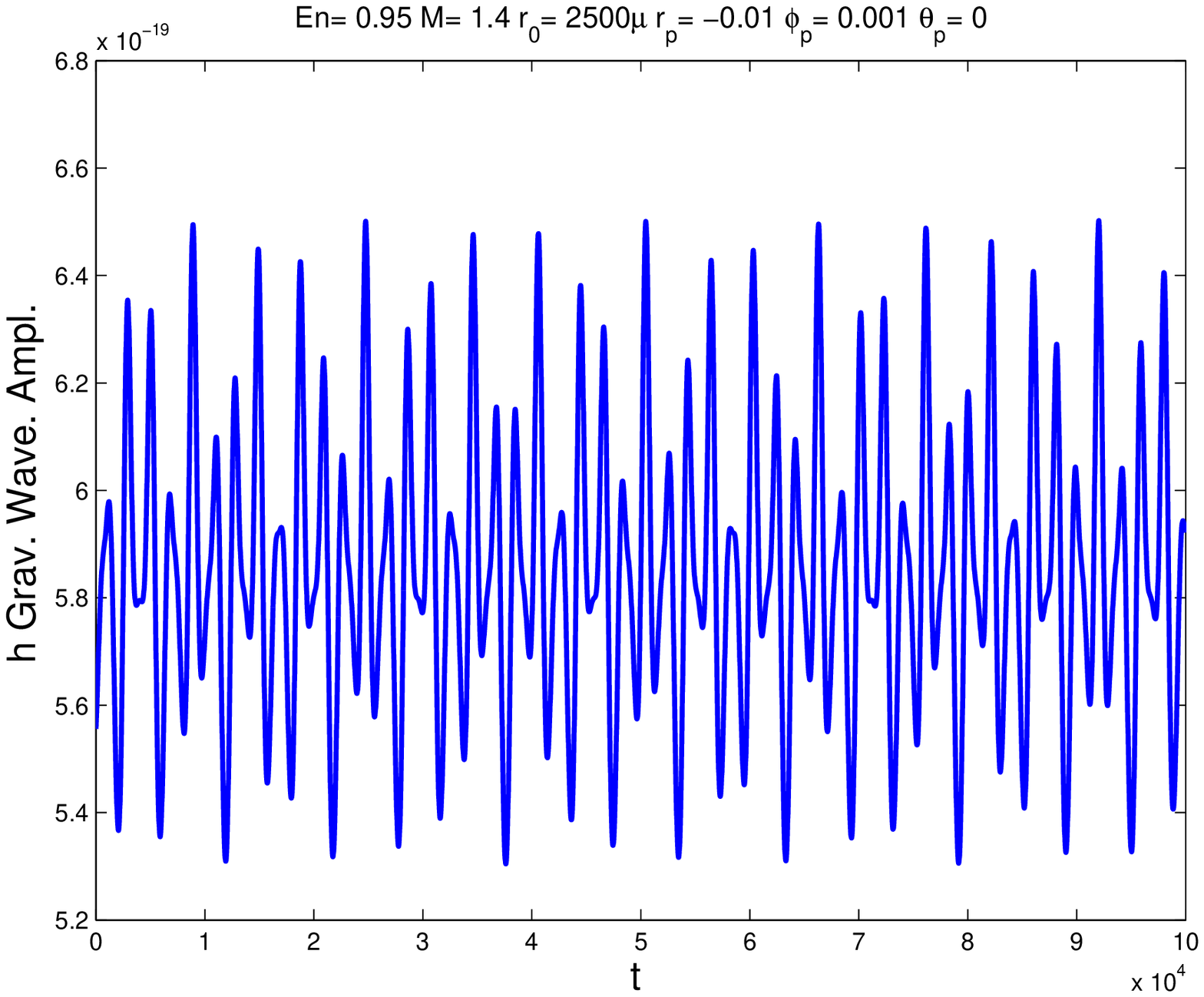}
\includegraphics[scale=0.25]{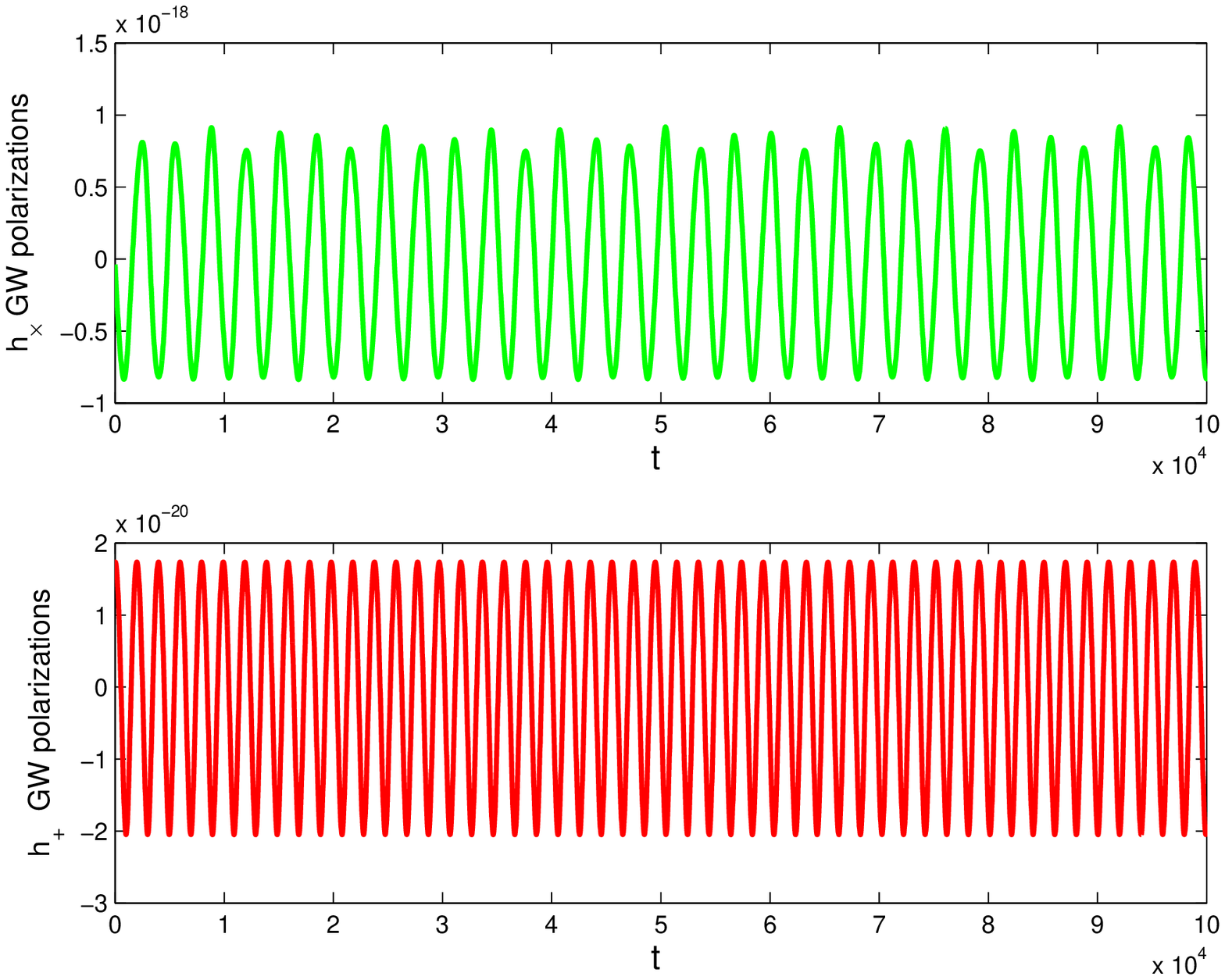}
 \tabularnewline
\hline
\end{tabular}
\caption {Plots along the panel  lines from left to right of field
velocities along the axes of maximum covariances, total
gravitational emission waveform $h$ and gravitational waveform
polarizations $h_{+}$ and $h_{\times}$ for a neutron star of $1.4
M_{\odot}$. The waveform has been computed for the Earth-distance
from Sagittarius A (the central Galactic Black Hole SgrA$^*$). The
plots we are showing have been obtained solving the system for the
following parameters and initial conditions: $\mu\approx1.4
M_{\odot}$, $E=0.95$, $\phi_{0}=0$, $\theta_{0}=\frac{\pi}{2}$,
$\dot{\theta_{0}}=0$, $\dot{\phi_{0}}=-\frac{1}{10}\dot{r}_{0}$
and $\dot{r}_{0}=-\frac{1}{100}$. From top to bottom of the
panels,  the  orbital radius is $r_0=20\mu,\,1500\mu,\,2500\mu$.
See also Table I.}\label{Fig:05}
\end{figure*}

\begin{figure*}[!ht]
\begin{tabular}{|c|c|c|}
\hline
\tabularnewline
\includegraphics[scale=0.25]{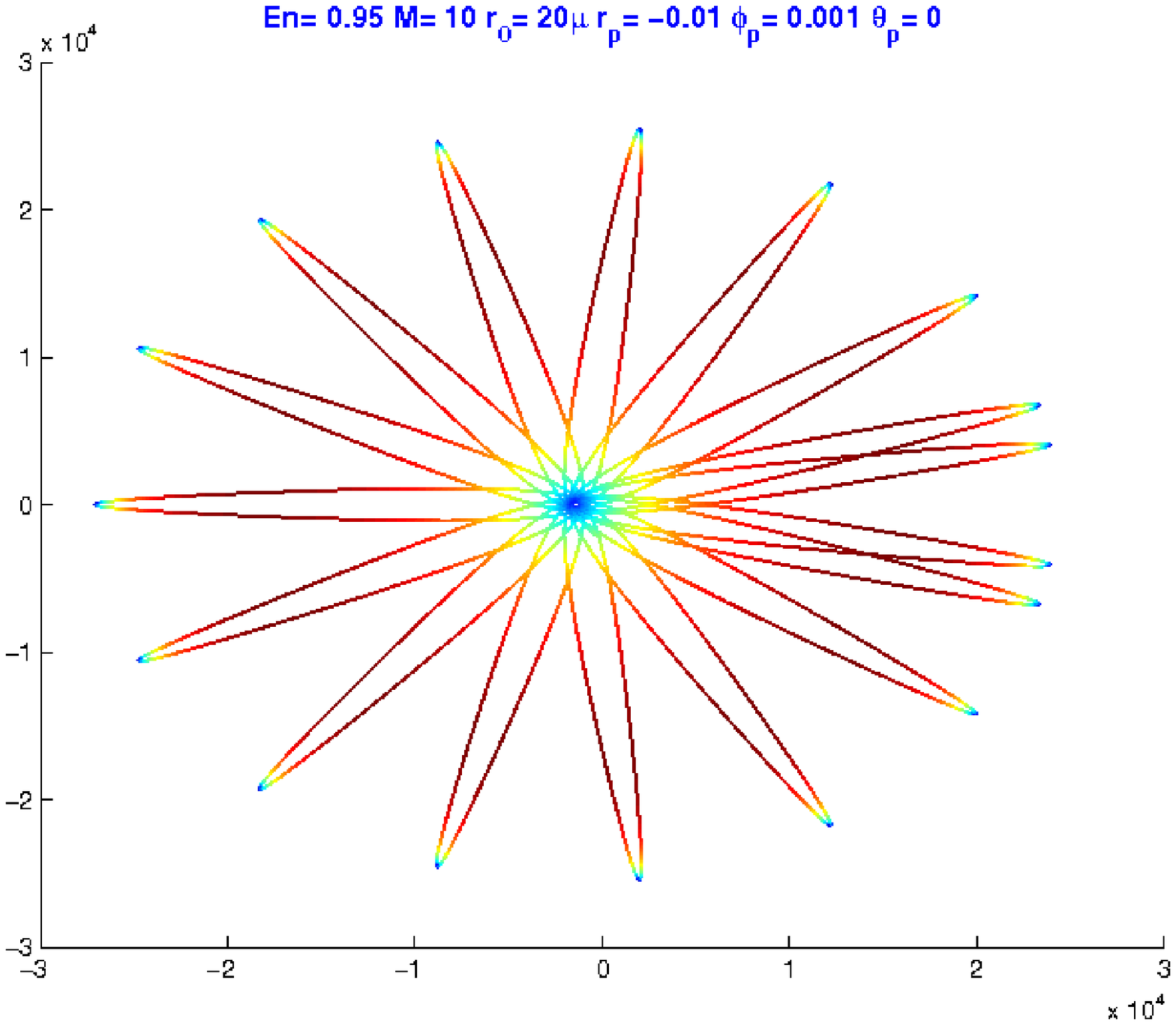}
\includegraphics[scale=0.25]{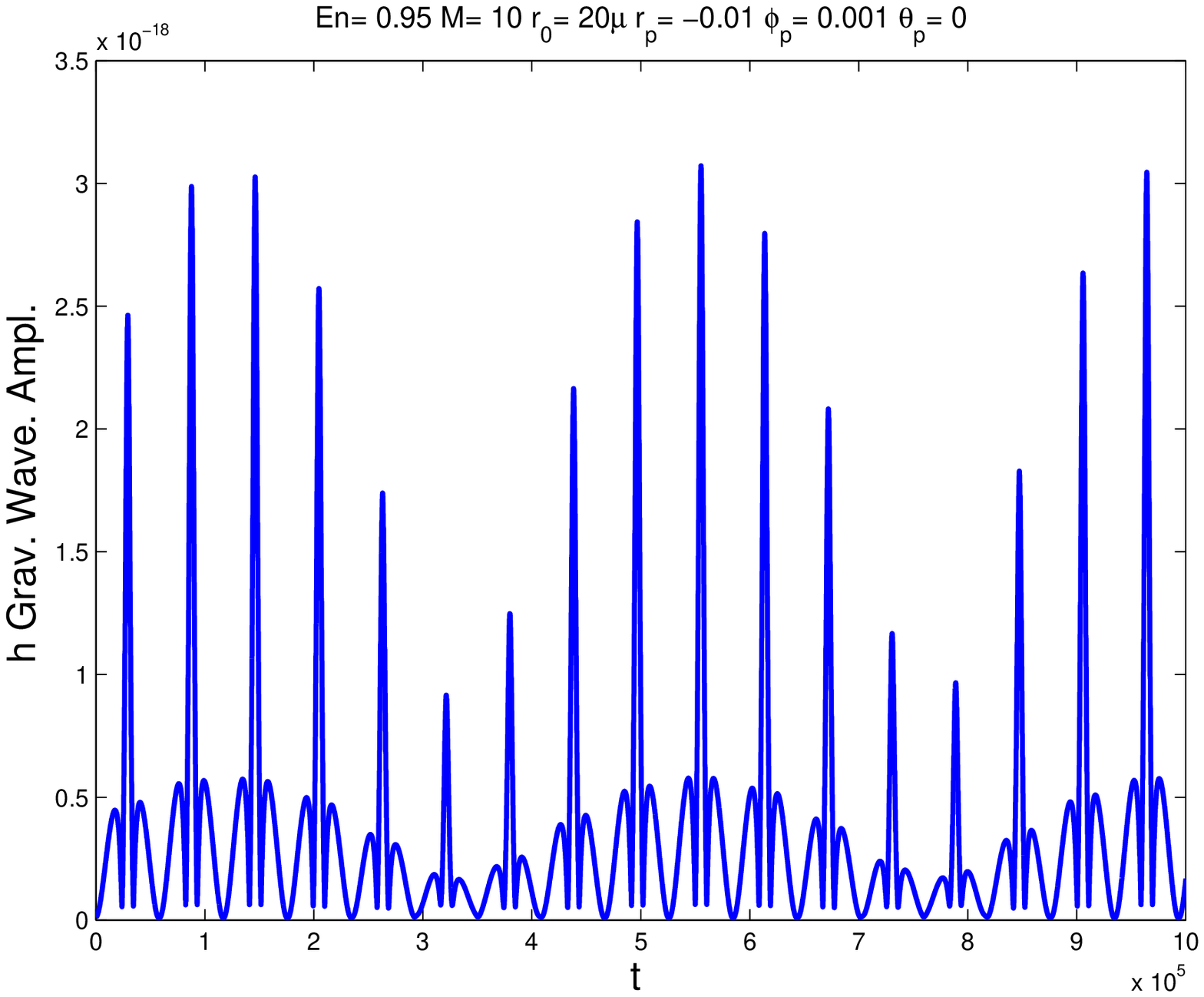}
\includegraphics[scale=0.25]{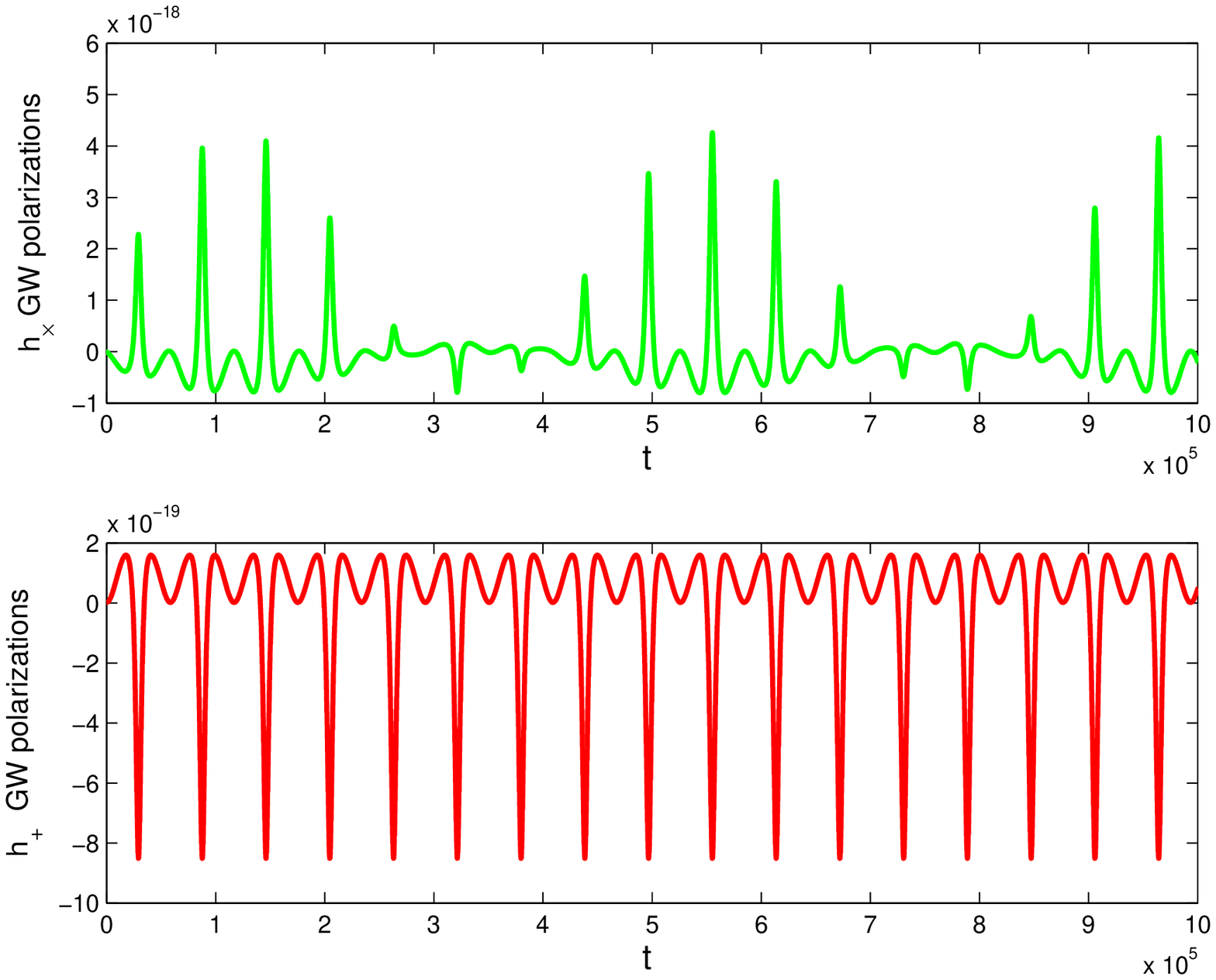}
 \tabularnewline
\hline
\includegraphics[scale=0.25]{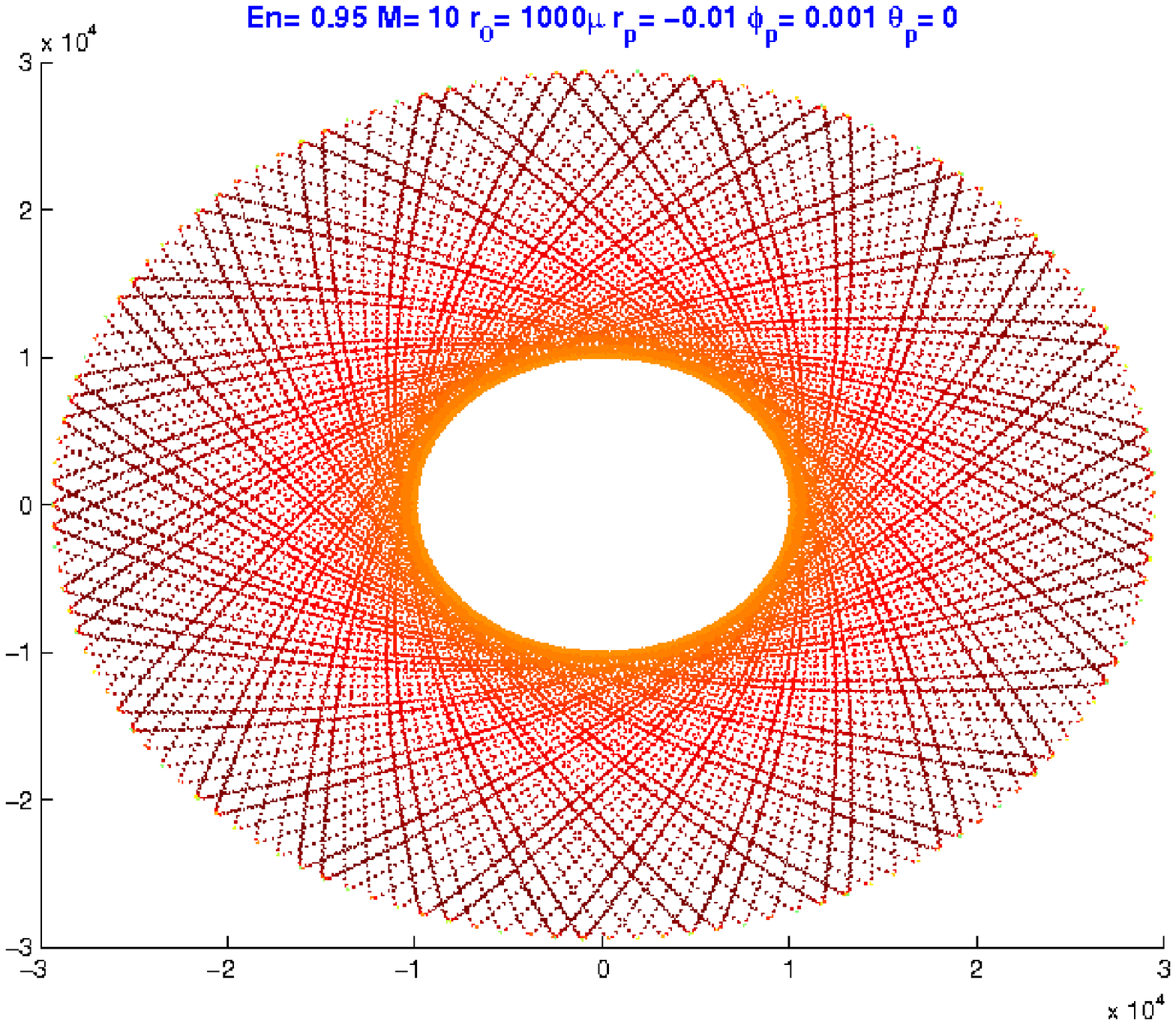}
\includegraphics[scale=0.25]{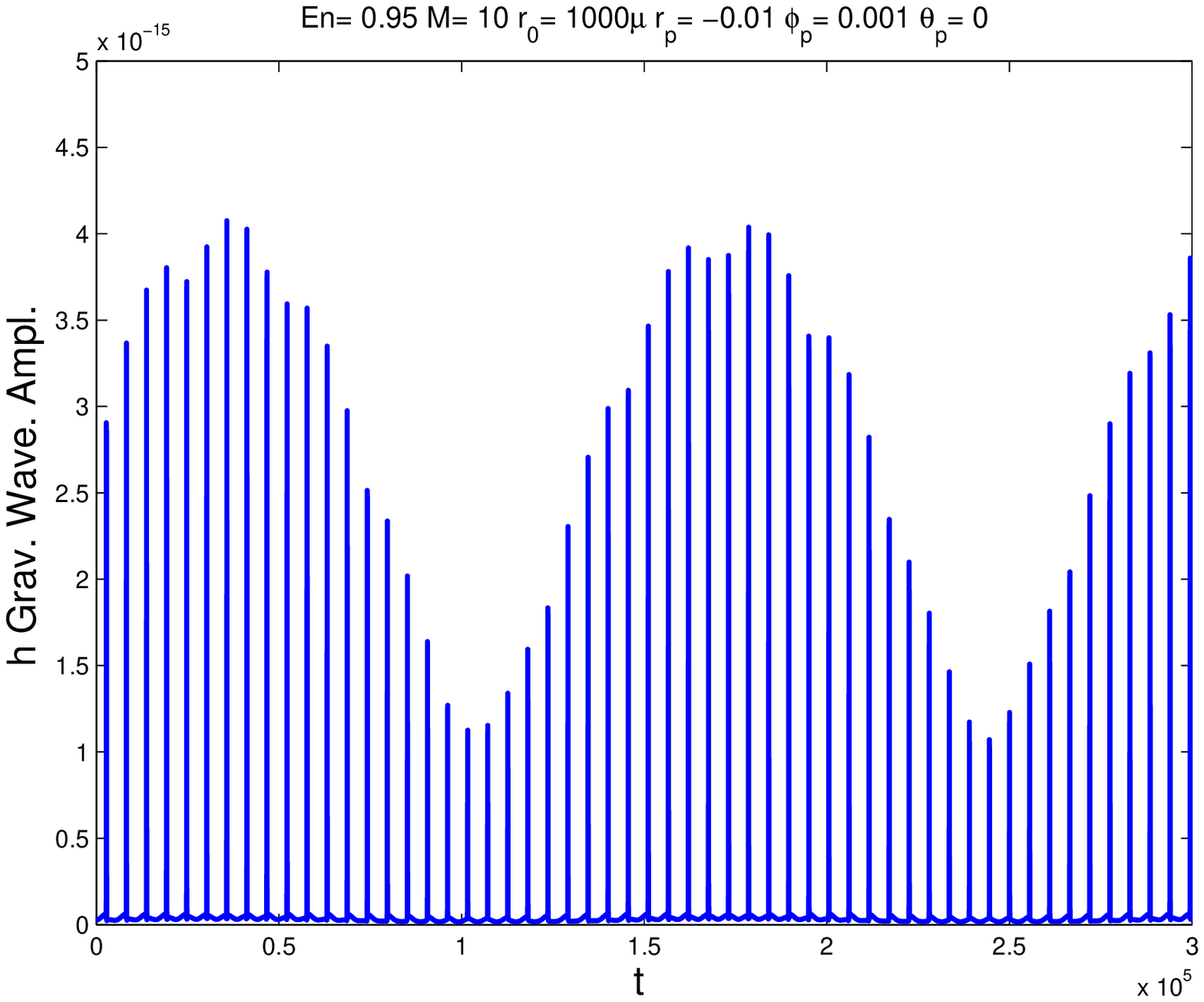}
\includegraphics[scale=0.25]{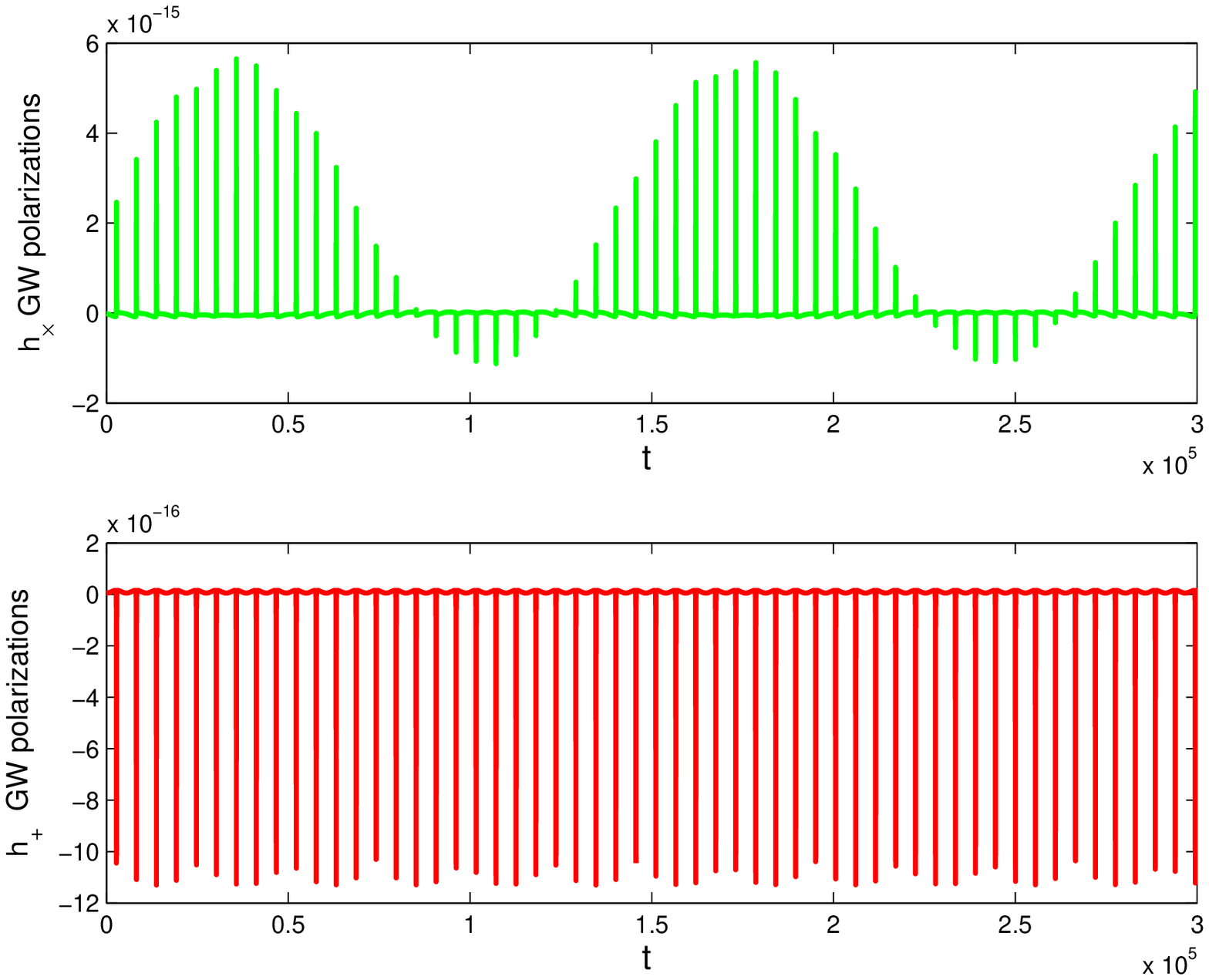}
\tabularnewline
\hline
 \tabularnewline
\includegraphics[scale=0.25]{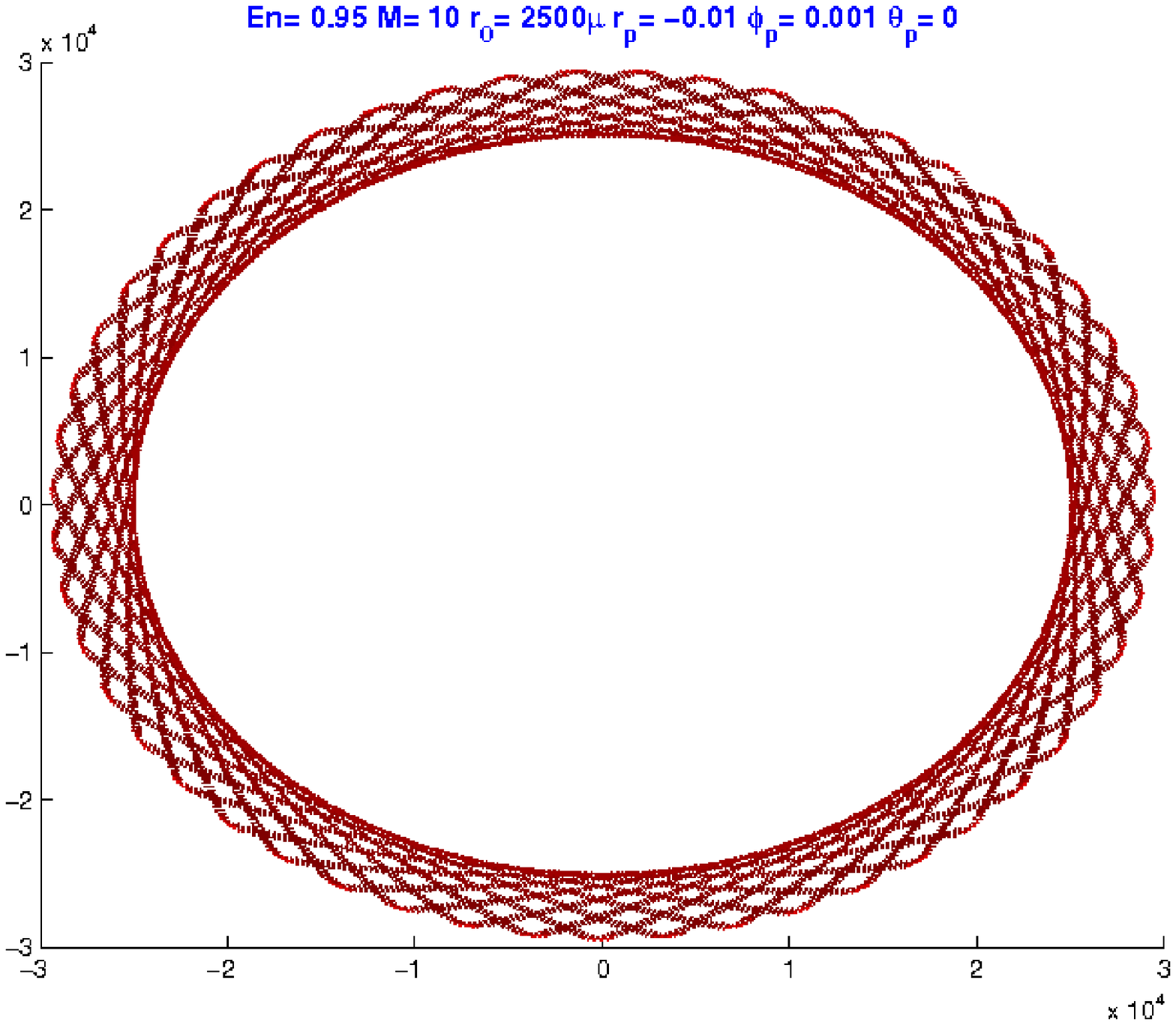}
\includegraphics[scale=0.25]{En_95_M_1.4_r0_2500_theta_p_0quadQ.eps}
\includegraphics[scale=0.25]{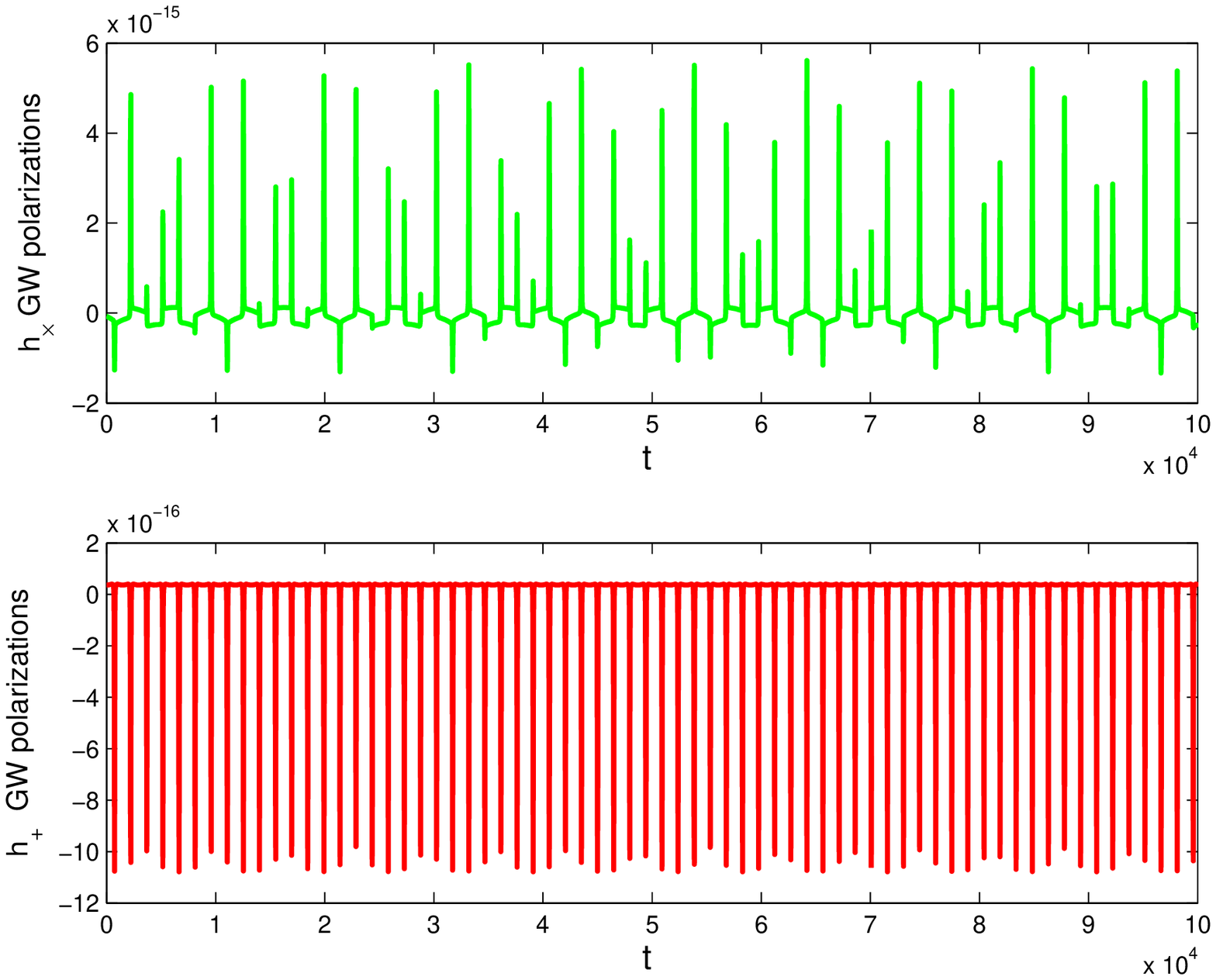}
\tabularnewline
\hline
\end{tabular}
\caption {Plots along the  panel lines from left to right of field
velocities along the axes of maximum covariances, total
gravitational emission waveform $h$ and gravitational waveform
polarizations $h_{+}$ and $h_{\times}$ for a Black Hole (BH) of
$10 M_{\odot}$. The waveform has been computed for the
Earth-distance to SgrA$^*$. The plots we are showing have been
obtained solving the system for the following parameters and
initial conditions: $\mu\approx10 M_{\odot}$,
$E=0.95$,$\phi_{0}=0$,
$\theta_{0}=\frac{\pi}{2}$,$\dot{\theta_{0}}=0$,$\dot{\phi_{0}}=-\frac{1}{10}\dot{r}_{0}$
and $\dot{r}_{0}=-\frac{1}{100}$. From top to bottom of the
panels, the  orbital radius is $r_0=20\mu,\,1000\mu,\,2500\mu$.
See also Table I}\label{Fig:07}
\end{figure*}

From Eq.(\ref{ampli1}), it is straightforward to show that, for a
binary system where $m\ll M$ and orbits have gravitomagnetic
corrections, the Cartesian components of GW-amplitude are
\begin{eqnarray*}
\nonumber
h^{xx}&=& 2 \mu [(3 \cos ^2\phi \sin ^2\theta-1)
   \dot{r}^2+6 r ( \dot{\theta}\cos ^2\phi \sin 2 \theta
   \\ &&-\dot{ \phi}\sin ^2\theta \sin2 \phi) \dot{r}
+r((3 \cos ^2\phi \sin ^2\theta-1)\ddot{r}\\ &&+3
   r (\dot{\theta}^2\cos2 \theta \cos ^2\phi -\dot{\phi} \dot{\theta} \sin2
   \theta \sin2\phi\\ &&
  -\sin\theta
   (\sin\theta(\dot{ \phi}^2\cos 2 \phi+\ddot{ \phi} \cos\phi \sin\phi)-\ddot{ \theta}\cos\theta \cos ^2\phi
)))],
      \end{eqnarray*}

\begin{eqnarray*}
\nonumber
h^{yy}&=&2 \mu[(3 \sin ^2\theta \sin ^2\phi-1)
    \dot{r}^2+6 r ( \dot{ \phi}\sin 2 \phi  \sin ^2\theta
   \\ &&+ \dot{\theta}\sin 2 \theta  \sin ^2\phi)
   \dot{r}+
   + r ((3 \sin ^2\theta  \sin ^2\phi
   -1) \ddot{r}\\ &&+3 r ( \dot{\theta}^2\cos 2 \theta  \sin ^2\phi
   +\dot{ \phi}
   \dot{\theta}\sin 2 \theta \sin 2 \phi
  \\ &&+ \sin \theta (\ddot{\theta}\cos \theta
   \sin ^2\phi+\sin \theta ( \dot{ \phi}^2\cos 2 \phi +\ddot\phi\cos \phi \sin \phi ))))],
\end{eqnarray*}

\begin{eqnarray*}
\nonumber
h^{xy}&=&h^{yx}=3 \mu[\cos 2 \phi \sin \theta(4\dot{\theta} \dot{ \phi} \cos \theta
    + \ddot{ \phi}\sin \theta )
   r^2\\ &&+2 \dot{r} (2 \dot{ \phi} \cos 2 \phi  \sin ^2\theta
   +\dot{\theta}\sin 2 \theta \sin 2 \phi )r
   \\ &&+\frac{1}{2} \sin 2 \phi (2 \ddot{r} \sin ^2\theta
   +r( 2\dot{\theta}^2 \cos 2 \theta-4  \dot{ \phi}^2\sin ^2\theta
   \\ &&+\ddot{\theta}\sin 2 \theta ))
   r+ \dot{r}^2\sin ^(\theta \sin 2 \phi],
\end{eqnarray*}

where we are assuming geometrized units. The above formulas have
been obtained from Eqs.(\ref{ddr}), (\ref{ddphi}), (\ref{ddtheta})
The  gravitomagnetic corrections give rise to signatures on the
GW-amplitudes that, in the standard Newtonian orbital motion, are
not present (see for example \cite{CDDIN,SF}). On the
other hand, as discussed in the Introduction, such corrections
cannot be discarded in peculiar situations as dense stellar
clusters or in the vicinity of galaxy central regions.
We are going to evaluate these
quantities and results are shown in Figs. \ref{Fig:03},
\ref{Fig:04}, \ref{Fig:05}, \ref{Fig:07}.

\subsection{Numerical results}
\begin{figure}[!ht]
\begin{tabular}{|c|}
\hline
\tabularnewline
\includegraphics[scale=0.5]{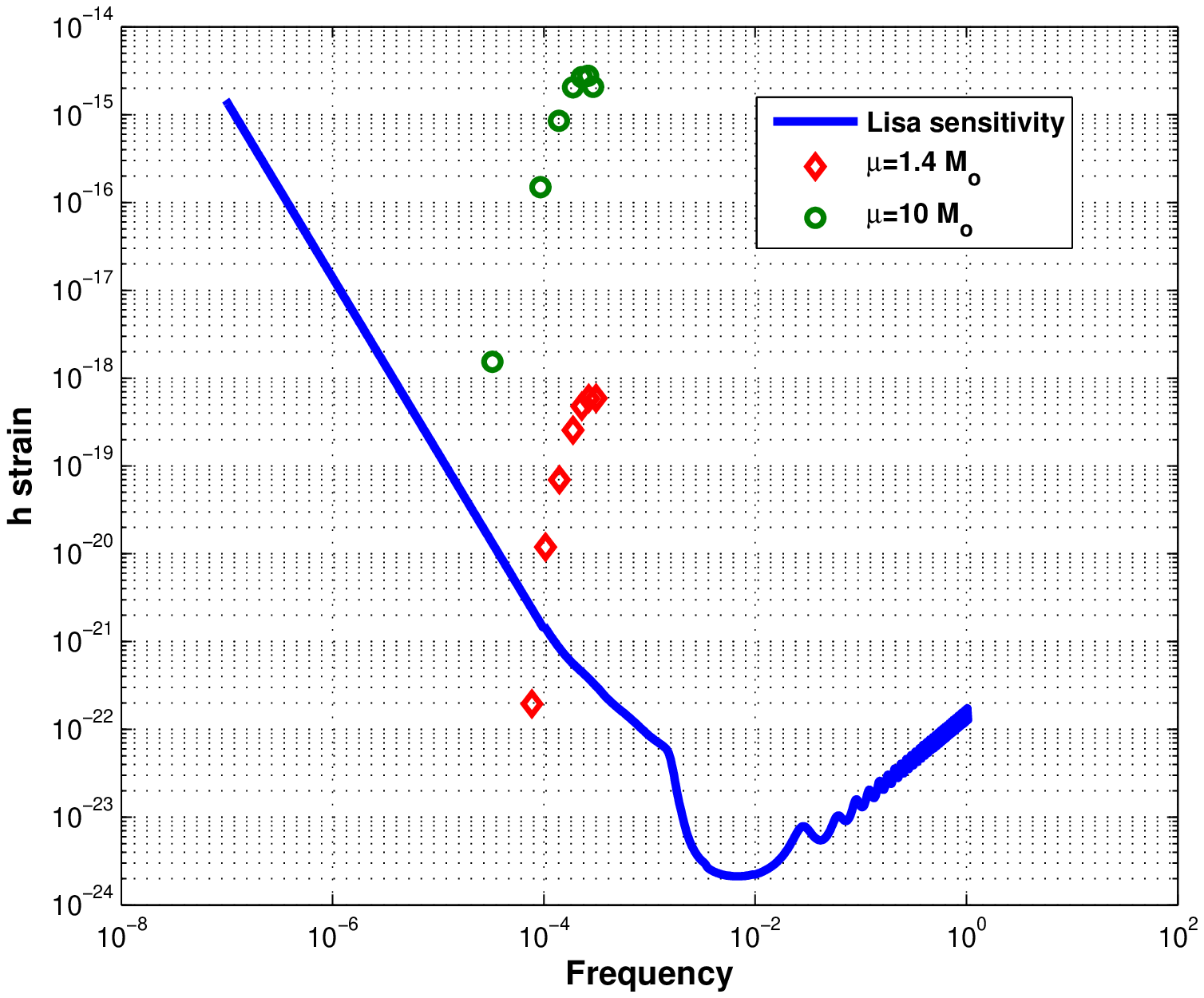}
 \tabularnewline
\hline
\end{tabular}
\caption {Plot of estimated  mean values of GW-emission in terms
of strain $h$ for two binary sources at the Galactic Center
SgrA$^*$ with reduced mass  $\mu\approx1.4M_{\odot}$ (red
diamonds) and $\mu\approx 10M_{\odot}$(green circles). The blue
line is the foreseen LISA sensitivity curve. The waveforms have
been computed for the Earth-distance to SgrA$^*$. The examples we
are showing have been obtained solving the systems for the
parameters and initial conditions reported in  Figs. \ref{Fig:05},
\ref{Fig:07}  and in Table I.}\label{Fig:09}
\end{figure}

Now we have all the ingredients to estimate the effects of
gravitomagnetic corrections on the GW-radiation. Calculations have
been performed in geometrized units in order to evaluate better
the relative corrections in absence of gravitomagnetism. For the
numerical simulations, we have assumed  the fiducial systems
constituted by a $m=1.4M_{\odot}$ neutron star or $m=10M_{\odot}$
massive stellar object orbiting around a MBH $M\simeq 3\times
10^6M_{\odot}$ as  SgrA$^*$. In the extreme mass-ratio limit, this
means that we can consider ${\displaystyle \mu=\frac{mM}{m+M}}$ of
about $\mu \approx 1.4M_{\odot}$  and $\mu \approx10M_{\odot}$.
The computations have been performed  starting with orbital radii
measured in the  mass unit and scaling the distance according to
the values shown in Table I. As it is possible to see in Table I,
starting from $r_{0}=20\mu$ up to $2500\mu$,   the orbital
eccentricity ${\displaystyle
\bm{e}=\frac{r_{max}-r_{min}}{r_{max}+r_{min}}}$ evolves towards a
circular orbit. In  Table I,  the GW-frequencies, in $mHz$, as
well as the $h$ amplitude strains and the two polarizations
$h_{+}$ and $h_{\times}$ are shown. The values are the mean values
of the GW $h$ amplitude strains ($h=\frac{h_{max}+h_{min}}{2}$)
and the maxima of the polarization waves (see Figs. \ref{Fig:05}
and \ref{Fig:07}). In Fig. \ref{Fig:09},   the fiducial LISA
sensitivity curve is shown \cite{LISA} considering the confusion
noise produced by White Dwarf binaries (blue curve). We show also
the $h$ amplitudes (red diamond and green circles for $\mu\approx
1.4 M_{\odot}$ and $\approx 10 M_{\odot}$ respectively). It is
worth noticing that, due to very high Signal to Noise Ratio, the
binary systems which we are considering  result extremely
interesting, in terms of probability detection, for the LISA
interferometer (see Fig. \ref{Fig:09}).

\begin{table*}[!ht]
\caption{GW-amplitudes and frequencies as function of eccentricity
$e$, reduced mass $\mu$, orbital radius $r_0$ for the  two cases
of fiducial stellar objects  $m\simeq 1.4 M_{\odot}$ and $m\simeq
10 M_{\odot}$ orbiting around a MBH of mass $M\simeq 3\times
10^6M_{\odot}$.}
\begin{tabular}{|c|c|} \hline
\textbf{$1.4 M_{\odot} $}  & \textbf{$10 M_{\odot} $}\\
\begin{tabular}{|c|c|c|c|c|c}
  \hline
    $\frac{r_{0}}{\mu} $ & $  e $ &  $   f(mHz) $ & $ h $ & $  h_{+} $ & $  h_{\times} $\\
   \hline
  \hline
      $20  $ & $   0.91 $ &  $   7.7\cdot 10^{-2} $ & $ 2.0\cdot 10^{-22} $ & $  5.1\cdot 10^{-23} $ & $  5.1\cdot 10^{-22} $\\
         $200  $ & $  0.79 $ &  $1.1\cdot 10^{-1} $ & $ 1.2\cdot 10^{-20} $ & $  2.2\cdot 10^{-21} $ & $  3.1\cdot 10^{-20} $\\
         $500  $ & $  0.64 $ & $1.4\cdot 10^{-1}$ & $  6.9\cdot 10^{-20}$ & $   8.7\cdot 10^{-21}$ & $   1.7\cdot 10^{-19}$\\
        $1000  $ & $ 0.44 $ & $  1.9\cdot 10^{-1} $ & $ 2.6\cdot 10^{-19} $ & $  6.4\cdot 10^{-20}  $ & $ 6.4\cdot 10^{-19} $\\
        $1500  $ & $ 0.28 $ & $  2.3\cdot 10^{-1} $ & $ 4.8\cdot 10^{-19} $ & $  3.6\cdot 10^{-20} $ & $  1.2\cdot 10^{-18} $\\
        $2000  $ & $ 0.14 $ & $  2.7\cdot 10^{-1} $ & $ 5.9\cdot 10^{-19} $ & $   4.9\cdot 10^{-20} $ & $  1.3\cdot 10^{-18} $\\
        $2500   $ & $ 0.01 $ & $   3.1\cdot 10^{-1} $ & $ 5.9\cdot 10^{-19} $ & $  1.7\cdot 10^{-20} $ & $  9.2\cdot 10^{-19} $\\
\end{tabular}
&
\begin{tabular}{c|c|c|c|c|}
  \hline
   $ e $ &  $   f(mHz) $ & $ h $ & $  h_{+} $ & $  h_{\times} $\\
   \hline
  \hline
      $0.98 $ &  $   3.2\cdot 10^{-2} $ & $ 1.5\cdot 10^{-18} $ & $  1.6\cdot 10^{-19} $ & $  4.3\cdot 10^{-18} $\\
      $0.87 $ & $   9.2\cdot 10^{-2} $ & $ 1.5\cdot 10^{-16} $ & $  2.5\cdot 10^{-18} $ & $  4.1\cdot 10^{-16} $\\
      $0.71 $ & $  1.4\cdot 10^{-1}$ & $  8.5\cdot 10^{-16}$ & $   7.0\cdot 10^{-18}$ & $   2.4\cdot 10^{-15}$\\
      $ 0.49 $ & $  1.9\cdot 10^{-1} $ & $ 2.0\cdot 10^{-15} $ & $  1.6\cdot 10^{-17}  $ & $ 5.6\cdot 10^{-15} $\\
      $ 0.32 $ & $  2.3\cdot 10^{-1} $ & $ 2.7\cdot 10^{-15} $ & $   2.5\cdot 10^{-17} $ & $  7.4\cdot 10^{-15} $\\
      $ 0.19 $ & $  2.6\cdot 10^{-1} $ & $ 2.8\cdot 10^{-15} $ & $   3.3\cdot 10^{-17} $ & $  7.6\cdot 10^{-15} $\\
       $ 0.08 $ & $   2.9\cdot 10^{-1} $ & $ 2.1\cdot 10^{-15} $ & $  4.0\cdot 10^{-17} $ & $  5.6\cdot 10^{-15} $\\

  \end{tabular}
  \\
 \hline
\end{tabular}
\end{table*}

\section{Rate and event number estimations in dense stellar systems}
At this point, it is important to give some estimates of the
number of events where gravitomagnetic effects could be a
signature for orbital motion and  gravitational radiation.  From
the GW emission point of view, close orbital encounters,
collisions and tidal interactions have to be dealt on average if
we want to investigate the gravitational radiation in a dense
stellar system. On the other hand, dense stellar regions are the
favored target for LISA interferometer \cite{freitag} so it is
extremely useful to provide suitable numbers before its launching.

To this end, it is worth giving the  stellar encounter rate
producing GWs in  astrophysical systems like dense globular
clusters or  the Galactic Center. In general, stars are
approximated as point masses. However, in dense regions of stellar
systems,  a star can pass so close to another that they raise
tidal forces which dissipate their relative orbital kinetic energy
and the Newtonian mechanics or the weak field limit of GR cannot
be adopted as  good approximations. In some cases, the loss of
energy can be so large that stars form binary (the situation which
we have considered here)  or multiple systems; in other cases, the
stars collide and coalesce into a single star; finally stars can
exchange gravitational interaction in non-returning encounters.

To investigate and parameterize all these effects, one has to
compute the collision time $t_{coll}$, where $1/t_{coll}$ is the
collision rate, that is, the average number of physical collisions
that a given star suffers per unit time. As a rough approximation,
one can restrict to stellar clusters in which all stars have the
same mass $m$.

Let us consider an encounter with initial relative velocity
$\mathbf{v}_{0}$ and impact parameter $b$. The angular momentum
per unit mass of the reduced particle is $L=bv_{0}$. At the
distance of closest approach, which we denote by $r_{coll}$, the
radial velocity must be zero, and hence the angular momentum is
$L=r_{coll}v_{max}$, where $v_{max}$ is the relative speed at
$r_{coll}$. It is easy to show that \cite{binney}

\begin{equation}
b^{2}=r_{coll}^{2}+\frac{4Gmr_{coll}}{v_{0}^{2}}\,.\label{eq:b}\end{equation}
If we set $r_{coll}$ equal to the sum of the radii of the two
stars,  a collision will occur if  the impact parameter is less
than the value of $b$, as determined by Eq.(\ref{eq:b}).

The function $f(\mathbf{v}_{a})d^{3}\mathbf{v}_{a}$ gives the
number of stars per unit volume with velocities in the range
$\mathbf{v}_{a}+d^{3}\mathbf{v}_{a}.$ The number of encounters per
unit time with impact parameter less than $b$, which are suffered
by a given star, is  $f(\mathbf{v}_{a})d^{3}\mathbf{v}_{a}$ times
the volume of the annulus with radius $b$ and length $v_{0}$, that
is,

\begin{equation}
\int f(\mathbf{v}_{a})\pi
b^{2}v_{0}d^{3}\mathbf{v}_{a},\label{eq:integrale}\end{equation}
where $v_{0}=\left|\mathbf{v-v}_{a}\right|$ and $\mathbf{v}$ is
the velocity of the considered star. The quantity in
Eq.(\ref{eq:integrale}) is equal to $1/t_{coll}$ for a star with
velocity $\mathbf{v}$: to obtain the mean value of $1/t_{coll}$,
we average over $\mathbf{v}$ by multiplying (\ref{eq:integrale})
by $f(\mathbf{v})/\nu$, where $\nu=\int
f(\mathbf{v})d^{3}\mathbf{v}$ is the number density of stars and
the integration is over $d^{3}\mathbf{v}$. 
Thus
\begin{eqnarray}
\frac{1}{t_{coll}}&=& \frac{\nu}{8\pi^{2}\sigma^{6}}\int
e^{-(v^{2}+v_{a}^{2})/2\sigma^{2}}\times\nonumber\\&&\left(r_{coll}\left|\mathbf{v-v}_{a}\right|+
\frac{4Gmr_{coll}}{\left|\mathbf{v-v}_{a}\right|}\right)d^{3}\mathbf{v}d^{3}\mathbf{v}_{a}\,.
\label{eq:invtcoll}\end{eqnarray} Replacing the variable
$\mathbf{v}_{a}$ by $\mathbf{V}=\mathbf{v-v}_{a}$, the argument of
the exponential is then
$-\left[\left(\mathbf{v}-\frac{1}{2}\mathbf{V}\right)^{2}+\frac{1}{4}V^{2}\right]/\sigma^{2}$,
and if we replace the variable $\mathbf{v}$ by ${\displaystyle
\mathbf{v}_{cm}=\mathbf{v}-\frac{1}{2}\mathbf{V}}$ (the center of
mass velocity), then one has

\begin{equation}
\frac{1}{t_{coll}}=\frac{\nu}{8\pi^{2}\sigma^{6}} \int
e^{-(v_{cm}^{2}+V^{2})/2\sigma^{2}}\left(r_{coll}V+
\frac{4Gmr_{coll}}{V}\right)dV\,.\label{eq:invtcoll1}\end{equation}
The integral over $\mathbf{v}_{cm}$ is given by

\begin{equation}
\int
e^{-v_{cm}^{2}/\sigma^{2}}d^{3}\mathbf{v}_{cm}=\pi^{3/2}\sigma^{3}\,.\label{eq:intint}\end{equation}
Thus

\begin{equation}
\frac{1}{t_{coll}}=\frac{\pi^{1/2}\nu}{2\sigma^{3}}\int_{\infty}^{0}e^{-V^{2}/4\sigma^{2}}
\left(r_{coll}^{2}V^{3}+4GmVr_{coll}\right)dV\label{eq:invtcoll2}\end{equation}
The integrals can be  easily calculated and then we find

\begin{equation}
\frac{1}{t_{coll}}=4\sqrt{\pi}\nu\sigma
r_{coll}^{2}+\frac{4\sqrt{\pi}\nu
Gmr_{coll}}{\sigma}\,.\label{eq:invtcooll3}\end{equation} The
first term of this result can be derived from the kinetic theory.
The rate of interaction is $\nu\Sigma\left\langle V\right\rangle$,
where $\Sigma$ is the cross-section and $\left\langle
V\right\rangle $ is the mean relative speed. Substituting
$\Sigma=\pi r_{coll}^{2}$ and $\left\langle V\right\rangle
=4\sigma/\sqrt{\pi}$ (which is appropriate for a Maxwellian
distribution with dispersion $\sigma$) we recover the first term
of (\ref{eq:invtcooll3}). The second term represents the
enhancement in the collision rate by gravitational focusing, that
is, the deflection of trajectories by the gravitational attraction
of the two stars.

If $r_{*}$ is the stellar radius, we may set $r_{coll}=2r_{*}$. It
is convenient to introduce the escape speed from stellar surface,
${\displaystyle v_{*}=\sqrt{\frac{2Gm}{r_{*}}}}$, and to rewrite
Eq.(\ref{eq:invtcooll3}) as

\begin{equation}
\Gamma=\frac{1}{t_{coll}}=16\sqrt{\pi}\nu\sigma
r_{*}^{2}\left(1+\frac{v_{*}^{2}}{4\sigma^{2}}\right)=16\sqrt{\pi}\nu\sigma
r_{*}^{2}(1+\Theta),\label{eq:invtcoll4}\end{equation}
where

\begin{equation}
\Theta=\frac{v_{*}^{2}}{4\sigma^{2}}=\frac{Gm}{2\sigma^{2}r_{*}}\label{eq:safronov}\end{equation}
is the Safronov number \cite{binney}. In evaluating the rate, we
are considering only those  encounters producing gravitational
waves, for example,  in the LISA range, i.e. between $10^{-4}$ and
$10^{-1}$ Hz (see e.g. \cite{Rub}). Numerically, we have
\begin{eqnarray}
&&\Gamma \simeq  5.5\times 10^{-10} \left(\frac{v}{10 {\rm km
s^{-1}}}\right) \left(\frac{\sigma}{UA^2}\right) \left(\frac{{\rm
10 pc}}{R}\right)^3 {\rm
yrs^{-1}}\nonumber\\ && \qquad\Theta<<1\label{eq:thetamin}
\end{eqnarray}
\begin{eqnarray}
&&\Gamma \simeq  5.5\times 10^{-10} \left(\frac{M}{10^5 {\rm
M_{\odot}}}\right)^2 \left(\frac{v}{10 {\rm km s^{-1}}}\right)
\left(\frac{\sigma}{UA^2}\right)\times\nonumber\\ && \left(\frac{{\rm 10
pc}}{R}\right)^3 {\rm yrs^{-1}}\qquad\Theta>>1\label{eq:thetamagg}
\end{eqnarray}
If $\Theta>>1$, the energy dissipated exceeds the relative kinetic
energy of the colliding stars, and the stars  coalesce into a
single star. This new star may, in turn, collide and merge with
other stars, thereby becoming very massive. As its mass increases,
the collision time is shorten and then there may be runaway
coalescence leading to the formation of a few supermassive objects
per clusters. If $\Theta<<1$, much of the mass in the colliding
stars may be liberated and forming new stars or a single
supermassive objects (see \cite{Belgeman,Shapiro}). Both cases are
interesting for LISA purposes.

Note that when one has the effects of quasi-collisions (where
gravitomagnetic effects, in principle, cannot be discarded) in an
encounter of two stars in which the minimal separation is several
stellar radii, violent tides will raise on the surface of each
star. The energy that excites the tides comes from the relative
kinetic energy of the stars. This effect is important for
$\Theta>>1$ since the loss of small amount of kinetic energy may
leave the two stars with negative total energy, that is, as a
bounded  binary system. Successive peri-center passages will
dissipates more energy by GW radiation, until the binary orbit is
nearly circular with a negligible or null GW radiation emission.

Let us apply these considerations to the Galactic Center which can
be modelled as a system of several compact stellar clusters, some
of them similar to very compact globular clusters with high
emission in X-rays \cite{townes}.

For a typical globular cluster  around the Galactic Center, the
expected event rate is of the order of $2\times 10^{-9}$
yrs$^{-1}$ which may be increased at least by a factor $\simeq
100$ if one considers the number of globular clusters in the whole
Galaxy. If the stellar cluster at the Galactic Center is taken
into account and assuming the total mass $M\simeq 3\times 10^6$
M$_{\odot}$, the velocity dispersion $\sigma\simeq $ 150 km
s$^{-1}$ and the radius of the object $R\simeq$ 10 pc (where
$\Theta=4.3$), one expects to have $\simeq 10^{-5}$ open orbit
encounters per year. On the other hand, if a cluster with total
mass $M\simeq 10^6$ M$_{\odot}$, $\sigma\simeq $ 150 km s$^{-1}$
and $R\simeq$ 0.1 pc is considered, an event rate number of the
order of unity per year is obtained. These values could be
realistically achieved by data coming from the forthcoming space
interferometer LISA. As a secondary effect, the above  wave-forms
could constitute the "signature"  to classify the different
stellar encounters  thanks to the differences of the shapes (see
Figs. \ref{Fig:05} and \ref{Fig:07}).

\section{Discussion, conclusions and perspectives}
We have considered  the two-body problem in Newtonian and relativistic theory of orbits in view of characterizing the gravitational radiation, starting from the motion of the sources.
We have reported several results concerning the equations
of motion, and the associated Lagrangian formulation, of compact
binary systems. These equations are necessary when constructing
the theoretical templates for searching and analyzing the GW
signals from inspiralling compact binaries in VIRGO-LIGO and LISA
type experiments. Considering the two-body problem, we mean the
problem of the dynamics of two structureless, non-spinning
point-particles, characterized by solely two mass parameters $m_1$
and $m_2$, moving under their mutual, purely gravitational
interaction. Surely this problem, because of its conceptual
simplicity, is among the most interesting ones to be solved within
any theory of gravity. Actually, there are two aspects of the
problem: the first sub-problem consists into obtaining the equation
of the binary motion, the second is to find the (hopefully exact)
solution of that equation. We referred to the equation of motion
as the explicit expression of the acceleration of each of the
particles in terms of their positions and velocities. It is well
known that in Newtonian gravity, the first of these sub-problems
is trivial, as one can easily write down the equation of motion
for a system of $N$ particles, while the second one is difficult,
except in the two-body case $N = 2$, which represents, in fact, the
only situation amenable to an exact treatment of the solution. In
GR, even writing down the equations of motion in the simplest case
$N = 2$ is difficult. Unlike in Newton's theory, it is impossible
to express the acceleration by means of the positions and
velocities, in a way which would be valid within the {\it exact}
theory. Therefore we are obliged to resort to approximation
methods. Let us feel reassured that plaguing the exact theory of
GR with approximation methods is not a shame. It is fair to say
that many of the great successes of this theory, when confronted
to experiments and observations, have been obtained thanks to
approximation methods. Furthermore, the beautiful internal wheels
of GR also show up when using approximation
methods, which often deserve some theoretical interest in their
own, as they require interesting mathematical techniques. Here we
have investigated the equation of the binary motion in the
post-Newtonian approximation, {\it i.e.} as a formal expansion
when the velocity of light $c$ tends to infinity. As a consequence
of the equivalence principle, which is incorporated by hand in
Newton's theory and constitutes the fundamental basis of GR, the acceleration of $particle1$ should not depend on $m_1$ (nor
on its internal structure), in the {\it test-mass} limit where the
mass $m_1$ is much smaller than $m_2$. This is, of course, satisfied
by the Newtonian acceleration, which is independent of $m_1$, but
this leaves the possibility that the acceleration of the $particle
1$, in higher approximations, does depend on $m_1$, via the
so-called self-forces, which vanish in the test-mass limit.
Indeed, this is what happens in the post-Newtonian  and
gravitomagnetic corrections, which show explicitly many
terms proportional to (powers of) $m_1$. Though the approximations
and corrections to the orbits are really a consequence of GR, they
should be interpreted using the common-sense language of Newton.
That is, having chosen a convenient general-relativistic
(Cartesian) coordinate system, like the harmonic coordinate system
adopted above, we have express the results in terms of the
coordinate positions, velocities and accelerations of the bodies.
Then, the trajectories of the particles can be viewed as taking
place in the absolute Euclidean space of Newton, and their
(coordinate) velocities as being defined with respect to absolute
time. Not only this interpretation is the most satisfactory one
from a conceptual point of view, but it represents also the most
convenient path for comparing the theoretical predictions and the
observations. For instance, the Solar System dynamics at the first
post-Newtonian level is defined, following a recent resolution of
the International Astronomical Union, in a harmonic coordinate
system, the Geocentric Reference System (GRS), with respect to
which one considers the {\it absolute} motion of the planets and
satellites. But because the equations come from GR, they are
endowed with the following properties, which make them truly {\it relativistic}.
 \begin{itemize}
 \item The one-body problem in GR corresponds to the Schwarzschild solution, so the
equations possess the correct {\it perturbative} limit, that given by the geodesics of the Schwarzschild
metric, when the mass of one of the bodies tends to zero.
 \item Because GR admits the Poincar\' e group as a global symmetry (in the case of
asymptotically flat space-times), the harmonic-coordinate equations of motion stay invariant when
we perform a global Lorentz transformation
\item Since the particles emit gravitational radiation there are some terms in the equations which
are associated with radiation reaction. These terms appear 
 at the order $2.5$PN or  $c^-5$
that we discarded in our discussion (where $5 = 2s + 1$, $s = 2$
being the helicity of the graviton). They correspond to an {\it
odd}- order PN correction, which does not stay invariant in a time
reversal. By contrast,  the {\it even}-orders, as
1PN,  correspond to a dynamics which is conservative.
\item GR  is a non-linear theory (even in vacuum), and some part of the gravitational
radiation which was emitted by the particles in the past scatters off the static gravitational field
generated by the rest-masses of the particles, or interacts gravitationally with itself.
 \end{itemize}
From all these considerations, the post-Newtonian equations were
also obtained, for the motion of the centers of mass of extended
bodies, using a technique that can be qualified as more {\it
physical} than the surface-integral method, as it takes explicitly
into account the structure of the bodies.

Particularly interesting is considering
gravitomagnetic effects in the geodesic motion. In particular, one can
consider the orbital effects of higher-order terms in $v/c$
which is the main difference with respect to the standard approach
to the gravitomagnetism. Such terms are often discarded but, as we
have shown, they could give rise to interesting phenomena in tight
binding systems as binary systems of evolved objects (neutron
stars or black holes). They could be important for objects falling
toward extremely massive black holes as those seated in the
galactic centers \cite{cutler,cutler1}. The leading parameter for
such correction is the ratio $v/c$ which, in several physical
cases cannot be simply discarded. For a detailed discussion see
for example \cite{capozzlamb,capozzre,sereno,sereno1}. A part the
standard periastron precession effects, such terms induce
nutations and are capable of affecting the stability basin of the
orbital phase space. As shown, the global structure of such a
basin is extremely sensitive to the initial angular velocities,
the initial energy and  mass conditions which can determine
possible transitions to chaotic behaviors. Detailed studies on the
transition to chaos could greatly aid in gravitational wave
detections in order to determine the shape, the spectrum and the
intensity of the waves (for  a discussion see \cite{levin,gair}).

In the second part of this review, we have summarized many of the
most important topics in the theory of GWs. Linearized theory as
described in  is adequate to describe the propagation of GWs and
to model the interaction of GWs with our detectors.  A variety of
formalisms have been developed.
\begin{itemize}

\item {\it Newtonian theory} The emission of gravitational waves from stellar encounters in Newtonian regime interacting on elliptical, hyperbolic and parabolic orbits is studied in the quadrupole approximation. Analytical expressions are then derived for the gravitational wave luminosity, the total energy output and gravitational radiation amplitude produced in tight impacts where two massive objects closely interact at an impact distance of $1AU$.

\item {\it Post-Newtonian theory.}  PN theory is one of the most
important of these formalisms, particularly for modeling binary
systems.  Roughly speaking, PN theory analyzes sources using an
iterated expansion in two variables: The ``gravitational
potential'', $\Phi \sim M/r$, where $M$ is a mass scale and $r$
characterizes the distance from the source; and velocities of
internal motion, $v$.  (In linearized theory, we assume $\Phi$ is
small but place no constraints on $v$.)  Newtonian gravity emerges
as the first term in the expansion, and higher order corrections
are found as the expansion is iterated to ever higher order.  Our
derivation of the quadrupole formula  gives the leading order term
in the PN expansion of the emitted radiation.  See  \cite{lb02}
and references therein for a comprehensive introduction to and
explication of this subject.

\item {\it Gravitomagnetic corrections.} The gravitomagnetic effect could give rise to interesting phenomena in tight
binding systems such as binaries of evolved objects (NSs or BHs).
The effects reveal particularly interesting if
$\displaystyle{\frac{v}{c}}$ is in the range
$\displaystyle{(10^{-1} \div 10^{-3})c}$. They could be important
for objects captured and falling toward extremely massive black
holes such as those at the Galactic Center. Gravitomagnetic
orbital corrections, after long integration time, induce
precession and nutation and then modification on the wave-form. In
principle, GW emission could present signatures of gravitomagnetic
corrections after suitable integration times in particular for the
on going LISA space laser interferometric GW antenna.
 \end{itemize}
 
 To conclude,  Henri Poincar\'e \cite{poincare} once remarked that
real problems can never be classified as solved or unsolved ones,
but that they are always {\it more and less solved}. This remark
applies particularly well to the problem of motion, which has had
chequered history. Even the Newtonian problem of motion, which
appeared to well understood after the development of the powerful
methods of classical mechanics \cite{tisserand} embarked on an
entirely new career after work of Poincar\'e which has led to many
further developments (see \cite{arnold,gallavotti}). The
Einsteinian problem of motion has not even reached a classical
stage where the basic problems appear as well understood. At first
sight the best developed approximation method in GR, the PN one,
would seem to constitute such classical stage, but the literature
on the PN problem of motion is full of repetitions, errors or
ambiguities. We was to conclude this review by giving a list of
issues that need to be clarified. We renounced this project
because, if one wishes to look at the work done with a critical
eye, nearly all aspects of the problem of motion and GWs need to
be thoroughly re-investigates for mathematical, physical or
conceptual reasons; so that the list of open problems would,
consistent with the remark of Poincar\' e. One thing is certain:
the  problem of motion and GWs is no longer a purely theoretical
problem, tanks to the  improvement in  the precision of positions
measurements in the solar systems, and to the discovery of the
binary pulsar 1913+16 which is a relativistic laboratory; the
problem has become an important tool of modern astrophysics. It is
therefore of some urgency, not only to complete and unify the work
already done, but also to develop new approaches in order to aim both
formal and conceptual clarification of the basic issues, and to obtain more accurate explicit results.

\section{Acknowledgments}
We thank S. Capozziello and L. Milano for fruitful discussions and for the useful suggestions  on the topics of this review.

\end{document}